\documentclass[%
 reprint,
 superscriptaddress,
 amsmath,amssymb,
 aps,pra,
]{revtex4-1}

\usepackage{graphicx}
\usepackage{dcolumn}
\usepackage{bm}
\usepackage{braket}
\usepackage{textcomp}
\usepackage{siunitx}
\usepackage{enumitem}
\usepackage{esvect}
\usepackage{xcolor}
\usepackage{bibunits}
\usepackage[version=4]{mhchem}

\newcommand\norm[1]{\left\lvert#1\right\rvert}

\begin{document}
\begin{bibunit}
\title{Towards long-distance quantum networks with superconducting processors and optical links}

\author{Sourabh~Kumar}
\email{sourabh.phys@gmail.com}
\affiliation{Institute for Quantum Science and Technology and Department of Physics
and Astronomy, University of Calgary, Calgary T2N 1N4, Alberta, Canada}
\author{Nikolai~Lauk}
\altaffiliation[Current address: ]{Division of Physics, Mathematics and Astronomy $\mathrm{\&}$ Alliance for Quantum Technologies, California Institute of Technology, 1200 East California Blvd., Pasadena, California 91125, USA}
\affiliation{Institute for Quantum Science and Technology and Department of Physics
and Astronomy, University of Calgary, Calgary T2N 1N4, Alberta, Canada}
\author{Christoph~Simon}
\affiliation{Institute for Quantum Science and Technology and Department of Physics
and Astronomy, University of Calgary, Calgary T2N 1N4, Alberta, Canada} 

\date{\today}%

\begin{abstract}
We design a quantum repeater architecture, necessary for long distance quantum networks, using the recently proposed microwave cat state qubits, formed and manipulated via interaction between a superconducting nonlinear element and a microwave cavity. These qubits are especially attractive for repeaters because in addition to serving as excellent computational units with deterministic gate operations, they also have coherence times long enough to deal with the unavoidable propagation delays. Since microwave photons are too low in energy to be able to carry quantum information over long distances, as an intermediate step, we expand on a recently proposed microwave to optical transduction protocol using excited states of a rare-earth ion ($\mathrm{Er^{3+}}$) doped crystal. To enhance the entanglement distribution rate, we propose to use spectral multiplexing by employing an array of cavities at each node. We compare our achievable rates with direct transmission and with two other promising repeater approaches, and show that ours could be higher in appropriate regimes, even in the presence of realistic imperfections and noise, while maintaining reasonably high fidelities of the final state. Thus, in the short term, our work could be directly useful for secure quantum communication, whereas in the long term, we can envision a large scale distributed quantum computing network built on our architecture. 
\end{abstract}

\maketitle


\section{\label{sec:intro}Introduction}

Quantum networks will enable applications such as secure quantum communication based on quantum key distribution \cite{84_Cryptography_Brassard, 02_Cryptography_Zbinden}, remote secure access to quantum computers based on blind quantum computation
\cite{09_Universal_Kashefi,12_Demonstration_Walther}, private database queries \cite{11_Practical_Zbinden}, more precise global timekeeping \cite{14_Quantum_Lukin} and telescopes \cite{12_Longer_Croke}, as well as fundamental tests of quantum non-locality and quantum gravity \cite{12_Fundamental_Terno}. If the total distance through which entanglement needs to be distributed is greater than a few hundred kilometres, then direct transmission through fibers leads to prohibitive loss. This can be overcome using quantum repeaters \cite{98_Quantum_Zoller,11_Quantum_Gisin}, which require small quantum processors and quantum memories at intermediate locations. The development of quantum computers has recently seen many impressive accomplishments \cite{16_Scalable_Martinis,17_Solving_Spiropulu,18_Fault_Schoelkopf,18_Quantum_Savage}. Networking quantum computers is of interest both in the medium term, when individual processors are likely to be limited in size due to technical constraints, and in the long term, where one can envision 
a full-fledged quantum internet \cite{08_Quantum_Kimble,18_Internet_Hanson,17_Towards_Simon}, which would be similar to the classical internet of today, but much more secure, and powerful in certain aspects. \par 

Out of different architectures for quantum computation being pursued \cite{05_Resource_Rudolph,02_Architecture_Wineland,07_SQ_Blais,99_QIP_Small,98_Silicon_Kane}, superconducting circuits are currently one of the leading systems. Easy addressability, high fidelity operations, and strong coupling strength with microwave photons \cite{04_Strong_Schoelkopf,10_CQED_Gross,17_Microwave_Nori,16_SQ_Semba} are some of their most attractive features. Until recently, one challenge for using superconducting processors in a network context were the relatively short coherence times of superconducting qubits (SQs) \cite{12_SQ_Steffen}. Long absolute coherence times are important for quantum networks because of the unavoidable time delays associated with propagation. However, a recent development could help circumvent this limitation. Instead of the states of a SQ, one can use the states of the coupled microwave cavity as logical qubits, and manipulate these cavity states using strong SQ-cavity interaction \cite{13_Hardware_Mirrahimi,14_Dynamically_Devoret,15_Confining_Devoret,16_Extending_Schoelkopf,16_Schrodinger_Schoelkopf,16_Universal_Goto,17_Engineering_Blais,16_Bifurcation_Goto,17_Robust_Tiwari,17_Quantum_Blais,18_CNOT_Schoelkopf}. The coherence time of these cavities can be much higher than 100 ms \cite{07_Ultrahigh_Visentin,13_Reaching_Schoelkopf,18_3D_Grassellino}, whereas that of the best SQ is only close to 0.1 ms \cite{12_SQ_Steffen}. Moreover, the infinite dimensional Hilbert space of a cavity can allow new error correction codes which are more resource efficient than the conventional multi-qubit codes \cite{01_Encoding_Preskill,14_Dynamically_Devoret,16_Extending_Schoelkopf,17_Degeneracy_Mirrahimi,18_Stabilized_Girvin}. Since photon loss is the only prominent 
error channel here, unlike other architectures, it is also relatively easier to keep track of these errors. \par 
 
Out of several cavity based architectures for quantum computation, the approach introduced in ref.\ \cite{16_Universal_Goto,17_Engineering_Blais} seems to be one of the most promising candidates because of the relative experimental ease in preparing and manipulating their system. They use coherent states of a microwave cavity as qubits. A nonlinear superconducting element comprising of Josephson junction(s) is placed inside the cavity, and the whole system is driven using a two-photon drive, which generates superpositions of coherent states, also known as cat states, and stabilises them against amplitude decay, even in the presence of single photon loss. Gates can be performed using detunings and additional single photon drives. Two or more cavities can be coupled easily, e.g.\ using a transmon \cite{14_Dynamically_Devoret,16_Schrodinger_Schoelkopf,18_CNOT_Schoelkopf, 19_Entanglement_Schoelkopf}.\par

\begin{figure*}[t]
\centering
\includegraphics[width=0.92\textwidth]{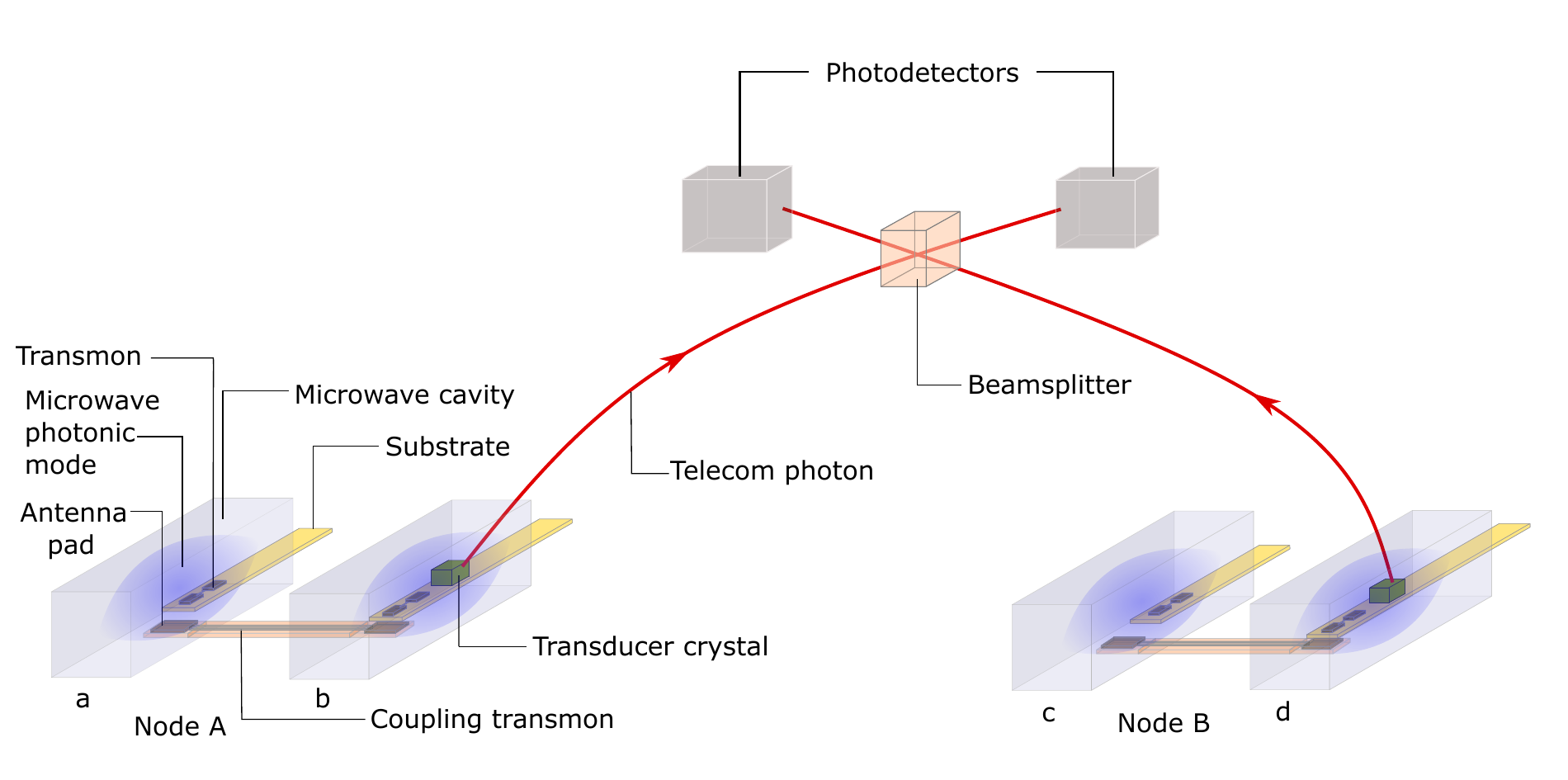}
\caption{(color online) Schematic diagram to demonstrate the generation of entanglement between distant 3-D microwave cavities. At each node, there is a pair of cavities, coupled using a transmon with two antenna pads, one in each cavity. All the cavities have a nonlinear superconducting element, e.g.\ another transmon, which provides the necessary Kerr nonlinearity for microwave photons. One of the pair of cavities at each node (cavity `a' in node A and cavity `c' in node B) acts as the stationary qubit, while microwave to optical transducers generate telecom wavelength flying qubits from the other cavity (cavity `b' in node A and cavity `d' in node B). The flying qubits are interfered on a beamsplitter located midway between the nodes, and single-photon detection events herald entanglement between the stationary qubits `a' and `c'.}\label{fig:schematic_entanglement}
\end{figure*}

Let us consider a possible scenario where the quantum processors at each node are based on the microwave cavity architecture discussed in ref.\ \cite{16_Universal_Goto,17_Engineering_Blais}. Since the energies of the microwave photons exiting these cavities are lower than the thermal noise at room temperature, they cannot be used to carry quantum information over long distances. We need quantum transducers to convert these microwave photons to telecom wavelengths, which can then be transported via optical fibres or in free space. In addition, we also need a way to encode and send information that is robust to the realistic noise and losses incurred in long distance transmission. In this paper, we propose a novel scheme to generate robust entanglement between cat state qubits of distant microwave cavities mediated via telecom photons (see Fig.\ \ref{fig:schematic_entanglement}). There has been a great deal of work related to microwave-to-optical transduction by hybridising different physical systems with superconducting circuits and microwave resonators, e.g.\ NV centers in diamond \cite{10_Strong_Esteve}, cold gases \cite{09_Strong_Schmiedmayer}, rare-earth ions \cite{12_Coupling_Johansson,14_Interfacing_Fleischhauer,14_Magneto_Longdell}, and optomechanical systems \cite{13_Nanomechanical_Cleland,14_Bidirectional_Lehnert,18_Harnessing_Regal,18_Microwave_Groblacher}.  Some of us recently developed a protocol to achieve transduction in a rare-earth ion ($\mathrm{Er^{3+}}$) doped crystal \cite{18_Electron_Thiel}, which has an advantage of easier integration with the current telecom fibers, mainly because of its addressable telecom wavelength transition. We use that protocol in our scheme and discuss it in greater detail here. \par

Unlike in classical communication, where one can amplify signals at intermediate stations, in quantum communication, the no cloning theorem prohibits this kind of amplification. One solution is to use quantum repeaters, where the total distance is divided into several elementary links and entanglement is first generated between the end points of each elementary link, and then swapped between neighbouring links in a hierarchical fashion to ultimately distribute it over the whole distance \cite{98_Quantum_Zoller,11_Quantum_Gisin}. Many repeater protocols rely on using atomic ensembles \cite{01_Long_Zoller,11_Quantum_Gisin} at the repeater nodes. But the success probability of their swapping operation based on linear optics is limited to a maximum of $\mathrm{50\%}$ \cite{99_Bell_Suominem}, which hampers the performance over long distances and complex networks. It is possible to increase the success probability using auxiliary photons \cite{11_Arbitrarily_Grice,16_Efficiency_Simon}. But this makes the system more complicated and error-prone, limiting its practical utility. Repeater approaches using deterministic swapping operations have been proposed in different systems to overcome this limitation, e.g.\ in trapped ions \cite{09_Quantum_Simon}, Rydberg atoms \cite{10_Efficient_Zoller,10_Quantumrepeaters_Simon}, atom-cavity systems \cite{15_Cavity-based_Rempe, 19_Telecom_Painter}, and individual rare-earth ions in crystals \cite{18_Quantum_Simon}. Some components of such a repeater have been implemented experimentally in a few systems, e.g.\ NV\ centers in diamond \cite{15_Loophole_Hanson}, trapped ions \cite{07_Entanglement_Monroe}, and quantum dots \cite{16_Generation_Imamoglu}.\par

Deterministic two-qubit gates are easily achievable in the microwave cavity architecture proposed in ref.\ \cite{16_Universal_Goto,17_Engineering_Blais}. So, instead of relying on one of the above systems to build a repeater, we leverage the quantum advancements of superconducting processors, and develop a new repeater scheme using microwave cavities and transducers. This could also be useful in allocating some resources of relatively nearby quantum computing nodes to serve as repeater links to connect more distant nodes. We calculate the entanglement distribution rates, and compare those with direct transmission, with the well-known ensemble-based Duan-Lukin-Cirac-Zoller (DLCZ) repeater protocol \cite{01_Long_Zoller}, and with a recently proposed single-emitter-based approach in rare-earth (RE) ion doped crystals \cite{18_Quantum_Simon}. We conclude that our approach could yield higher rates in suitable regimes. We also estimate the fidelities of our final entangled states in the presence of realistic noise and imperfections, and we find them to be sufficiently high to perform useful quantum communication tasks, even without entanglement purification or quantum error correction. The latter protocols are likely to be needed for more complex tasks such as distributed quantum computing, and we anticipate that it should be possible to incorporate them in the present framework. 
 
The paper is organised as follows: In Sec.\ \ref{sec:qubit}, we briefly describe the qubit and the gates, following ref.\ \cite{16_Universal_Goto, 17_Engineering_Blais}. In Sec.\ \ref{sec:scheme}, we discuss the entanglement generation scheme between distant qubits, which includes a description of the transduction protocol. Sec.\ \ref{sec:repeater} deals with our proposal for a quantum repeater with the same architecture. In Sec.\ \ref{sec:additional}, we provide some additional implementation details pertinent to our proposal. In Sec.\ \ref{sec:rate_and_fidelity}, we estimate the rates and fidelities of our final entangled states, and make pertinent comparisons with other schemes. In Sec.\ \ref{sec:conclusion}, we draw our conclusions and enumerate a few open questions and avenues for future research.

\begin{figure*}
\centering
\includegraphics[width=0.95\textwidth]{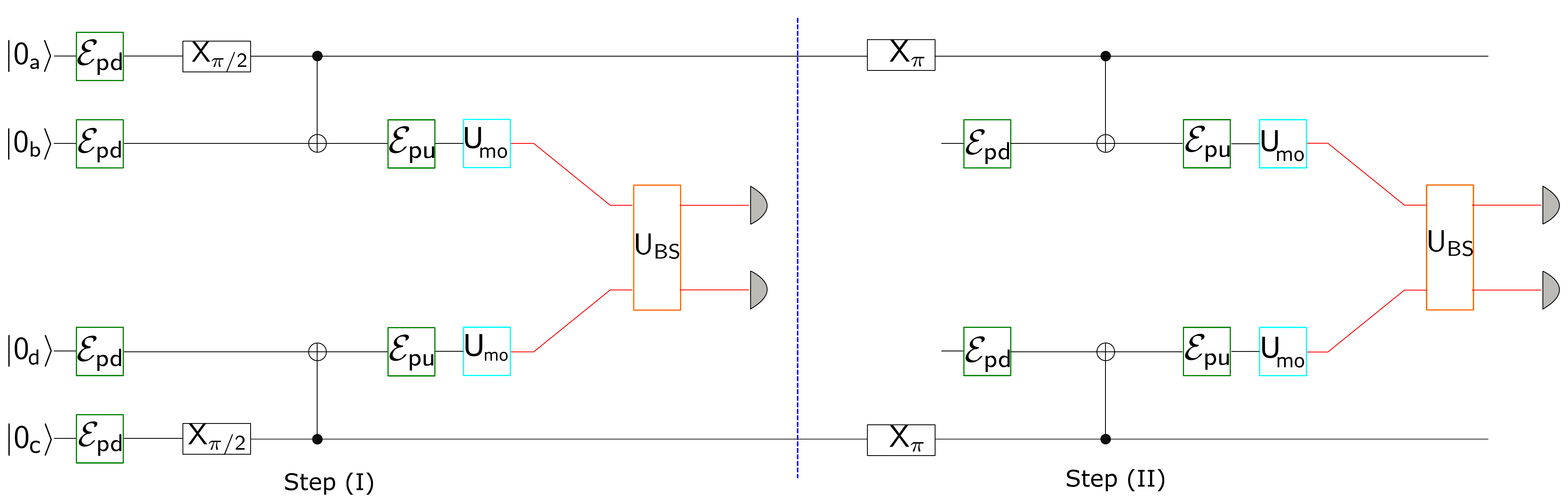}
\caption{(color online) Quantum circuit to explain the entanglement generation protocol. In addition to the standard symbols for quantum operations, we denote the cavity driving operation by $\mathcal{E}_{pd}$, the undriving operation by $\mathcal{E}_{pu}$, microwave to optical transduction by $\mathrm{U_{mo}}$ and the beamsplitter operation by $\mathrm{U_{BS}}$. The driving operation takes the Fock states $\ket{0}$ and $\ket{1}$ to the cat states $\ket{C_{\alpha}^+}\equiv \ket{\bar{0}}$ and $\ket{C_{\alpha}^-}\equiv \ket{\bar{1}}$ respectively. The protocol is divided into two steps; the second step is undertaken to discard the noise terms and to stabilize the protocol against path length fluctuations of the fiber, following Barrett-Kok's original proposal \cite{05_Efficient_Kok}. The state at the end of the protocol is either $\frac{1}{\sqrt{2}}(\ket{\bar{1}_\mathrm{a} \bar{0}_\mathrm{c}} + \ket{\bar{0}_\mathrm{a} \bar{1}_\mathrm{c}})$ or $\frac{1}{\sqrt{2}}(\ket{\bar{1}_\mathrm{a} \bar{0}_\mathrm{c}} - \ket{\bar{0}_\mathrm{a} \bar{1}_\mathrm{c}})$ depending on whether the same or different detector(s) clicked during each step.}\label{fig:schematic_protocol}
\end{figure*}

\section{\label{sec:qubit} The qubit and the universal set of gates} 

\subsection{Theoretical model}
An individual physical unit of the quantum processor is a nonlinear microwave cavity. The nonlinearity arises because of the interaction of the cavity with a nonlinear superconducting element, e.g.\ a transmon placed inside (see Fig.\ \ref{fig:schematic_entanglement}). The system is driven by a two-photon drive. The Hamiltonian of the system with two photon driving, in a frame rotating at the resonance frequency of the cavity is  

\begin{equation}\label{eqn:single_qubit}
H_0(t)=-K {a^{\dagger}}^2 a^2 + (\mathcal{E}_p(t) {a^{\dagger}}^2 +\mathcal{E}_p^*(t) a^2)
\end{equation}

Here, $a$ is the annihilation operator of the cavity mode, $K$ is the magnitude of Kerr nonlinearity, and $\mathcal{E}_p(t)$ is the time dependent pump amplitude of the two photon parametric drive. \par

The coherent states $\ket{\alpha}$ and $\ket{-\alpha}$, where $\alpha=\sqrt{\mathcal{E}_p/K}$, are instantaneous eigenstates of this Hamiltonian. Under an adiabatic evolution of the  pulse amplitude $\mathcal{E}_p(t)$, the photon number states (Fock states) $\ket{0}$ and $\ket{1}$ are mapped to the states $\ket{C_{\alpha}^+}$ and $\ket{C_{\alpha}^-}$ respectively. These so called cat states are equal superpositions of coherent states: $\ket{C_{\alpha}^\pm}=\mathcal{N}_{\alpha}^\pm (\ket{\alpha}\pm\ket{-\alpha})$, where the normalization constant $\mathcal{N}_{\alpha}^\pm=1/\sqrt{2 (1\pm e^{-2|\alpha|^2})}$. Faster non-adiabatic evolution is possible with high fidelity using a transitionless driving approach \cite{09_Transitionless_Berry,17_Engineering_Blais} (see Sec.\ \ref{sec:driving} for a detailed discussion). \par

There are several advantages of forming and manipulating cat states this way using Kerr nonlinearity, as compared to other techniques, e.g.\ using the engineered dissipation approach \cite{14_Dynamically_Devoret,15_Confining_Devoret}. The system is easier to implement experimentally, has better stabilization against losses, has effectively longer coherence times, and since all the processes are unitary, it is possible to trace back a path in phase space, e.g.\ between cat states and Fock states, which is an important requirement for our entanglement generation scheme discussed in Sec.\ \ref{sec:scheme}. For conciseness, we shall refer to the mapping from photon number states to cat states as `driving' and from cat to photon number states as `undriving'.\par

We pick $\ket{C_{\alpha}^+}$ and $\ket{C_{\alpha}^-}$ as our logical qubits, denoted by $\ket {\bar{0}}$ and $\ket {\bar{1}}$ respectively. We pick $\alpha$ large enough such that the coherent states $\ket{\alpha}$ and $\ket{-\alpha}$ have negligible overlap. Continuous rotation around X axis can be implemented using an additional single photon drive, described by the Hamiltonian 

\begin{equation}\label{eqn:H_x}
H_x=H_0 + \mathcal{E}_x (a+a^{\dagger})
\end{equation}

Here $\mathcal{E}_x$ is the amplitude of the additional drive. Rotation around Z axis can be achieved by switching off the two photon drive, and evolving the system under the Hamiltonian 

\begin{equation}\label{eqn:H_z}
H_z=-K ({a^{\dagger} a})^2
\end{equation}

To complete a universal set of gates, we need a two qubit entangling operation. The Hamiltonian for the system of two linearly coupled cavities is

\begin{equation}\label{eqn:two_qubit}
H_{c}=H_{01}+H_{02}+\mathcal{E}_{c}(a_1^{\dagger}a_2+a_1 a_2^{\dagger})
\end{equation}

$H_{01}$ and $H_{02}$ are the Hamiltonians of the individual cavities (given by Eqn.\ \ref{eqn:single_qubit}), $a_1$ and $a_2$ are the annihilation operators for their respective microwave photonic modes, and $\mathcal{E}_{c}$ is the coupling strength. The cavities can be coupled using a transmon with two antenna pads, one in each of these cavities, overlapping with the microwave fields inside \cite{16_Schrodinger_Schoelkopf,18_CNOT_Schoelkopf, 19_Entanglement_Schoelkopf} (see Fig.\ \ref{fig:schematic_entanglement}). \par

\section{\label{sec:scheme}Entanglement generation scheme between two distant nodes}
Next, we discuss our scheme to generate entanglement between distant cat state qubits, which involves application of a few of the gates discussed in the previous section. The scheme is inspired from Barrett-Kok's (BK) original theoretical proposal \cite{05_Efficient_Kok} for entanglement generation between distant atomic spins. The protocol is robust to path length fluctuations of the fiber connecting the distant spins. Minor variants of the BK protocol have been implemented in a few recent experiments \cite{13_Heralded_Hanson, 15_Loophole_Hanson, 16_Robust_Devoret}; probably the most notable one was the demonstration of loophole-free violation of Bell's inequality \cite{15_Loophole_Hanson}. The novelty of our work is in  bringing a promising contender for quantum computation, i.e.\ microwave cat qubits, and probably the only feasible carrier for long distance quantum communication, i.e.\ optical photons, together in a single platform, with a robust BK like entanglement generation protocol. We rely on microwave cavity cat states for computational and storage tasks, and on optical Fock states for communication. \par

\subsection{\label{sec:protocol}Entanglement generation protocol}
Our goal is to entangle one qubit at one node, let's call it node `A', with another qubit at another node, let's call it node `B'. For this, we start with a set of two coupled cavities at each node, e.g.\ cavities `a' and `b'  at node `A' and cavities `c' and `d' at node `B', as shown in Fig.\ \ref{fig:schematic_entanglement}. We shall perform a sequence of operations such that at the end, one of these cavities at each node (cavity `a' at node `A' and cavity `c' at node `B') are entangled. Our approach for remote entanglement generation is somewhat similar in spirit to those suggested for NV centers in diamond \cite{07_Quantum_register_Lukin, 16_Robust_Markham}, where the aim is to entangle distant long-lived nuclear spins of \ce{^{13} C} (analogous to our cavities `a' and `c'), by performing spin-photon operations on the optically addressable electronic spins of NV centers and using the hyperfine interaction between the nearby electronic and nuclear spins. A similar approach has also been proposed recently in a rare-earth ion based repeater architecture\cite{18_Quantum_Simon}. Here, individual $\mathrm{Er^{3+}}$ spins in crystals are manipulated to generate spin-photon entanglement, and the state of $\mathrm{Er^{3+}}$ is then mapped to the long lived nuclear spins of the nearby $\mathrm{Eu^{3+}}$ ions, which serve as the storage qubits.\par

Following the BK's proposal, we incorporate a two step protocol, depicted in Fig.\ \ref{fig:schematic_protocol}. Step (II) helps in discarding the noise terms and in stabilizing against path length fluctuations. In step (I) of the protocol shown in Fig. \ref{fig:schematic_protocol}, we first set out to generate entanglement between the two neighbouring cavities at each node. Let's focus on cavities `a' and `b'  at node `A'. The cavities are initially in vacuum state. They are driven using suitable microwave pulses and an $X_{\pi/2}$ rotation on cavity `a' prepares the system in the state  $\frac{1}{\sqrt{2}}(\ket {\bar{0}_\mathrm{a}} + \ket {\bar{1}_\mathrm{a}})\otimes \ket {\bar{0}_\mathrm{b}}$. As a reminder, $\ket {\bar{0}}$ and $\ket {\bar{1}}$ are the cat states $\ket{C_{\alpha}^+}$ and $\ket{C_{\alpha}^-}$ respectively. We then perform a CNOT operation (see Sec.\ \ref{sec:gates} for the sequence of gates needed), with the first qubit as the control and the second as the target, to reach the entangled state $\frac{1}{\sqrt{2}} (\ket {\bar{0}_\mathrm{a}} \otimes \ket {\bar{0}_\mathrm{b}} + \ket {\bar{1}_\mathrm{a}} \otimes \ket {\bar{1}_\mathrm{b}})$. Cavity `b' is then undriven coherently (see Sec.\ \ref{sec:driving}), such that $\ket {\bar{0}_\mathrm{b}} \rightarrow \ket {0_\mathrm{b}}$ and $\ket {\bar{1}_\mathrm{b}} \rightarrow \ket {1_\mathrm{b}}$, where $\ket {0_\mathrm{b}}$ and $\ket {1_\mathrm{b}}$ are the microwave Fock states of cavity `b'. A quantum transducer is then used to convert microwave photons to propagating optical photons at telecom wavelength. We discuss one possible transduction protocol in the next subsection. To avoid usage of more complicated notations, telecom Fock states obtained after transduction shall also be represented by $\ket {0_\mathrm{b}}$ and $\ket {1_\mathrm{b}}$; the distinction between microwave and telecom Fock states should be clear from the context. \par

The same procedure is followed in a distant set of two cavities `c' and `d' at node `B'.  The cat states stored in the cavities `a' and `c' shall be referred to as the `stationary qubits', and the telecom Fock states as the `flying qubits'. The combined state of the system, including the stationary and flying qubits at this point is $\frac{1}{2}(\ket{\bar{0}_\mathrm{a}\bar{0}_\mathrm{c}} \ket{0_\mathrm{b}0_\mathrm{d}} + \ket{\bar{0}_\mathrm{a}\bar{1}_\mathrm{c}} \ket{0_\mathrm{b}1_\mathrm{d}}+\ket{\bar{1}_\mathrm{a}\bar{0}_\mathrm{c}} \ket{1_\mathrm{b}0_\mathrm{d}}+\ket{\bar{1}_\mathrm{a}\bar{1}_\mathrm{c}} \ket{1_\mathrm{b}1_\mathrm{d}})$. The telecom photons are directed to a beamsplitter placed midway between those distant nodes. Detection of only a single photon in either of the output ports of the beamsplitter would imply measuring the photonic state $\frac{1}{\sqrt{2}}(\ket{1_\mathrm{b}0_\mathrm{d}} + e^{-i \phi} \ket{0_\mathrm{b}1_\mathrm{d}})$ or $\frac{1}{\sqrt{2}}(\ket{1_\mathrm{b}0_\mathrm{d}} - e^{-i \phi} \ket{0_\mathrm{b}1_\mathrm{d}})$. Here $\phi$ is the phase difference accumulated because of the different path lengths of those photons. The detection event projects the stationary qubits into either of the maximally entangled odd Bell states $\frac{1}{\sqrt{2}}(\ket{\bar{1}_\mathrm{a} \bar{0}_\mathrm{c}} + e^{-i \phi} \ket{\bar{0}_\mathrm{a} \bar{1}_\mathrm{c}})$ or $\frac{1}{\sqrt{2}}(\ket{\bar{1}_\mathrm{a} \bar{0}_\mathrm{c}} - e^{-i \phi} \ket{\bar{0}_\mathrm{a} \bar{1}_\mathrm{c}})$.\par

If the detectors can't resolve the number of photons, or if both the nodes emitted a photon each and one of those photons was lost in transmission, then instead of the above pure entangled state of stationary qubits, the projected state would be a mixture, with contribution from the term $\ket{\bar{1}_\mathrm{a}\bar{1}_\mathrm{c}}$. To get rid of this term, and to make the protocol insensitive to phase noise which is greatly detrimental to the fidelity of the final state, we follow BK's approach \cite{05_Efficient_Kok}, and perform step (II).\par

In step (II), we flip the stationary qubits ($\mathrm{X_{\pi}}$ rotation), drive the cavities `b' and `d' from vacuum states to $\ket{\bar{0}_\mathrm{b}}$ and $\ket{\bar{0}_\mathrm{d}}$ respectively, and then perform the same operations as in step (I), i.e.\ CNOT operation, followed by undriving of the cavities `b' and `d', followed by transduction to telecom photons which leave the cavity, and finally single photon detection in one of the output ports of the beamsplitter. Successful single photon detection heralds the entanglement, and the state of the stationary qubits is either $\frac{1}{\sqrt{2}}(\ket{\bar{1}_\mathrm{a} \bar{0}_\mathrm{c}} + \ket{\bar{0}_\mathrm{a} \bar{1}_\mathrm{c}})$ or $\frac{1}{\sqrt{2}}(\ket{\bar{1}_\mathrm{a} \bar{0}_\mathrm{c}} - \ket{\bar{0}_\mathrm{a} \bar{1}_\mathrm{c}})$ depending on whether the same or different detector(s) clicked during each step. Once the two nodes A and B share an entangled state, it can also be used to teleport a quantum state from one node to the other using only classical bits and local operations \cite{93_Teleporting_Wootters}.\par

The main reason that we undrive one of the two cavities at each node to generate Fock states, which we then transduce to telecom wavelengths, instead of the more obvious route of directly transducing these microwave cat states to telecom cat states and then transporting it over a fiber is that cat states dephase because of photon loss and it's complicated to correct for those errors over long-distances where fiber loss could be greater than $90 \%$ \cite{17_Cat_Jiang, 16_Concurrent_Jiang}.\par 

\subsection{\label{sec:transduction}Transduction protocol}
Microwave to telecom transduction has attracted significant attention in recent times for various applications involving hybridising different systems \cite{13_Hybrid_Nori}. Coupling superconducting circuits with spin based systems e.g.\ NV centers \cite{10_Strong_Esteve}, cold gases \cite{09_Strong_Schmiedmayer} and rare-earth ions \cite{12_Coupling_Johansson,13_Anisotropic_Bushev,14_Interfacing_Fleischhauer,15_Interfacing_Morigi}, or optomechanical systems \cite{13_Nanomechanical_Cleland,14_Bidirectional_Lehnert,18_Harnessing_Regal,18_Microwave_Groblacher} are a few possible ways to achieve desired frequency conversions. Our entanglement generation protocol does not rely on a specific implementation of the transducer. However as a concrete example, we elaborate on the protocol some of us proposed in ref. \cite{18_Electron_Thiel} using a rare-earth ion ($\mathrm{Er^{3+}}$) doped crystal. High efficiency conversion due to ensemble enhanced coupling strengths, and easy integration with the current fiber optics owing to its available telecom wavelength transition, are a few of the attractive features of this system. \par

\begin{figure*}
\centering
\includegraphics[width=0.95\textwidth]{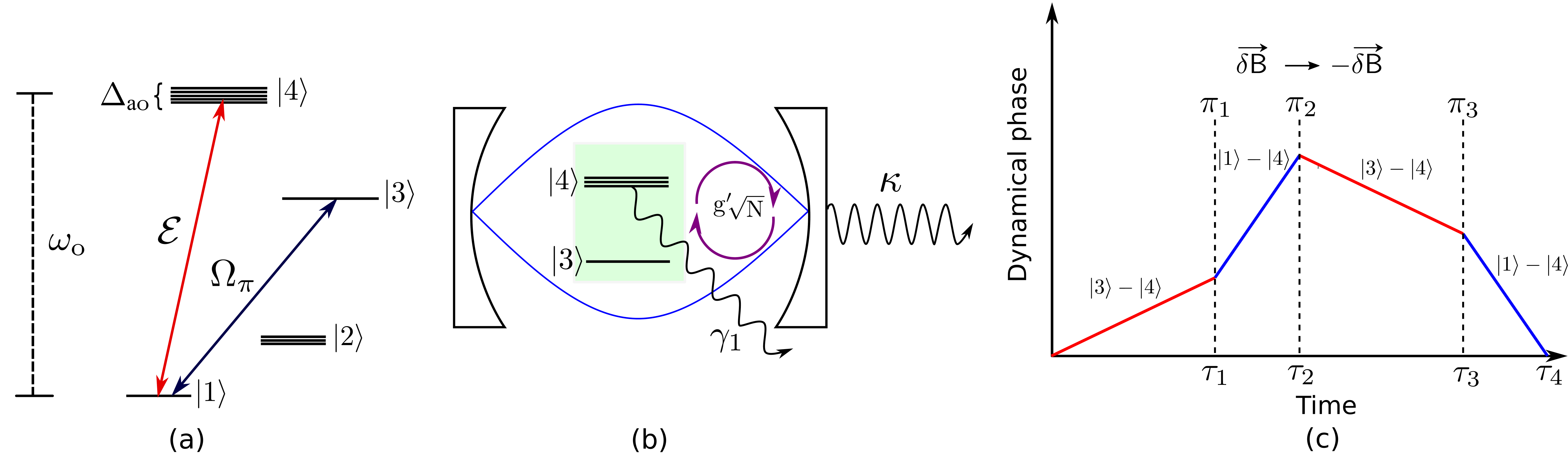}
\caption{(color online) Reversible microwave to telecom transduction.(a) Energy level diagram of an ensemble of $\mathrm{Er^{3+}}$ ions in  $\mathrm{Y_2 Si O_5}$ crystal. A narrow spectral feature is selected from the naturally broadened $\ket{1}$ to $\ket{4}$ transition, which is later artifically broadened by $\Delta_{ao}$ using a magnetic field gradient, $\vec{\delta B}$. The spin transition $\ket{3}$ to $\ket{4}$ is used for the microwave side of the transduction process. A sequence of $\pi$ pulses ($\Omega_{\pi}$) is applied between $\ket{1}$ to $\ket{3}$ at different times to transfer the population and control the dephasing and rephasing of the collective dipole. An optical photon of frequency $\omega_o$, described by the field $\mathcal{E}$ is emitted during the final step of transduction from microwave to optical wavelength. (b) Transfer of a microwave photon to a collective spin transition. On resonance, the collective spin transition ($\ket{3}-\ket{4}$) couples strongly to a single mode of the microwave cavity. The enhanced coupling strength is $g' \sqrt{N}$, where $N$ is the number of participating spins in the ensemble, and $g'$ is the single spin-cavity coupling strength. The spin decay rate ($\gamma_1$) and the cavity decay rate ($\kappa$) are also shown. (c) Variation of the dynamical phase of the collective dipole as a function of time to illustrate the role of various $\pi$ pulses applied between $\ket{1}$ and $\ket{3}$. The dipoles dephase because of the induced inhomogeneous broadenings (spin broadening between $\ket{3}-\ket{4}$ and optical broadening between $\ket{1}-\ket{4}$). As the direction of the magnetic field gradient is reversed ($\vec{\delta B}\rightarrow -\vec{\delta B}$), the dipoles rephase leading to a collective emission of a single optical photon. Although, in the figure, we do not show the effect of the intrinsic inhomogeneous broadening, the decay linewidth, and the homogeneous linewidth of the spin transition for simplicity, we do consider them for our calculation of the transduction efficiency.} \label{fig:schematic_transduction}
\end{figure*}

In this section, we shall explicitly describe only the one-way microwave to telecom frequency conversion, since we only need this in our entanglement generation scheme. The opposite conversion can be achieved by running the pertinent steps of our transduction protocol in reverse. The protocol is implemented in an $\mathrm{Er^{3+}}$ doped $\mathrm{Y_2 Si O_5}$ crystal, which could be placed inside a microwave cavity (see cavities `b' and `d' in Fig.\ \ref{fig:schematic_entanglement}). $\mathrm{Er^{3+}}$ is a Kramers ion and as such has doubly degenerate energy eigenstates. An external constant magnetic field, $\vv{B_0}$, splits the ground state (\ce{^{4}I_{15/2}}) into the Zeeman levels $\ket{1}$ and $\ket{2}$, and the excited state (\ce{^{4}I_{13/2}}) into $\ket{3}$ and $\ket{4}$. The energy level diagram of the ensemble of $\mathrm{Er^{3+}}$ ions is shown in Fig.\ \ref{fig:schematic_transduction}. The optical transitions between ground and excited states are at telecom wavelengths (around 1536 nm), whereas the Zeeman splitting is in GHz range for magnetic field strength ($|\vv{B_0}|$) of the order of tens to hundreds of mT depending on the field's directions \cite{18_Electron_Thiel,13_Anisotropic_Bushev}.

The transduction approach is inspired from the Controlled Reversible Inhomogeneous Broadening (CRIB) memory protocol \cite{01_Complete_Kroll,07_Analysis_Gisin} and is similar to the previous transducer proposal by O'Brien et al.\ \cite{14_Interfacing_Fleischhauer}. The idea is to first map the microwave photon to a spin excitation and then map this spin excitation to an optical excitation followed by its eventual read-out. We assume that initially all the ions are in the ground state $\ket{1}$. As in the CRIB memory protocol, a narrow absorption line from the inhomogeneously broadened $\ket{1}-\ket{4}$ transition is prepared by transferring the rest of the population to an auxiliary level, e.g.\ level $\ket{2}$. At a later stage of the protocol, this narrow absorption line is artificially broadened using, for example, a one-dimensional magnetic field gradient, $\vv{\delta B} (z)$, where the `$z$' direction can be suitably chosen.

In contrast to the original proposal \cite{14_Interfacing_Fleischhauer}, we propose to use the excited state Zeeman levels $\ket{3}$ and $\ket{4}$ (see Fig.\ \ref{fig:schematic_transduction} (b)) to transfer a microwave photon from the cavity into a collective spin excitation. The excited state spin transition of $\mathrm{Er^{3+}}$ is less susceptible to the various spin-dephasing sources, such as spin flip-flop and instantaneous spectral diffusion, and can thus have considerably longer coherence lifetimes than the ground state spin transition \cite{18_Electron_Thiel}. The spin transition $\ket{3}-\ket{4}$ is initially detuned from the microwave cavity by an amount $\delta$. A $\mathrm{\pi}$ pulse is applied to transfer all the population from $\ket{1}$ to $\ket{3}$. The spin transition is now brought in resonance with the cavity. The dynamics of the system can be described by the following Hamiltonian:
\begin{equation}
H=\sum_{j} \hbar \frac{\omega_j}{2} \sigma_z^{(j)}+\sum_j \hbar g'(a^\dagger \sigma_-^{(j)} + a \sigma_+^{(j)}) + \hbar \omega_m a^\dagger a
\end{equation}
where $\sigma_-^{(j)}$ is the spin flip operator, $\sigma_z^{(j)}$ is the spin population operator and $\omega_j$ is the spin transition frequency of the $j^{th}$ spin, and $a$ is the annihilation operator of the cavity mode and $\omega_m$ is the resonance frequency of the cavity. The collective spin $S_- =\frac{1}{\sqrt{N}}\sum_{j=1}^{N} \sigma_-^{(j)}$ strongly couples to the cavity mode with an enhanced coupling strength $g' \sqrt{N}$, where $N$ is the total number of participating spins. The free evolution of the coupled cavity-spin system transfers the cavity excitation into the collective spin excitation in time $T_S=\pi/(2 g'\sqrt{N})$. The efficiency of the process is primarily limited by the natural spin inhomogeneous broadening, denoted by $\Delta_{ns}$. For realistic values of $g'\sqrt{N}=2 \pi \times 34$ MHz, $\Delta_{ns}=2 \pi \times 10$ MHz, the homogeneous linewidth of spin transition $\gamma_2=2 \pi \times 100$ kHz, the spin decay rate $\gamma_1=2 \pi \times 160$ Hz, and the cavity decay rate $\kappa= 2 \pi \times 10$ Hz \cite{18_Electron_Thiel}, the efficiency of transfer is $99.04\%$.\par

The emission of a telecom photon is controlled by the dephasing and rephasing of the collective spin and optical dipoles. The change of the dynamical phase as a function of time is depicted in Fig.\ \ref{fig:schematic_transduction} (c). After the microwave photon has been transferred to the spin excitation, the spin transition is detuned from the cavity, and the field gradient, $\vv{\delta B} (z)$, is applied. This artificially broadens the optical and spin transitions by $\Delta_{ao}$ and $\Delta_{as}$ respectively. After some spin-dephasing time $\tau_1$, a $\mathrm{\pi}$ pulse ($\pi_1$ pulse in Fig.\ \ref{fig:schematic_transduction} (c)) is applied to transfer all the population back from $\ket{3}$ to $\ket{1}$. The collective excitation now dephases at a rate $\Delta_{ao}$ on the optical transition. To initiate the rephasing process required for emission of a photon, the direction of the magnetic field gradient is reversed ($\vv{\delta B}(z)\rightarrow -\vv{\delta B}(z)$). The population from $\ket{1}$ is again taken to $\ket{3}$ ($\pi_2$ pulse) and is kept there for some time to allow spin rephasing. Afterwards, the population is brought back to $\ket{1}$ ($\pi_3$ pulse) and the optical dipoles rephase, leading to a collective emission of a single photon corresponding to the $\ket{1}-\ket{4}$ transition frequency, $\omega_0$. The efficiency of the overall transduction process also depends on the total time spent in the rephasing and dephasing operations, the finite optical depth of the crystal, and the coupling of the generated photon to the fiber. The overall efficiency can be greater than $85 \%$ for realistic values \cite{14_Interfacing_Fleischhauer}. Higher efficiencies are possible with a more optimized $\pi$ pulse sequence. It is worth mentioning here that even though these transduction operations may look complicated, they can be compatible with the kind of 3-D microwave cavities we have in mind. See Sec.\ \ref{sec:compatibility_transduction} for a discussion on this. \par

The fidelity of the final entangled state between the nodes depends on the indistinguishability of the photons interfering at the beamsplitter. Interestingly, for photon-echo based quantum memories in big atomic ensembles, this value can approach unity, as decoherence predominantly affects only the efficiency of the photon generation process, but not the purity of the generated photons \cite{07_Interference_Gisin}. Possible imperfections in the setup, e.g.\ scattering of the pump laser, might cause some optical noise, but these could be almost entirely filtered out, as was shown in similar systems \cite{15_Solid_Riedmatten,13_Demonstration_Sellars}. We realistically assume that the temperature in the dilution refrigerator is low enough ($\approx 20$ mK) to neglect contribution of thermal microwave photons \cite{16_Schrodinger_Schoelkopf,18_CNOT_Schoelkopf}. However generation of spurious microwave photons possibly because of the noisy microwave circuit components and heating caused by lasers would adversely affect the overall process. This needs to be overcome with better experimental control or alternative theoretical approaches \cite{18_Harnessing_Regal,18_Microwave_Groblacher}. \par

\begin{figure*}
\centering
\includegraphics[width=0.9\textwidth]{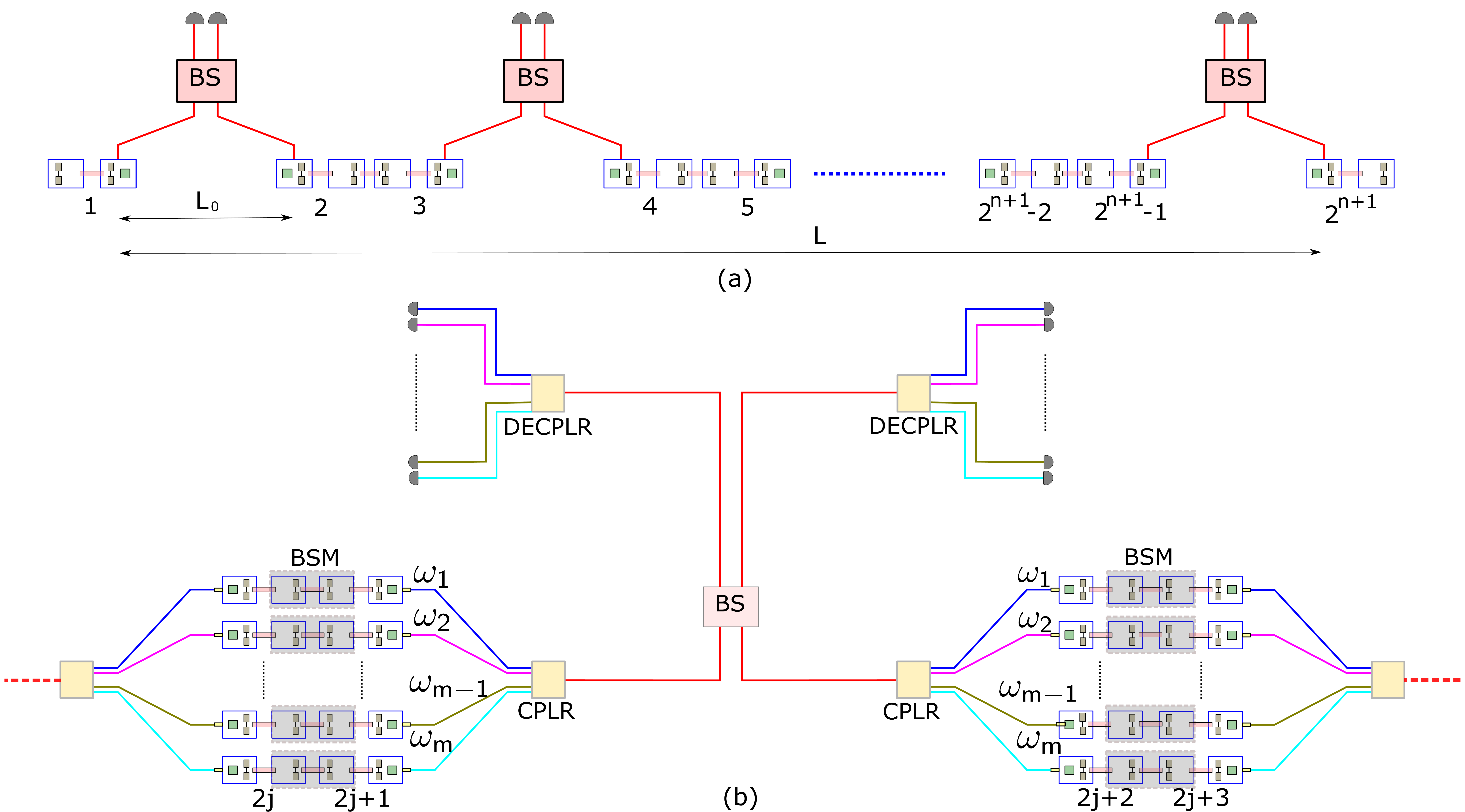}
\caption{(color online) Schematic diagram of our repeater architecture. (a) Repeater architecture using microwave cavities at each node. Independent attempts to generate entanglement are made in $2^{n}$ elementary links, each of length $L_0$. Each pair of coupled cavities are denoted by the same number. The stationary qubits shall have the subscript `s' while the flying qubits shall have the subscript `f'. In the first step, the neighbouring sets of stationary qubits, e.g.\ `$1_\mathrm{s}$' and `$2_\mathrm{s}$', `$3_\mathrm{s}$' and `$4_\mathrm{s}$', and so on, are entangled. Not all the links need to be entangled at the same time. As soon as the neighbouring links are entangled, we go to the second step, where entanglement is swapped using Bell state measurement (BSM) and classical communication, such that the qubits `$1_\mathrm{s}$' and `$4_\mathrm{s}$', `$5_\mathrm{s}$' and `$8_\mathrm{s}$', and so on are entangled. It takes $n$ steps to distribute entanglement over the total length $L=2^n L_0$. (b) Part of a spectrally multiplexed repeater architecture. An array of cavities is placed at each node. A frequency translator is placed at the output port of each cavity generating the flying qubits, and independent attempts are made to distribute entanglement in $m$ spectrally distinct channels in each elementary link. The coupler (CPLR) couples light from those $m$ channels to the same fiber, while the decoupler (DECPLR) decouples it, followed by $2 m$ single photon detectors for heralding entanglement. The greyish shaded boxes represent the possible BSM between neighboring storage cavities (for the entanglement swap operation), needed to distribute entanglement hierarchically over the total length, $L$, in the same set of identical channels (same frequency). There are $m$ sets of $2^n$ identical channels. The protocol is successful when entanglement is distributed over the total length in at least one set of identical channels.}\label{fig:schematic_repeater}
\end{figure*}

\section{\label{sec:repeater}Quantum repeater architecture based on microwave cavities}
Quantum repeaters are essential for distributing entanglement over long distances \cite{98_Quantum_Zoller,11_Quantum_Gisin} for quantum computation and quantum communication applications. Many repeater proposals involve using atomic ensembles as stationary qubits mainly because good optical quantum memories can be built using them \cite{10_Quantum_Young,09_Optical_Tittel,12_Efficient_Pan,16_Efficient_Pan}. However, one drawback is their probabilistic swapping operation, which hampers the entanglement distribution rate, and this prompted several single-emitter based approaches \cite{09_Quantum_Simon,10_Efficient_Zoller,10_Quantumrepeaters_Simon,15_Cavity-based_Rempe, 19_Telecom_Painter,18_Quantum_Simon,15_Loophole_Hanson,07_Entanglement_Monroe,16_Generation_Imamoglu}. Since nonlinear microwave cavities have sufficiently long coherence times, are easy to address, and have deterministic two-qubit gates, in addition to a quantum computing architecture, one can also envision a quantum repeater architecture based on them. Moreover, this also allows relatively nearby quantum computing nodes to serve as repeater nodes to connect more distant quantum computers. Here we discuss a ground based quantum repeater architecture with microwave cavities and transducers.\par

\subsection{Single set of cavities in an elementary link} \label{sec:repeater_single}
A repeater architecture with a few elementary links using microwave cavities and transducers, connected via optical fibers, is shown in Fig.\ \ref{fig:schematic_repeater} (a). Entanglement is distributed in multiple steps in a hierarchical way. In the first step, entanglement is generated between individual links of length $L_0$, possibly after several independent attempts, e.g.\ after the first step, stationary qubits `$1_\mathrm{s}$' and `$2_\mathrm{s}$', `$3_\mathrm{s}$' and `$4_\mathrm{s}$', and so on, are entangled. However, not all the links need to be successfully entangled at the same time. We keep trying until the neighbouring links are entangled, and in the subsequent steps, entanglement is distributed over the whole length by performing the entanglement swapping operations. For instance, in the second step, stationary qubits `$1_\mathrm{s}$' and `$4_\mathrm{s}$', `$5_\mathrm{s}$' and `$8_\mathrm{s}$', and so on are entangled by performing Bell state measurement (BSM) on qubits `$2_\mathrm{s}$' and `$3_\mathrm{s}$', `$6_\mathrm{s}$' and `$7_\mathrm{s}$', and so on. It takes `n' such steps (known as nesting levels) to distribute entanglement over the total length, $L$, comprising of $\mathrm{2^n}$ elementary links, $L=2^n L_0$.\par 

\subsection{Multiplexed repeater architecture}\label{sec:repeater_multiplexed}
An important metric for quantifying the usefulness of a repeater architecture is the rate of distribution of entangled states between distant nodes. One way to increase this rate is to use a spectrally multiplexed architecture \cite{07_Multiplexed_Kennedy,11_Quantum_Gisin, 14_Spectral_Tittel,18_Quantum_Simon}. A part of that architecture is shown in Fig.\ \ref{fig:schematic_repeater} (b). An array of cavities is employed at each end of the elementary link. There are `$m$' number of the pair of coupled cavities (the storage cavity and the cavity generating the flying qubit) in this array. Each of the cavity generating the flying qubits at each node is coupled to a corresponding cavity at the next node independently via a quantum channel. To distinguish each of these $m$ channels spectrally, a frequency translator which spans tens of GHz can be optically coupled to the output ports of the cavities generating the flying qubits \cite{17_Heralded_Tittel}. These photons with different carrier frequencies are then coupled to the same spatial mode of a fiber using a tunable ridge resonator filter with $\approx 1$ MHz resonance linewidths \cite{18_Bridging_Vahala}, thus allowing a very large multiplexing in principle. At the measurement station, there is a beamsplitter, two sets of ridge resonators for demultiplexing, and $2m$ single photon detectors.\par

Independent attempts are made to distribute entanglement along each set of identical channels of frequency $\omega_j$. There are $m$ such sets and each set comprises of $2^n$ identical channels. Our approach requires entangling operations (for entanglement swap) only between neighbouring storage cavities, as shown in Fig.\ \ref{fig:schematic_repeater} (b)). A more complicated multiplexing approach that requires all-to-all connectivity between the storage qubits at each node would yield higher rates \cite{07_Multiplexed_Kennedy}, but we do not adopt that approach here because all-to-all connectivity is difficult to be realised experimentally with good fidelities \cite{17_Quantum_Blais,17_Robust_Tiwari}. Our protocol is successful if entanglement is successfully distributed over the total distance $L$ in at least one set of identical channels.   

In sec.\ \ref{sec:rate_and_fidelity}, we discuss our achievable rates for distribution of entanglement, compare those rates with direct transmission and with two other repeater approaches, and provide the fidelities for our final entangled states. Before proceeding to that section, we list the realistic values adopted for some of the experimental parameters and calculate the fidelities of the relevant operations in the next section. This would be necessary for a proper discussion on the achievable rates and fidelities, and hence the slight detour. 

\section{Some implementation details}\label{sec:additional}

\begin{table*}																	
\centering
\includegraphics[width=0.75\linewidth]{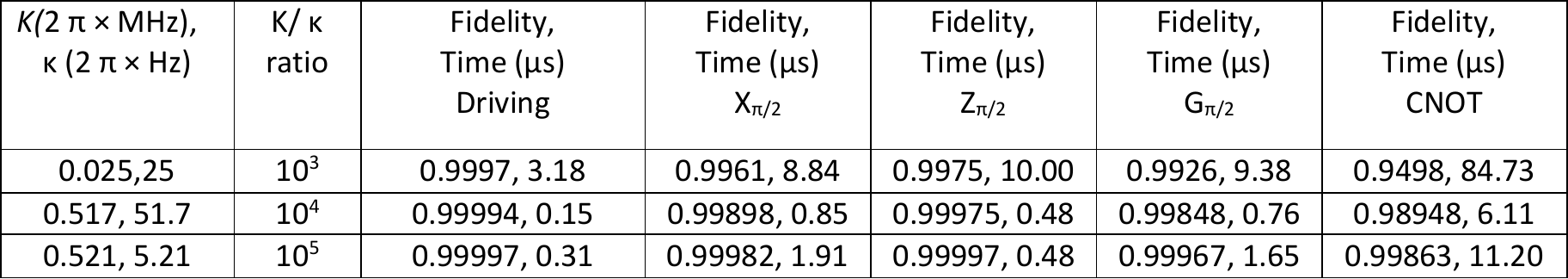}
\caption{Time and fidelities of different operations for different values of Kerr nonlinearity $K$ and cavity decay rate $\kappa$. The driving operation takes the Fock state $\ket{0}$ ($\ket{1}$) to the cat state $\ket{C_{\sqrt{2}}^+}$ ($\ket{C_{\sqrt{2}}^-}$). $X_{\pi/2}$, $Z_{\pi/2}$ , and $G_{\pi/2}$ are $\pi/2$ rotations about X axis, Z axis, and a suitable axis in the two qubit Hilbert space respectively. The values of $K$ and $\kappa$ are chosen such that the ratio $K/\kappa$ is $10^3$, $10^4$, and $10^5$ in the three rows. For conciseness, we shall often refer to these ratios instead of specifying the explicit values.}
\label{tab:gates}
\end{table*} 

\subsection{The qubit and the gates}\label{sec:exp_qubit}
One way to realise the Hamiltonian \eqref{eqn:single_qubit} using 3-D cavities is to place a non-linear element, e.g.\ a transmon inside the cavity (see Fig.\ \ref{fig:schematic_entanglement}), and drive it using suitable microwave tones \cite{17_Engineering_Blais, 14_Dynamically_Devoret, 15_Confining_Devoret}. In 2-D, one could use a  $\mathrm{\lambda/4}$ transmission line resonator (TLR) terminated by a flux-pumped SQUID \cite{17_Engineering_Blais, 08_Flux_Tsai}. \par

Next, we quantify the performance of the qubits in the presence of realistic losses. Most of the simulations incorporating such losses have been carried out using an open source software, QuTiP (Quantum Toolbox in Python) \cite{12_Qutip_Nori}. The most prominent loss mechanism for the cavities is individual photons leaking out of the cavity at a rate $\kappa$. \par

Through most of the paper, we have tried to illustrate our points with the help of explicit examples. We have chosen 3 sets of experimentally realisable values for Kerr nonlinearity $K$ \cite{12_Josephson_Blais}, and cavity decay rate $\kappa$ \cite{07_Ultrahigh_Visentin,13_Reaching_Schoelkopf,18_3D_Grassellino}, which are listed in Table \ref{tab:gates}-\ref{tab:kappa}. Although the individual values of $K$ and $\kappa$ are all different, these are chosen such that the ratios of $K/\kappa$ are $10^3$, $10^4$ and $10^5$ in increasing order. These will often represent the different rows of our tables. The fidelities of most of the operations depend on these ratios, rather than their individual magnitudes. For the sake of brevity, we'll often refer to these values by referring to the ratios. \par

\subsubsection{Driving and undriving of the cavity} \label{sec:driving}

The pulse amplitude $\mathcal{E}_p(t)$ can be increased adiabatically to evolve the state of cavity from the Fock states $\ket{0}$, and $\ket{1}$ to the cat states $\ket{C_\alpha^+}$ and $\ket{C_\alpha^-}$ respectively, preserving parity, where $\alpha=\sqrt{\mathcal{E}_p/K}$. As an example suggested in ref.\ \cite{17_Engineering_Blais}, we take the pulse $\mathcal{E}_p(t)=\mathcal{E}_p^0 (1-e^{(-t/\tau)^4})$, $\mathcal{E}_p(t=0)=0$, $\mathcal{E}_p(t)= \mathcal{E}_p^0 =2 K$ for $t\gg \tau$ to create a cat state with $\alpha=\sqrt{2}$. For $K/\kappa=10^3$, $\ket{0}$ is mapped to $\ket{C_{\alpha}^+}$ with $\mathrm{99.62\%}$ fidelity in the duration $1.3 \tau=0.04$ ms. To undrive the cavity, we simply apply the time reversed control pulse, to get back to $\ket{0}$ with the same fidelity. \par  

\begin{figure}
\centering
\includegraphics[width=0.4\textwidth]{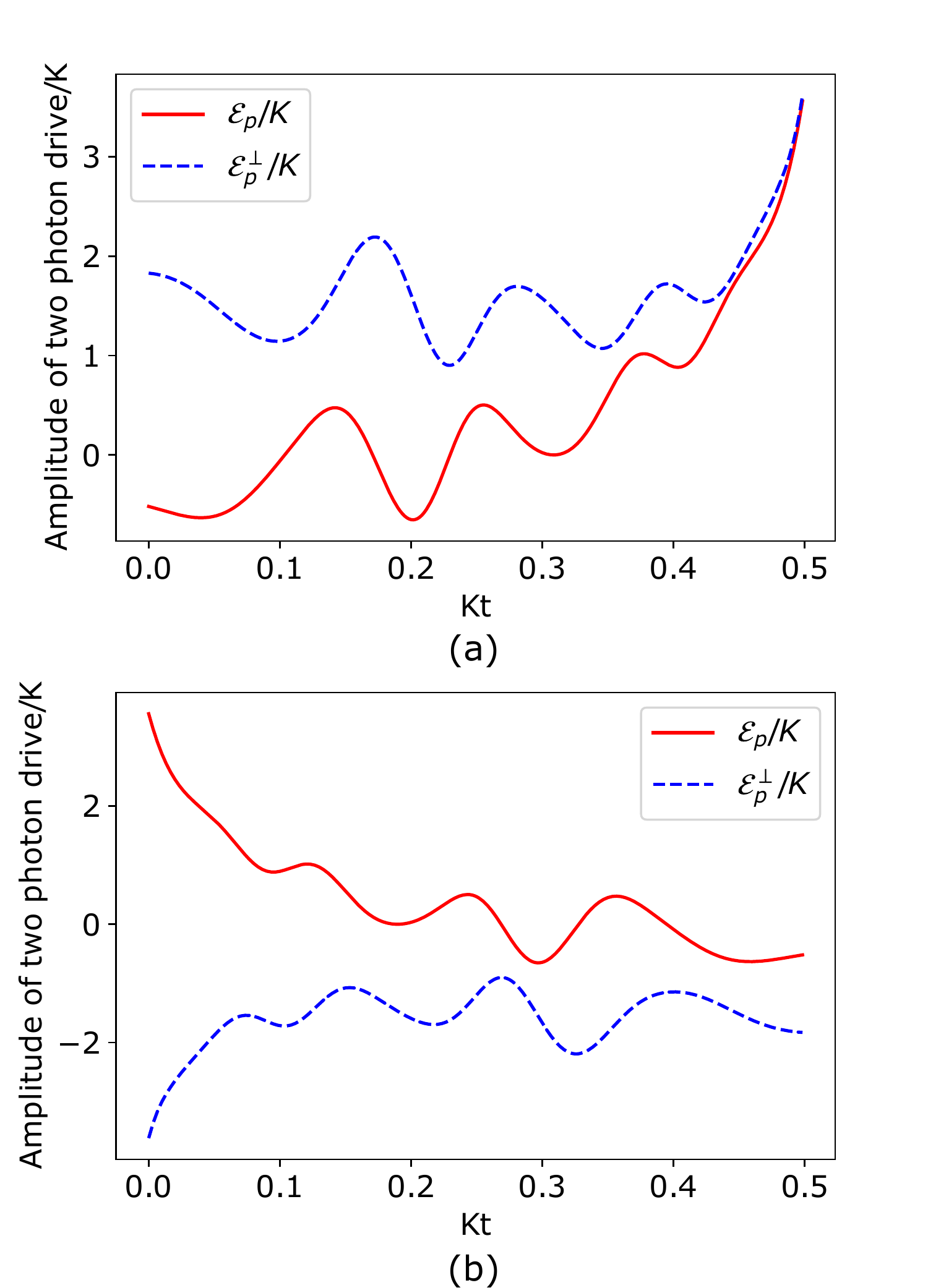}
\caption{(color online) Optimised GRAPE pulses for driving and undriving the cavities. (a) The pulses needed to drive the cavity from $\ket{0}$ to $\ket{C_{\sqrt{2}}^+}$ in time $0.5/K$ with $99.97 \%$ fidelity for $K/\kappa=10^3$. The red solid curve is the original two photon drive ($\mathcal{E}_p(t)$) while the blue dashed curve is the additional orthogonal drive ($\mathcal{E}_p^\perp(t)$). (b) The pulses needed to undrive the cavity from $\ket{C_{\sqrt{2}}^+}$ to $\ket{0}$ in time $0.5/K$ with $99.97 \%$ fidelity for $K/\kappa=10^3$.} \label{fig:grape_driving_undriving}
\end{figure}

Since this was just a smooth pulse obtained after some guess work, it need not be optimized to reach the final state with the highest possible fidelity in the fastest possible way. Faster mapping with higher fidelity is possible with an additional orthogonal two photon drive $i \mathcal{E}_p^\perp(t) ({a^{\dagger}}^2 - a^2)$ \cite{17_Engineering_Blais} using the non-adiabatic transitionless driving approach propopsed in ref.\ \cite{09_Transitionless_Berry}. The shape of both of these pulses ($\mathcal{E}_p(t)$ and $\mathcal{E}_p^\perp(t)$) is optimized with Gradient Ascent Pulse Engineering (GRAPE) algorithms \cite{05_Optimal_Glaser,11_Comparing_Herbruggen}. Fig.\ \ref{fig:grape_driving_undriving} (a) shows the two orthogonal drive pulses applied for mapping $\ket{0}$ to $\ket{C_{\sqrt{2}}^+}$ in $t=0.5/K$ with $99.97 \%$ fidelity for $K/\kappa=10^3$. Fig.\ \ref{fig:grape_driving_undriving}(b) shows the pulses required for undriving the cavity from $\ket{C_{\sqrt{2}}^+}$ to $\ket{0}$ in $t=0.5/K$. Table \ref{tab:gates} gives the drive times and fidelities for different ratios of $K/\kappa$. \par

Values of $K \mathrm{\geq 2\pi\times 10}$ MHz, much higher than those we have considered, are possible, and would make the process much faster \cite{12_Josephson_Blais}. However, the ideal platforms for very large $K$ are usually the 2-D TLRs. But, at present, since the quality factors of 3-D cavities are reported to be much higher than those of 2-D TLRs, they seem to be a better choice for long distance entanglement distribution as long coherence time is more important than the speed of operations (see Sec.\ \ref{sec:rate_and_fidelity}). Our entanglement generation scheme, however, is independent of the type of cavity.\par 

\subsubsection{Single and two qubit gates}\label{sec:gates}
Rotation around X axis can be implemented with an additional single photon drive $\mathcal{E}_x$ (see Eqn.\ \eqref{eqn:H_x}). The drive amplitude $\mathcal{E}_x$ should however be much smaller than the two photon drive drive strength $\mathcal{E}_p^0=\norm{\alpha}^2 K$ to stay within the qubit space. Under that approximation, $X_{\pi/2}$ is accomplished in $t=\pi/(8 \norm{\alpha} \mathcal{E}_x )$. In the absence of noise due to single photon loss, taking a very small value of $\mathcal{E}_x$ would yield a very high fidelity, but noise forces one to be fast to minimise decoherence, even if it means going a little bit out of the qubit space. Thus one has to optimise between reduction in fidelity because of noise and because of leaking out of the qubit space, keeping the operations sufficiently fast. See Table \ref{tab:gates} for the fidelities and gate times, taking $\mathcal{E}_x=\mathcal{E}_p^0/10$, $\mathcal{E}_x=\mathcal{E}_p^0/20$, and $\mathcal{E}_x=\mathcal{E}_p^0/45$ for the three $K/\kappa$ ratios respectively. \par

$Z_{\pi/2}$ is achieved in time $\pi/(2 K)$ by the free evolution under the Kerr Hamiltonian in Eqn.\ \ref{eqn:H_z}. Tables \ref{tab:gates} shows the fidelities and times for our parameters.\par

For a two qubit entangling gate (see Eqn.\ \eqref{eqn:two_qubit}), we start with the state $\mathrm{\ket{\bar{0}} \otimes \ket{\bar{0}}} $ and evolve it under the Hamiltonian described by Eqn.\ \ref{eqn:two_qubit}. The rotation in the two-qubit space is denoted by $G_\theta$, where $\theta$ is the same angle as that disscussed in ref.\ \cite{16_Universal_Goto}. When $\theta=\pi/2$, we arrive at the maximally entangled state $\mathrm{\frac{(1+i)\ket{\bar{0}} \otimes \ket{\bar{0}}+(1-i)\ket{\bar{1}} \otimes \ket{\bar{1}}}{2}} $ in $t=\pi/(8 \norm{\alpha}^2 \mathcal{E}_{c})$, where $\mathcal{E}_{c}$ is the coupling strength between the two cavities. As was the case for $X_\phi$ rotation, to remain in the Hilbert space of the two qubits, a small $\mathcal{E}_{c}$ is preferable, but time and noise considerations pushes us to be fast. See Table \ref{tab:gates} for the times and fidelities for $\mathcal{E}_{c}=\mathcal{E}_p^0/15$, $\mathcal{E}_{c}=\mathcal{E}_p^0/25$, and $\mathcal{E}_{c}=\mathcal{E}_p^0/55$ for the three ratios of $K/\kappa$ respectively.\par

This set of values for the single photon driving amplitude $\mathcal{E}_{x}$ and the coupling strength $\mathcal{E}_{c}$ has been chosen after running many simulations, and taking different values in each of these simulations to arrive at a set which gave reasonably good fidelities. But optimum amplitudes, possibly in the form of time dependent functions, can be obtained using GRAPE, and they might yield slightly higher fidelities.\par

In addition to the two-qubit rotations $G_\theta$, we need additional single qubit operations to perform a CNOT gate in our chosen basis of $\ket{C_\alpha^+}$ and $\ket{C_\alpha^-}$. One of the sequence of operations which yields a CNOT gate with the first qubit as the control and the second qubit as the target is 
\begin{equation}
\mathrm{X_{\pi/2}^2 X_{-\pi/2}^1 Z_{\pi/2}^1 G_{\pi/2} X_{-\pi/2}^1 Z_{-\pi/2}^1 X_{\pi/2}^1}
\end{equation}
Here $\mathrm{X_\phi}$ and $\mathrm{Z_\phi}$ are the respective single qubit rotations around those axes. The superscript denotes the qubit on which those gates act. Table \ref{tab:gates} shows the resulting fidelities and time taken for different parameters. \par

\subsection{Compatibility of 3-D microwave cavities with additional elements and operations}\label{sec:compatibility}

\begin{table*}
\centering
\includegraphics[width=0.9\linewidth]{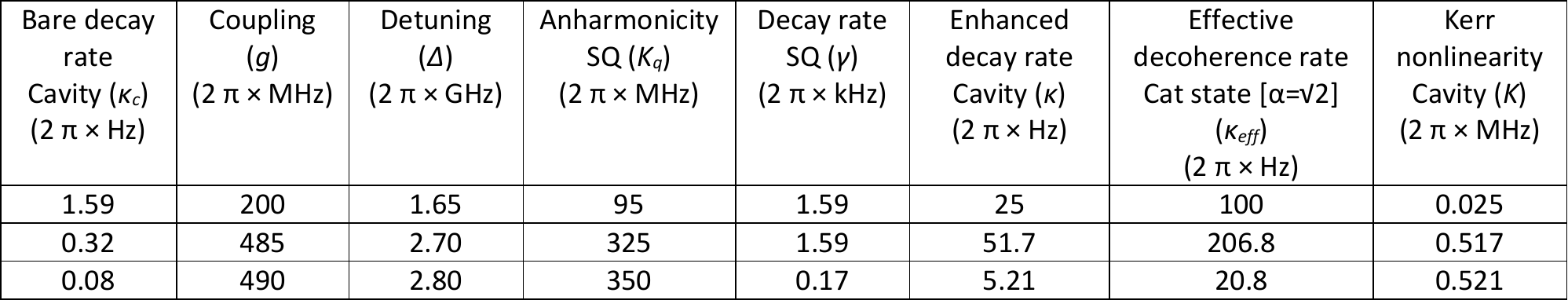}
\caption{Estimating the effect of a SQ placed inside the cavity on its coherence time and on the Kerr nonlinearity for photons. The decay rate of the cavity ($\kappa$) is affected by the SQ-cavity coupling strength ($g$), the detuning ($\Delta$), and the decay rate of the SQ ($\gamma$). The decay rate of the bare cavity ($\kappa_c$) is shown in column 1. The effective coherence time of the cat state ($\kappa_{eff}$), with $\alpha=\sqrt{2}$, is tabulated in the second-last column. The last column provides the estimated magnitudes of Kerr nonlinearity $K$, which is related to the above parameters and the anharmonicity of the SQ ($K_q$).}
\label{tab:kappa}
\end{table*}

\subsubsection{Effect of a Kerr nonlinear element on the decay rate of a microwave cavity}\label{sec:decay_rate}

The Hamiltonian of a weakly anharmonic SQ coupled to a cavity can be expressed in the form 
\begin{equation}\label{eqn:Hamil_SQ_cavity}
H=\omega_c a^{\dag}a+\omega_q b^{\dag}b - K_q {b^{\dagger}}^2 b^2 + g(a^{\dag}b+a b^{\dag}) 
\end{equation}

Here, $a$ is the annihilation operator for the cavity mode with frequency $\omega_c$, and $b$ is the annihilation operator for the SQ mode with $\omega_q$ as the transition frequency between its ground state and the first excited state. $K_q$ is the anharmonicity of the SQ mode, and $g$ is the coupling strength between the cavity and qubit modes. \par

The qubit-cavity interaction hybridizes these modes, and in the dispersive regime, where the coupling $g$ is much less than the detunings between the cavity and qubit transition frequencies, the cavity acquires a bit of the SQ component and vice-versa. Consequently, a part of the nonlinearity of the SQ mode ($K_q$) is also inherited by the cavity and in the dressed picture, there appears a Kerr nonlinearity term of the form $-K {a^{\dagger}}^2 a^2$ (see the Hamiltonian in Eqn.\ \ref{eqn:single_qubit}), where the mode $a$ now has a small SQ component. $K$ can be estimated by observing the energy spectrum of the eigenstates of the full Hamiltonian given in Eqn.\ \ref{eqn:Hamil_SQ_cavity}. We do so numerically by diagonalising the above Hamiltonian and observing the energy difference between consecutive cavity levels $\omega_{\ket{i+1,0}}-\omega_{\ket{i,0}}$. Here, $\ket{i,j}$ denotes an eigenstate of the above Hamiltonian with $i$ and $j$ number of excitations in the dressed cavity and SQ mode respectively. We verify that the energy difference between consecutive cavity levels follow the trend expected from a Kerr nonlinearity term in the Hamiltonian. Our way of obtaining the Kerr nonlinearity numerically agrees well with those seen in experiments \cite{13_Observation_Schoelkopf}. 

The coherence times mentioned in ref.\ \cite{07_Ultrahigh_Visentin,13_Reaching_Schoelkopf,18_3D_Grassellino} are for a bare cavity. But due to the hybridization of the cavity-SQ modes, the lifetimes of both the cavity and the SQ change. Since a SQ has a shorter coherence time than our desired kind of cavities, the SQ-cavity interaction will shorten the lifetime of the cavity. We therefore need to take this effect into account. \par 

To the lowest order in $g/\Delta$, the `inverse-Purcell' enhanced decay rate of the cavity $\kappa \approx (1-(g/\Delta)^2)\kappa_c+(g/\Delta)^2 \gamma$ \cite{07_Generating_Schoelkopf,16_Quantum_Schoelkopf,09_Dispersive_Blais}. Here, $g$ is the qubit-cavity coupling strength, $\Delta=\omega_q-\omega_c$ is the detuning between the cavity and the qubit, and $\kappa_c$ and $\gamma$ are the decay rates of the bare cavity and the SQ respectively. Taking the above $\kappa$, the effective coherence time of a cat state in the cavity in the presence of the two photon drive, $\kappa_{eff}$, is numerically found by fitting the evolution of coherence as a decaying exponential. It is approximately equal to $2 \kappa \norm{\alpha}^2$, where $\norm{\alpha}$ is the size of the coherent state. 

For experimentally conceivable values of $\kappa_c$ \cite{07_Ultrahigh_Visentin,13_Reaching_Schoelkopf,18_3D_Grassellino}, $\gamma$ \cite{19_SQs_Oliver,12_SQ_Steffen}, $g$ \cite{07_cQED_Schoelkopf,13_Observation_Schoelkopf,17_Microwave_Nori}, $\Delta$ \cite{13_Observation_Schoelkopf,16_Quantum_Schoelkopf}, and $K_q$ \cite{12_Black_Girvin, 12_Josephson_Blais, 13_Autonomously_Devoret, 16_Quantum_Schoelkopf}, the calculated $K$ and $\kappa$ are listed in Table \ref{tab:kappa}.

We note that due to the difference of as much as four orders of magnitude between $\kappa_c$ and $\gamma$, $\kappa$ is practically determined by $\gamma$. On the one hand, this relaxes the requirement on the quality factor of the bare cavity that is needed for our implementation. As an example, for $K/\kappa=10^4$, instead of taking $\kappa_c=2 \pi \times 0.32$ Hz, if we assume an order of magnitude larger $\kappa_c=2 \pi \times 3.2$ Hz, the resulting $\kappa$ stays almost the same (aprox.\ $5 \%$ variation). But on the other hand, large $\gamma$ makes achieving large $K/\kappa$ ratios difficult, as both $K$ and (the effect of $\gamma$ on) $\kappa$ increase simultaneously, though not exactly in the same way. Since the dependence of $K$ and $\kappa$ on the experimental parameters is not identical, it is still possible to obtain high $K/\kappa$ ratios with appropriate values for these parameters (see Table \ref{tab:kappa}). It should be pointed out that our selection of these values is based on running multiple simulations, ensuring that we stay in the dispersive regime, and from the intuition derived indirectly from approximate perturbative expressions in ref.\ \cite{07_Charge_Schoelkopf, 12_Black_Girvin, 13_Observation_Schoelkopf}. It is likely that a proper optimization of the relevant parameters might yield slightly better values for $K$ and $\kappa$ than the ones we have used. However, even with the values we have taken, we still manage to obtain high entanglement generation rates and fidelities, as discussed in the next section (Sec.\ \ref{sec:rate_and_fidelity}).\par

It is worth noting that improving $\gamma$ of the SQs will significantly improve the achievable $K/\kappa$ ratios. If the SQs were as long-lived as the cavities, then $K/\kappa$ ratios of the order of $10^7$ should be achievable with the parameters in Table \ref{tab:kappa}.\par

\subsubsection{Compatibility of 3-D microwave cavities with the transduction operations}\label{sec:compatibility_transduction}
Here we discuss the compatibility of high quality factor (high Q) 3-D superconducting cavities with the transduction operations, specifically those involving magnetic fields necessary for our protocol.\par

There is an increasing effort towards designing 3-D microwave cavities that would permit magnetic field manipulation of SQs \cite{16_3D_Fedorov,18_Fast_Kirchmair}, and spin systems \cite{18_Applying_Wallraff,18_Loop-gap_Kubo} inside them. The two most common elements used to build high Q superconducting cavities are Aluminium (Al) and Niobium (Nb) \cite{07_Ultrahigh_Visentin,13_Reaching_Schoelkopf,18_3D_Grassellino}. The demand for magnetic field at least in tens of mT for our transduction protocol \cite{18_Electron_Thiel,13_Anisotropic_Bushev} is difficult for cavities made out of pure Al, which loses its superconductivity at a criticial magnetic field strength, $H_c$ $\approx 10$ mT \cite{54_Superconducting_Eisenstein}, unless one includes a pathway for the magnetic field lines to escape without interacting much with the Al walls, e.g.\ by making a hybrid cavity including Copper (Cu) \cite{16_3D_Fedorov}. However, many of these designs would still result in slightly lower Q cavities for high magnetic fields. As a better alternative, one can opt for cavities made out of Nb, whose $H_c$ $\approx 200$ mT \cite{54_Superconducting_Eisenstein,01_Critical_Saito} can sustain the magnetic field strengths needed in our protocol. Another concern here is that the SQs (placed inside these cavities) themselves are predominantly made of Al, and it is important to shield them from strong magnetic fields. The fields generated in some experiments, e.g.\ in ref.\ \cite{18_Applying_Wallraff} have a gradient like character, and we can therefore place SQs close to regions of weak magnetic field while the transducer close to the more intense regions. \par 

It should be pointed out that since the transducer is placed in the cavity generating the flying qubits, and not the one which is supposed to act as a memory, the requirement of long coherence time is less stringent on this cavity. If, however, these operations turn out to be too difficult to be performed in the cavity while maintaining reasonably good coherence, one can adopt an alternative route. After the undriving operation to a microwave Fock state, one could transfer the microwave photon to a lower finesse cavity, possibly even a planar resonator, and perform transduction there. To make the operation faster, it should be possible to tune the decay rate of the original cavity to approximately 4 orders of magnitude higher than its intrinsic value \cite{13_Catch_Martinis, 18_On-demand_Schoelkopf}.  

\section{Entanglement distribution rates and fidelities}\label{sec:rate_and_fidelity}  

In the previous section, we presented some experimental values which will be relevant here to calculate the possible entanglement distribution rates for our repeater protocol, and to estimate the fidelities of the final states. We shall compare the rates with those possible with direct transmission, with the popular ensemble based DLCZ repeater approach \cite{01_Long_Zoller}, and with a recently proposed repeater approach using single rare-earth (RE) ions in crystals \cite{18_Quantum_Simon}. \par

The average time to distribute entanglement over length $L$ using a quantum repeater scheme with $2^n$ elementary links, each of length $L_0$ is  \cite{09_Quantum_Simon, 11_Quantum_Gisin, 07_Quantum_repeaters_Gisin, 18_Quantum_Simon} (see the Supplementary Information for a brief explanation of the formula)
\begin{equation}\label{eqn:time_entanglement_oper}
\left\langle T \right\rangle_L= (\frac{3}{2})^n (\frac{L_0}{c}+T_{o}) \frac{1}{P_0 P_1 P_2....P_n}
\end{equation}
Here, $T_{o}$ is the time taken for the local operations at one node for one elementary link, $P_0$ is the success probability of generation of entanglement in an elementary link, and $P_i$ is the success probability of entanglement swapping in the $i^{th}$ nesting level. The local operations include the gates and the driving, undriving, and transduction operations (see Fig.\ \ref{fig:schematic_protocol}) at each node. For our protocol, $P_0 = \frac{1}{2}\eta_t p {\eta_o}^2$. The transmission efficiency in the fiber is $\eta_t=e^{-\frac{L_0}{L_{att}}}$, where $L_{att}$ is the attenuation length. The probability of emitting a telecom photon from a cavity into the fiber mode is denoted by $p$. For our system, it is primarily the transduction efficiency.  And $\eta_o$ is the single (optical) photon detection efficiency. $P_i$ is mainly governed by the (stationary) qubit readout efficiency. Cat states can be read out in a Quantum Non-Demolition (QND) fashion with close to unit efficiency by measuring the parity \cite{14_Tracking_Schoelkopf,16_Extending_Schoelkopf} (see section S1.G in the Supplementary Information for a brief discussion on parity measurements). Another way to read-out cat states, is to un-drive them to Fock states, and then read out the number of photons non-destructively, using a similar technique as before \cite{10_QND_Schoelkopf}. Since two such measurements are needed for entanglement swap, $P_i={\eta_m}^2$, where $\eta_m$ is the QND efficiency. Values close to unity are expected for our long storage-time cavities.

\begin{figure}
\centering
\includegraphics[width=0.45\textwidth]{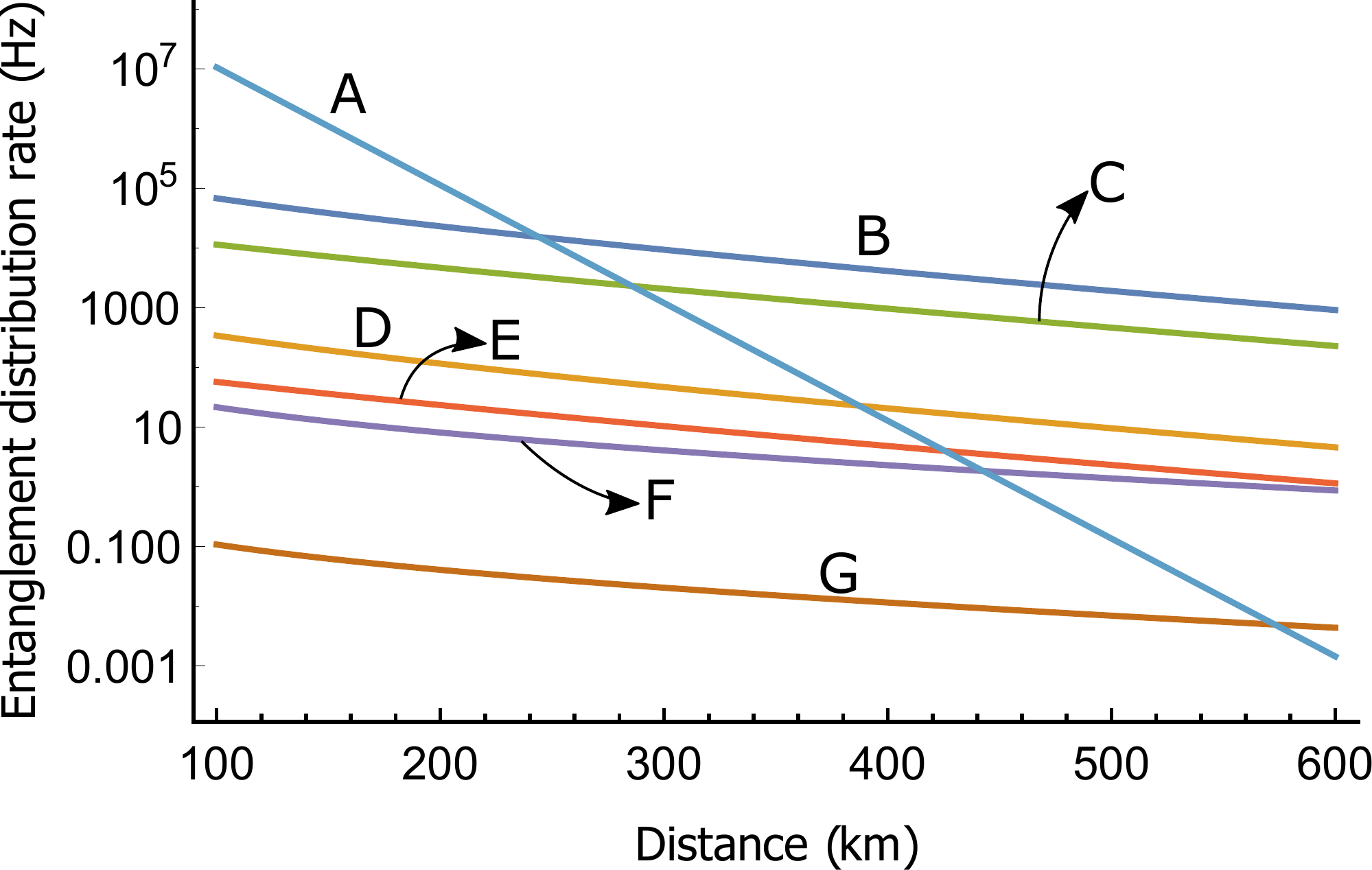}
\caption{(color online) Entanglement distribution rate as a function of distance for different schemes. `A' is direct transmission in a fiber with a 1 GHz entangled photon source, `B' and `C' are the multiplexed ($m=200$) versions of our cat scheme and the rare-earth (RE) scheme \cite{18_Quantum_Simon} respectively, and `D' and `E' are the non-multiplexed versions ($m=1$) of these schemes. `F' and `G' are the multiplexed and non-multiplexed versions of DLCZ scheme. The number of nesting levels, $n=3$ for all the repeater schemes.}\label{fig:entangle_distribn_rate}
\end{figure}

The entanglement distribution rate is the inverse of the average time needed to distribute entanglement. For the multiplexed architecture the rate is multiplied by $m$, since there are $m$ independent sets of channels operating in parallel.

In Fig.\ \ref{fig:entangle_distribn_rate}, we plot entanglement distribution rates for our proposed protocol as a function of the total distance $L$ and compare these with the rates possible using direct transmission and a source producing entangled photons at 1 GHz \cite{18_Highly_Schmidt}, with the rates achieved using the DLCZ protocol \cite{01_Long_Zoller,05_Measurement_Kimble}, and with those achieved using single RE ions (RE) in crystals \cite{18_Quantum_Simon}. We choose $K/\kappa=10^5$ and $n=3$ for our repeater protocol. In the Supplementary Information, we discuss the rationale behind choosing this particular nesting level, and also show the rates and fidelities for other nesting levels and other ratios of $K/\kappa$. We take realistic values for other pertinent experimental parameters, e.g.\ $p=0.8$, $P_i=0.9$, $\eta_d=0.9$, and $L_{att}=22$ km. The relevant parameters for DLCZ include the single photon generation probability and the memory efficiency, which too have been taken realistically as 0.01 and 0.9 respectively \cite{01_Long_Zoller,05_Measurement_Kimble,11_Quantum_Gisin}. Since the operation times for DLCZ protocol can in principle be fast enough to be ignored \cite{05_Measurement_Kimble}, we do not include those here. For the single RE ion based approach, apart from the parameters already considered for the rate calculation in ref.\ \cite{18_Quantum_Simon}, we have taken into account a realistic operation time of 0.1 ms. The figure also depicts the increased rates obtained by multiplexing ($m=200$) for all three repeater protocols. Our protocol yields higher rates than the other two repeater protocols. \par

Quantum repeaters would be practically more useful when the rate of distribution of entanglement surpasses that of direct transmission at the so called `cross-over' point. From the plot, the cross-over points for our protocol for $m=1$ and $m=200$ are 387 km and 244 km respectively. The fidelity of the final entangled state can be estimated by subsequently multiplying the fidelities of different operations and the residual coherence of the storage cavities. 

\begin{equation} \label{eqn:fidelity}
F_{tot}=  (F_{elem})^{l} \times (F_{swap})^{l-1} \times C_R
\end{equation}

Here, $F_{tot}$ is the fidelity of the final entangled state, $F_{elem}=\prod_{i} F_{O_i}$ is the fidelity of generation of entanglement in an elementary link, where $O_i$ are the various local operations, e.g.\ the gates, the driving and undriving of the cavities, and the transduction from microwave to optical frequencies. $F_{swap}$ is the fidelity of entanglement swap, and $l=2^n$ is the number of elementary links, where $n$ is the nesting level; the number of entanglement swaps needed to distribute entanglement over the whole length, $L$, is $l-1$. $C_R$ is the residual coherence of the storage cavities, only taking into account the decoherence because of the non-zero waiting times. \par

The residual coherence ($C_R$) for cat states is approximated as $e^{-\kappa_{eff} T}$, where $k_{eff}$ is the cat's decoherence rate and $T$ is the average time needed to distribute entanglement. One relatively simple way to have a larger $C_R$ is by undriving the storage cavities from the cat basis to the Fock basis before the waiting times. The decoherence rate in the Fock basis is approximately $\kappa$, which is less than the decoherence rate ($k_{eff}\approx 2\kappa|\alpha|^2$) of our cat states. The cavities would then be driven back to the cat basis when the additional operations need to be performed. Moreover, since even $\kappa$ is much greater than the decay rate of the bare cavity, $\kappa_c$, because of the relatively lossy SQ coupled to it, one could increase $C_R$ further by transferring the Fock state to an effectively longer lived microwave cavity for storage \cite{13_Catch_Martinis, 18_On-demand_Schoelkopf, 18_Deterministic_Wallraff, 18_Perfect_Sun}. The residual coherence can also be enhanced with appropriate error-correction codes \cite{01_Encoding_Preskill, 13_Hardware_Mirrahimi, 14_Dynamically_Devoret,15_Confining_Devoret, 16_Extending_Schoelkopf,17_Degeneracy_Mirrahimi,18_Coherent_Devoret, 18_Fault_Schoelkopf, 18_Stabilized_Girvin}. For suitably encoded cat states, error-correction has been experimentally implemented to increase the coherence time by a few times \cite{16_Extending_Schoelkopf}, and error detection has been performed in a fault-tolerant way \cite{18_Fault_Schoelkopf}. Note that our present protocol would require quantum error correction for qubits encoded in states of different parity. See the Supplementary Information for a discussion on different approaches to increase the residual coherence. 

In addition to the fidelities of the different operations calculated explicitly using QuTip (see the Supplementary Information for a list of all the operations considered), if we take the fidelity for transduction as $99.95\%$, comparable to the fidelities of several of the gate operations in Table \ref{tab:gates}, the highest fidelities of the final entangled states calculated for $m=1$ and $m=200$ at the cross-over points are approximately $91\%$ and $92\%$ for $m=1$ and $m=200$ respectively (see the Supplementary Information for details). \par

Considering only factors such as a non-zero single photon generation probability, and imperfect memory and detection efficiencies, and using the already mentioned realistic values for them (0.01, 0.9 and 0.9 respectively), the fidelity of the final state for DLCZ repeater protocol with $n=3$ would be roughly $75 \%$ \cite{11_Quantum_Gisin}. However this should be treated as kind of an upper bound, since we have not considered the finite fidelities of other operations, e.g.\ the read and write processes, the finite coherence times of the memories, and the effect of the phase fluctuations of the fibers, which would all contribute towards bringing down the final fidelity. For the RE ion based approach, an upper bound on fidelity, estimated considering only entanglement swapping and spin mapping ($\mathrm{Er^{3+}-Eu^{3+}}$) operations, is roughly $80\%$.\par

Using established protocols of entanglement purification, the fidelity of our final states can be significantly increased as they cross the threshold above which purification is possible \cite{07_Entanglement_Briegel}. Here, however, we do not incorporate error-correction or entanglement purification in our architecture, but we believe that those should be possible (see the Supplementary Information for a brief discussion on error-correction).\par 

\section{Conclusion and Outlook}\label{sec:conclusion}
Recently, orthogonal cat states in microwave cavities \cite{16_Universal_Goto, 17_Engineering_Blais} have been proposed as promising qubits for quantum computation. Their long coherence times also make them attractive in a quantum network context. Here, we proposed a robust scheme to generate entanglement between distant cat-state qubits, mediated by telecom photons in conventional optical fibers. As an important step of this entanglement generation scheme, we expanded on a specific microwave to optical transduction protocol some of us recently proposed using a rare-earth ion ($\mathrm{Er^{3+}}$) doped crystal \cite{18_Electron_Thiel}. We also designed a quantum repeater architecture using these cat state qubits and rare-earth ion based transducers. Our calculations show that higher entanglement distribution rates than that possible with direct transmission or with two other repeater approaches, including the well-estalished DLCZ approach, can be achieved, while maintaining a high fidelity for the final state in the presence of realistic noise and losses, even without entanglement purification or quantum error correction. By increasing the size of the local quantum processing nodes, our proposed approach can naturally be extended to a full-fledged distributed quantum computing architecture, thus paving the way for the quantum internet \cite{08_Quantum_Kimble,18_Internet_Hanson,17_Towards_Simon}.\par

We hope that our results will stimulate new experimental and theoretical work involving these cat state qubits and transducers. It would be useful to think about ways to generate higher Kerr nonlinearities and to diminish the effect of the SQ on the decay rate of the cavity, possibly by engineering novel Hamiltonians and by designing novel forms of filters respectively \cite{19_Fast_Buisson, 12_Josephson_Blais, 19_Quantum_Oliver}. Regarding cat state qubits, future work may include developing and incorporating efficient quantum error-correction codes \cite{15_Confining_Devoret, 17_Degeneracy_Mirrahimi, 18_Coherent_Devoret}, and possibly envisioning schemes to protect against phase decoherence. It is worth mentioning that although we have chosen microwave cat-state qubits, rather than the more obvious (microwave) Fock state qubits as our computational units, since quantum computation with error correction seems to be currently more mature with cat states, our entanglement generation protocol, and hence the overall repeater framework can straightforwardly be applied to Fock state qubits. Recently, new error correction codes have been proposed for Fock state qubits \cite{16_New_Girvin, 18_Performance_Jiang}. There have consequently been a few experiments to demonstrate error-correction \cite{18_Demonstration_Sun}, as well as single and two qubit gates \cite{18_CNOT_Schoelkopf, 19_Entanglement_Schoelkopf, 18_Demonstration_Sun} on suitably encoded Fock state qubits. The recent work thus tightens their competition with cat state qubits, and it will be interesting to see how the field progresses.\par 

Within the areas of modular and distributed quantum computing, it is of interest to think about how best to connect different computational units to efficiently distribute an arbitrary computational task \cite{13_Efficient_Stather,15_Efficient_Brierley,14_Freely_Simon, 14_Large-scale_Monroe}. Regarding the quantum communication aspect, one would need to figure out efficient repeater architectures for a truly global quantum network, possibly using both ground based and (quantum) satellite links \cite{05_Experimental_Pan,15_Entanglement_Simon,17_Towards_Simon,17_Satellite_Pan}.\par 

\begin{acknowledgments}
We thank F.\ Kimiaee Asadi, M.\ Falamarzi Askarani, S.\ Barzanjeh, A.\ Blais,  S.\ Goswami, J.\ Moncreiff, D.\ Oblak, S.\ Puri, A.\ Saxena, W.\ Tittel, and S.\ Wein for useful discussions. We are further thankful to A.\ Blais for providing detailed comments on an earlier version of the manuscript, to S.\ Puri for providing some of the codes used in ref. \cite{17_Engineering_Blais}, and to A.\ Saxena for his help in the initial phase of the project. S.K.\ was supported by an Eyes High Doctoral Recruitment Scholarship (EHDRS) and an Alberta Innovates Technology Futures (AITF) Graduate Student Scholarship. N.L.\ was supported by an Eyes High Post-Doctoral Fellowship (EHPDF) and by the Alliance for Quantum Technologies' (AQT's) Intelligent Quantum Networks and Technologies (INQNET) research program. C.S. acknowledges funding from the Natural Sciences and Engineering Research Council (NSERC) through its Discovery Grant and CREATE grant ``Quanta'', and from Alberta Innovates through ``Quantum Alberta''. 
\end{acknowledgments}
\putbib{}
\end{bibunit}

\begin{bibunit}
\pagebreak
\clearpage
\widetext
\begin{center}
\textbf{\large Supplementary Information for ``Towards long-distance quantum networks with superconducting processors and optical links''}
\end{center}

\setcounter{equation}{0}
\setcounter{figure}{0}
\setcounter{table}{0}
\setcounter{section}{0}
\setcounter{page}{1}
\makeatletter
\renewcommand{\theequation}{S\arabic{equation}}
\renewcommand{\thefigure}{S\arabic{figure}}
\renewcommand{\thetable}{S\arabic{table}}
\renewcommand{\thesection}{S\arabic{section}}
\renewcommand{\bibnumfmt}[1]{[S#1]}
\renewcommand{\citenumfont}[1]{S#1}

\section{Entanglement distribution rates and fidelities}\label{supp_sec:rates_and_fidleities}
In this section, we present the entanglement distribution rates and fidelities for different ratios of $K/\kappa$ and for different nesting levels $n$. We will try to build an intuition for the realistic distances over which entanglement can be distributed, and how best to beat direct transmission, for different experimental parameters. We shall consequently provide a justification for choosing $n=3$ for $K/\kappa=10^5$ in the main text.\par

\subsection{Description of the rate formula}
For a quantum repeater, it is intuitive to expect the average time required to distribute entanglement over the total distance $L$ to be of the form \cite{11_Quantum_Gisin} 

\begin{equation}\label{eqn:time_entanglement_no_oper}
\left\langle T \right\rangle_L= (\frac{L_0}{c}) \frac{f_0 f_1 f_2....f_n}{P_0 P_1 P_2....P_n}
\end{equation}

when the time needed to perform the operations is negligibly small compared to the communication time $L_0/c$ to and from the beamspitter placed in between the neighbouring nodes. $P_0$ is the probability that entanglement is generated in one elementary link in a single attempt, determined primarily by the losses in the fiber and the inefficiencies of the detectors. $P_i$ is the probability of successful entanglement swap in the $i^{th}$ nesting level. If all the components are ideal and lossless, and the operations are all deterministic, then $P_0=P_1=P_2=....P_n=1$. Non-unit efficiency of realistic operations obviously increases the average time. The factors $f_i$ arise because one has to have two neighbouring links entangled in the nesting level $i-1$ before attempting the swap operation for the nesting level $i$. They should all thus be in the range $1 \leq f_i \leq 2$. To our knowledge, an analytic expression for a general $f_i$ has not been derived so far, but to the lowest order in $P_0$, all the $f_i$'s can be well approximated to be 3/2 \cite{11_Quantum_Gisin}.\par

Now, if one takes into account the finite times for performing all the local operations, then in addition to the communication time $L_0/c$, one has to spend an additional time $T_o$ every time one makes an entanglement generation attempt in an elementary link. Thus, we obtain the formula  used in the main text of the paper (Eqn.\ 7 in the main text), i.e.\

\begin{equation*}
\left\langle T \right\rangle_L= (\frac{3}{2})^n (\frac{L_0}{c}+T_{o}) \frac{1}{P_0 P_1 P_2....P_n}
\end{equation*}

The entanglement distribution rate is the inverse of the average time needed to generate entanglement. For a multiplexed architecture, where `$m$' spectrally distinct sets of channels operate in parallel, the entanglement distribution rate is simply $m$ times the rate obtained in a single set of channels.

\subsection{Entanglement fidelity}
The fidelity of the final entangled state, $F_{tot}$, can be estimated by subsequently multiplying the fidelities of all the individual operations and the residual coherence of the memories, as shown in the following equation. 

\begin{equation*}
F_{tot}=  (F_{elem})^{l} \times (F_{swap})^{l-1} \times C_R
\end{equation*}

Here, $F_{elem}=\prod_{i} F_{O_i}$ is the fidelity of generation of entanglement in an elementary link, where $O_i$ are the various local operations, e.g.\ the gates, the driving and undriving of the cavities, and the transduction from microwave to optical frequencies. $F_{swap}$ is the fidelity of entanglement swap, and $l=2^n$ is the number of elementary links, where $n$ is the nesting level; the number of entanglement swaps needed is $l-1$. $C_R$ is the residual coherence of the storage cavities, taking into account the decoherence because of the non-zero waiting times. \par

\subsubsection{Residual coherence of the storage cavities}\label{sec:residual_coherence}
While waiting for entanglement to be established in neighbouring links, the cat states in the storage cavities decohere at the rate $k_{eff}=2 \kappa \alpha^2$, where $\kappa$ is the single photon decay rate and $\alpha$ is the amplitude of the coherent state. If the average time needed to distribute entanglement is $T$, then the residual coherence left in the storage cavities is approximately $C_R=e^{-\kappa_{eff} T}$. For $\alpha=\sqrt{2}$, $k_{eff}=4 \kappa$. Although error-correction can, in principle, correct these errors (see Section \ref{sec:error-correction}), currently an easier way to increase the residual coherence ($C_R$) is to undrive the cavity from the cat basis to Fock basis while waiting. Once entanglement is confirmed to be generated in the neighbouring links, one can then drive the cavity back to the cat basis to perform the logical operations. The trade-off in doing so would be a slight reduction in $F_{elem}$ and a slight increase in $T_{oper}$ because of the additional driving and undriving operations. But since these operations are quite fast and proceed with very high fidelity (see Table I in the main text), we find that, as compared to storing the photonic states in the cat basis, taking this route increases the final fidelity in almost all the cases. Therefore, unless stated otherwise, we shall adopt this route here to estimate our final fidelities. However, this interconversion is not a necessity, and keeping the photonic states in the cat basis will yield good fidelities for many cases as well.\par

We saw in the main paper that because of the difference of as much as four orders of magnitude in the loss rates of the cavity and the SQ coupled to it, $\kappa$ is essentially determined by the decay rate of the SQ, $\gamma$ (see Table II in the main text). We need strong dispersive coupling to the SQ to generate high Kerr nonlinearities, which makes the dependence of $\kappa$ on $\gamma$ stronger too. Under such a scenario, $\kappa \gg \kappa_c$, where $\kappa_c$ is the decay rate of the bare cavity without the SQ. However, for storage in the Fock basis without error-correction, in principle, we do not need the qubit cavity interaction ($g$) at all. So, ideally we would want to switch off the interaction during storage to keep $\kappa \approx \kappa_c$ during the waiting times, and switch it on when the gates need to be performed. But, turning $g$ on and off is likely not possible in the (circuit-QED) experimental setup. One could work with the detunings between the SQ and the cavity, and that will reduce $\kappa$ to a certain extent. But significant detuning is also experimentally challenging. Another solution could be to transfer the photons (in Fock basis) from these cavities used for computation to other long-lived cavities, whose $\kappa \approx \kappa_c$, using the techniques available \cite{13_Catch_Martinis, 18_On-demand_Schoelkopf, 18_Deterministic_Wallraff, 18_Perfect_Sun}. As an illustration of how much better the final fidelity gets if the transfer happens perfectly to and from these cavities, in the tables below, we shall also list the final fidelities assuming storage in the best currently available 3-D microwave cavities.\par
One could enhance the residual coherence further by integrating our storage cavities with other solid-state quantum memories, e.g.\ phosphorus spins in silicon \cite{13_Room_Thewalt}, or rare-earth ensembles \cite{15_Optically_Sellars}, where coherence times up to hours are possible. Alternatively, one could still use our entanglement generation scheme to implement a more complicated multiplexed repeater protocol which is mostly insensitive to the short coherence times of the storage cavities \cite{07_Multiplexed_Kennedy}. It however needs more resources and more complicated operations, which would be experimentally challenging. \par

\subsection{One elementary link ($n=0$)}\label{supp_sec:n_0}
For entanglement generation in one elementary link, we need 6 driving operations, 2 $\mathrm{X_{\pi/2}}$ rotations, 4 CNOT gates, 4 undriving operations, 4 transduction operations, and 2 $\mathrm{X_{\pi}}$ rotations (see Fig.\ 2 in the main text). Inter-converting the photons from cat basis to Fock basis for storage adds 4 more undriving and undriving operations. Taking the fidelity for a single transduction operation to be $99.95\%$, and calculating the others numerically with QuTiP, the fidelities for an elementary link of length $\SI{50}{\kilo\meter}$ for the three sets of values of $K$ and $\kappa$ are presented in Table \ref{tab:n_0}(a)-(b). Table \ref{tab:n_0}(a) is for the case when we store in the Fock basis in the same cavity, and Table \ref{tab:n_0}(b) is when we transfer the photon perfectly to another cavity, whose lifetime is 10 s, which is comparable to the highest quality factor (Q) 3-D cavities available currently \cite{18_3D_Grassellino}.\par

The first column of Table \ref{tab:n_0} gives the values for the Kerr nonlinearity and the decay rate of the cavity. The next two columns have broken down the fidelity of the final state into its two important parts: (i) the fidelity of the local operations, which includes all the gates, driving and transduction operations, and (ii) the residual coherence of the storage cavities as they wait for local operations and communication tasks to be finished. We have the final fidelity in the next column, followed by one showing the possible entanglement generation rates. There are two sets of the above values, one for multiplicity $m=1$ and the other for $m=200$. We observe that for all these values, the fidelity exceeds the entanglement purification bound $50 \%$ by a significant margin, such that even with imperfect local operations, the fidelity of the distributed entangled states can be further enhanced, however at the expense of the distribution rate \cite{07_Entanglement_Briegel}. We see that though the transfer of photons to a long lived cavity improves the coherence and hence the final fidelity, the final fidelity even without doing this is sufficiently high, and thus this additional step is not necessary here. This table might be pertinent if one were to design an architecture to distribute entanglement within a relatively short distance, e.g.\ a city.\par

\begin{table*}[h]
\centering
\includegraphics[width=0.9\linewidth]{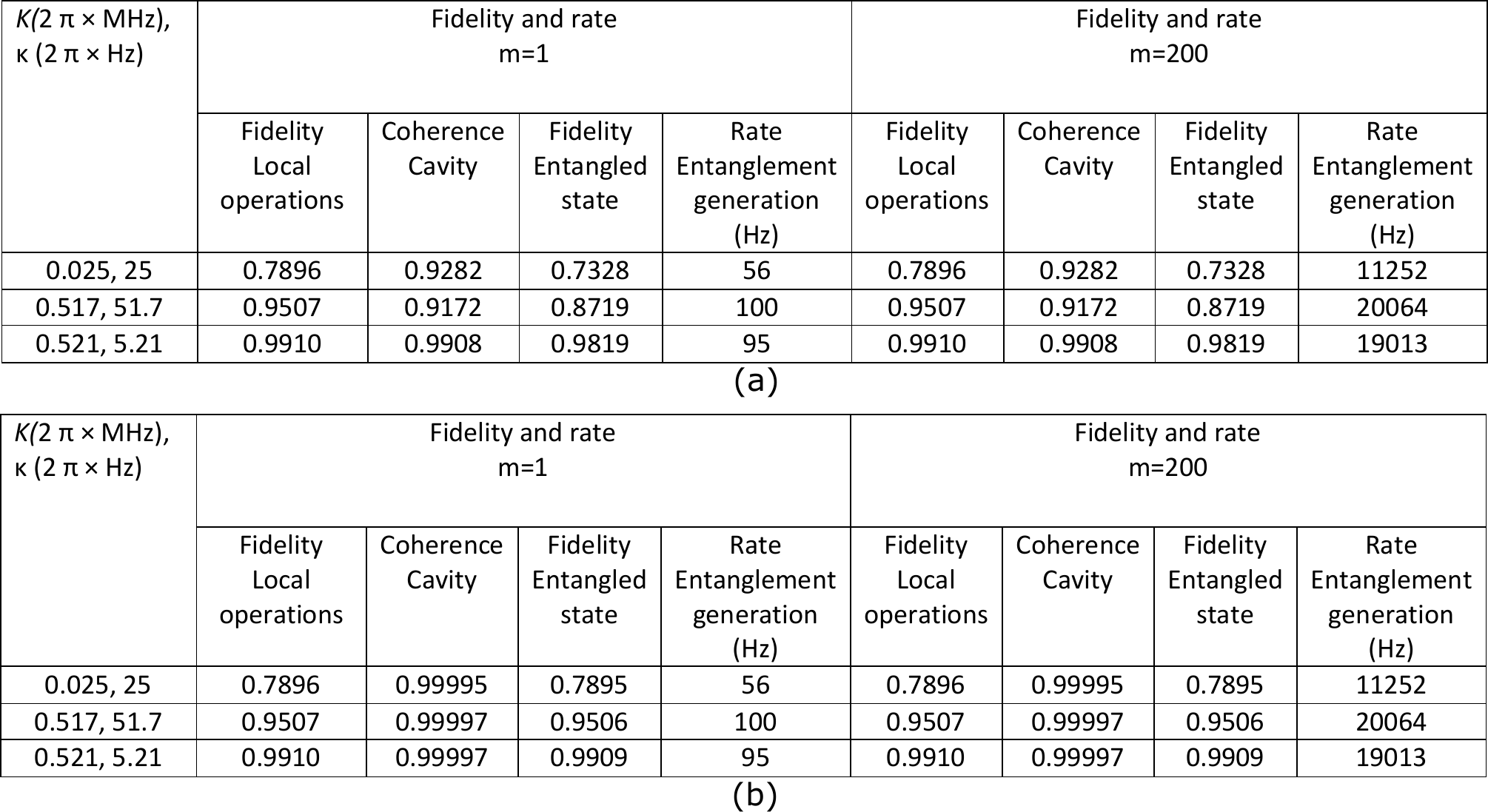}
\caption{Fidelities and entanglement distribution rates in an elementary link of length \SI{50}{\kilo\meter} for different values of $K$ and $\kappa$.(a) The photonic states are stored in the Fock basis in the same cavities used for computation. (b) Additional longer-lived cavities, whose lifetimes are \SI{10}{\second}, are used to store the photonic states (in Fock basis) during the waiting times. The fidelity of the final entangled state is estimated to be the product of the coherence of the cavity and the fidelity of local operations. One set of reading is for the non-multiplexed architecture ($m=1$), and another set is for the multiplexed architecture with multiplexing $m=200$.}
\label{tab:n_0}
\end{table*}

\subsection{Two elementary links ($n=1$)}\label{supp_sec:n_1}
The next table, i.e.\ Table \ref{tab:n_1} shows the fidelities for nesting level $n=1$, keeping the length of an elementary link the same, i.e.\ \SI{50}{\kilo\meter}. Here too, Table \ref{tab:n_1}(a) is for the case when the photons are stored in the same cavity, and  Table \ref{tab:n_1}(b) is for the case when they are transferred to effectively longer lived cavities for storage. In addition to a couple of sets of the above local operations required for an elementary link, we need to perform an entanglement swap, which includes a CNOT and a Hadamard at the sender and a rotation at the receiver. We include the fidelities of all these operations. We assume an X-Z rotation at the receiver as a worst case scenario. As compared to the previous table (Table \ref{tab:n_0}), we notice a drastic drop in coherence of the storage cavity for $m=1$ in Table \ref{tab:n_1}(a). This difference in coherence is because of the difference in the time duration for which the storage cavity needs to store entanglement. Even though the communication times remain the same, in a repeater, one has to wait till entanglement has been successfully established in a neighbouring link before one can perform the swap. For the previous case ($n=0$), there was no such waiting time required. This highlights the role of a good memory for repeaters. This also highlights the importance of multiplexing, where because of several parallel attempts, this wait time is reduced, and we get reasonable fidelities, for the same values of $K$ and $\kappa$ (see the column for $m=200$ in the first row). Since the cavities in Table \ref{tab:n_1}(b) have effectively much longer coherence times, they have good residual coherence even for $m=1$. Thus, if one wants high fidelity entangled states using multiplexing, one need not use the additional longer lived cavities, but if one wants high fidelity for $m=1$ as well, one would need to store the states in additional high-Q cavities. \par

\begin{table*}
\centering
\includegraphics[width=0.9\linewidth]{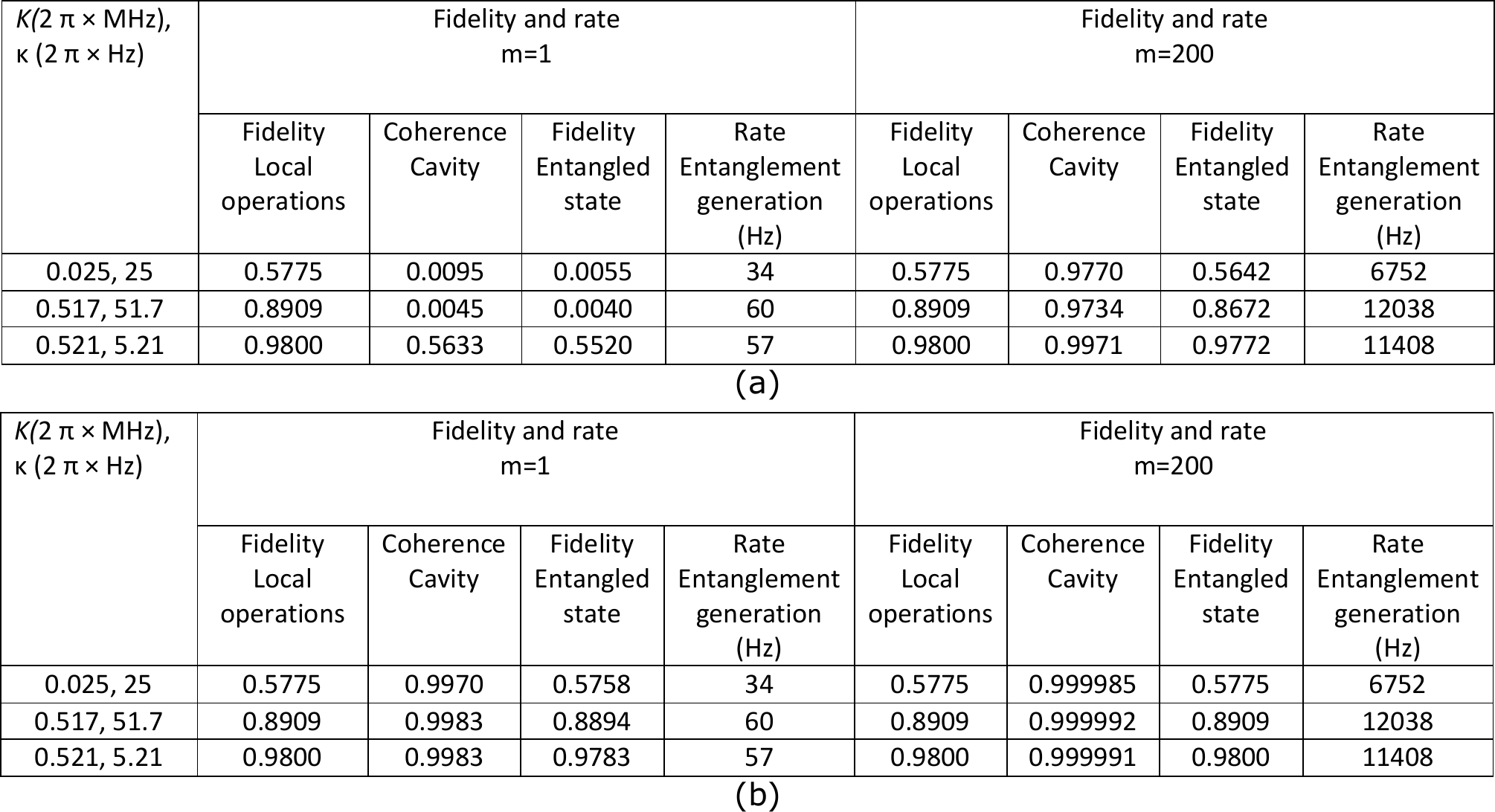}
\caption{Fidelities and entanglement distribution rates for nesting level $n=1$, taking  different values of $K$ and $\kappa$. The length of an elementary link is \SI{50}{\kilo\meter}. (a) The photonic states are stored in the Fock basis in the same cavities used for computation. (b) Additional longer-lived cavities are used to store the photonic states during the waiting times.  The first set of reading is for the non-multiplexed architecture ($m=1$), and the second set is for the multiplexed architecture with multiplexing $m=200$.}
\label{tab:n_1}
\end{table*}

\begin{table*}
\centering
\includegraphics[width=0.9\linewidth]{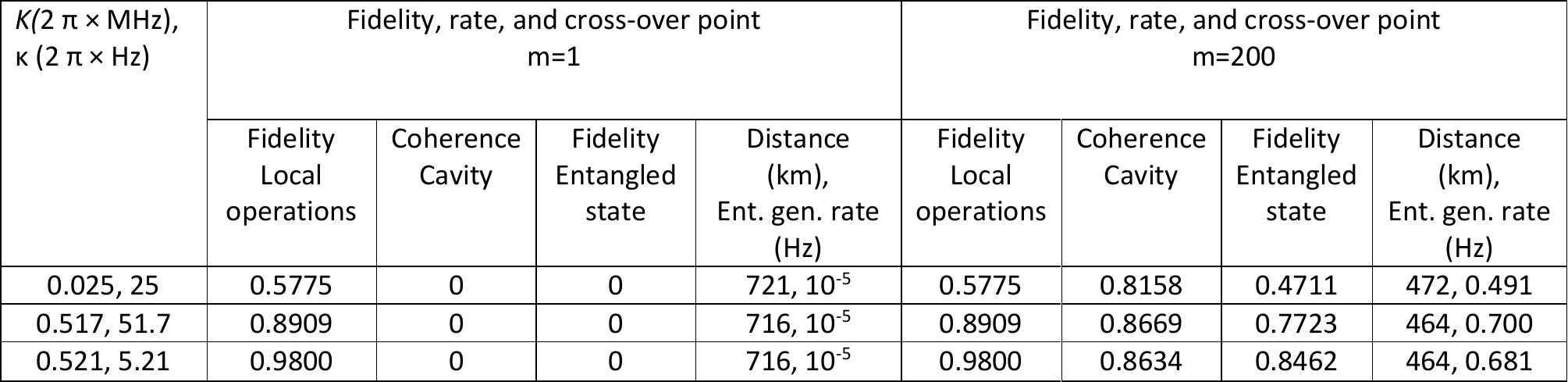}
\caption{Fidelities and entanglement distribution rates for nesting level $n=1$, taking  different values of $K$ and $\kappa$. The cross-over distances are listed in the fifth and ninth columns for different values of $K$ and $\kappa$. Additional longer-lived cavities are used to store the photonic states during the waiting times.}
\label{tab:n_1_dir_tra}
\end{table*}

A quantum repeater would practically be more useful when the rate of entanglement distribution using repeaters exceeds that of direct transmission using, for example, a \SI{1}{\giga\hertz} source of entangled photons. The distance at which this occurs is known as the cross-over point. For $n=1$, we numerically calculate that value to be close to \SI{700}{\kilo\meter} for $m=1$ and \SI{450}{\kilo\meter} for $m=200$ (see Table \ref{tab:n_1_dir_tra}). However, we find that if we store the states in the same cavities that we use for computation, then even the cavity with our best set of values of $K$ and $\kappa$ would decohere by the time we finish the operation. Instead, if we transfer the states to the longer lived cavities for storage, then the multiplexed version still yields usable rates and fidelities for higher $K/\kappa$ ratios. Table \ref{tab:n_1_dir_tra} shows the fidelities for that scenario. In the tables, fidelities smaller than $10^{-4}$ are written as 0.\par

For (low enough nesting levels in) our architecture, as we increase the number of nesting levels, this cross-over distance becomes smaller and so does the waiting time. Therefore we need to increase the nesting level and hope that we might be able to beat direct transmission with good fidelities. However, since the fidelity for local operations for the first set of $K$ and $\kappa$ is $57.75 \%$ for $n=1$, adding another nesting level would drop it below $50 \%$, rendering the generated entangled state unuseful for most quantum communication tasks. Therefore, we focus only on our second and third set of values of $K$ and $\kappa$ from now on. \par

\subsection{Four elementary links ($n=2$)}\label{supp_sec:n_2} 
Table \ref{tab:n_2} shows the fidelities for $n=2$ at the cross-over distances for the two sets of values of $K$ and $\kappa$. In addition to the fidelities, the table also shows the different cross-over distances. Without multiplexing ($m=1$), the average time needed to distribute entanglement (over the cross-over distance) for all $K/\kappa$ ratios considered is so large that the storage cavities decohere almost completely, yielding very low residual coherence and final fidelities (see Table \ref{tab:n_2}(a)). But with multiplexing ($m=200$), we see that the fidelity of the final entangled state for even our moderately good values of $K$ and $\kappa$ ($K/\kappa=10^4$) at the cross-over point (\SI{291}{\kilo\meter}) is $65.31 \%$, which is well above the purification threshold. For the $K/\kappa$ ratio of $10^5$, the final fidelity at the cross-over point (\SI{292}{\kilo\meter}) is  $94.01\%$. \par
If we transfer the photonic states to additional longer-lived cavities, then we find that the coherence of the cavities is not negligible anymore (see Table \ref{tab:n_2}(b)) for $m=1$. Even for the non-multiplexed version, one can beat direct transmission with useful fidelities. 
 
\begin{table*}
\centering
\includegraphics[width=0.9\linewidth]{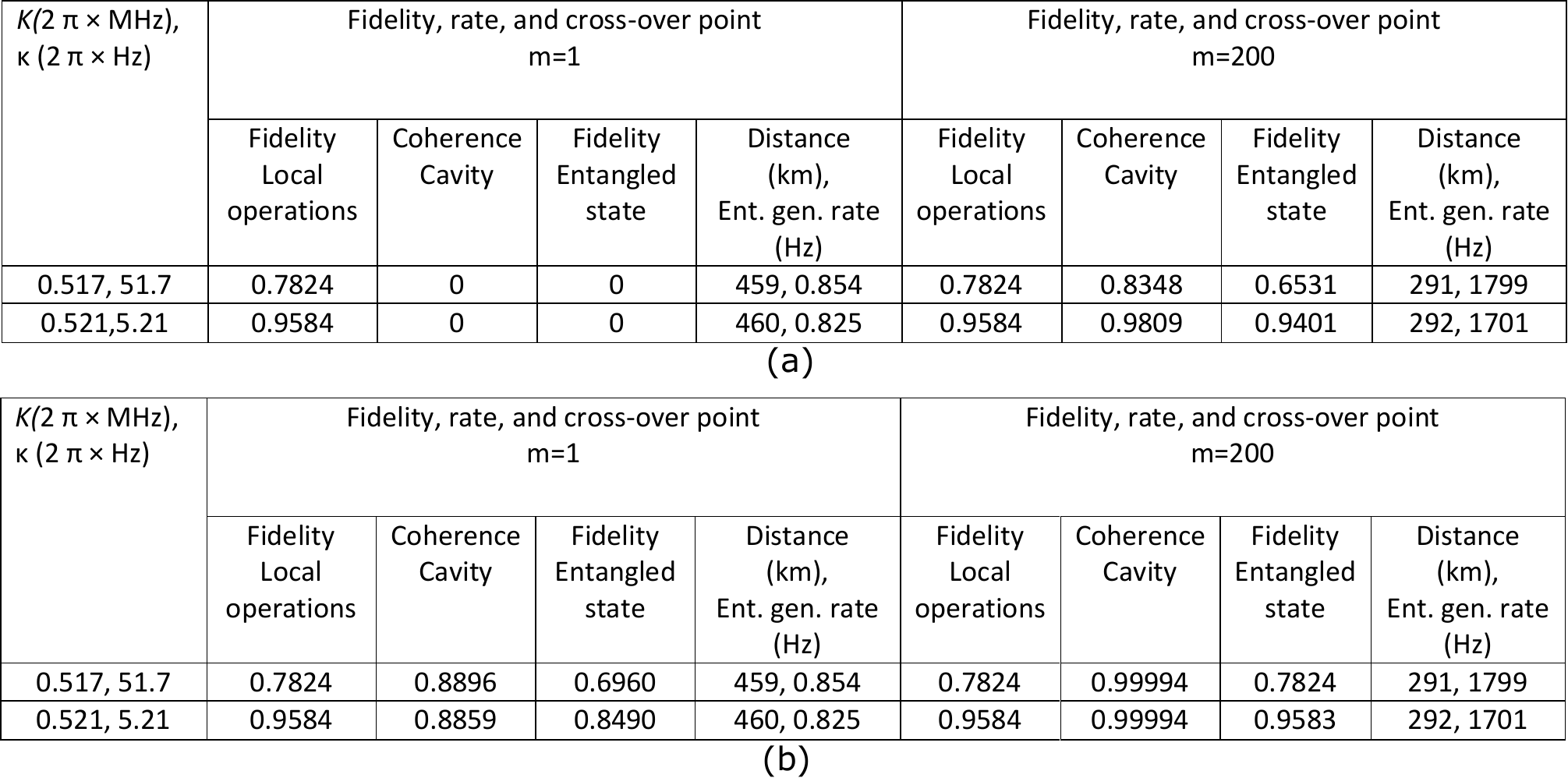}
\caption{Fidelities and entanglement distribution rates for different cross-over points for nesting level $n=2$. (a) The photonic states are stored in the same cavities used for computation. (b) Additional longer-lived cavities are used to store the photonic states during the waiting times. The cross-over distances are listed in the fifth and ninth columns for different values of $K$ and $\kappa$. One set of reading is for the non-multiplexed architecture ($m=1$), and another set is for the multiplexed architecture with multiplexing $m=200$.}
\label{tab:n_2}
\end{table*}

\subsection{Eight elementary links ($n=3$)}\label{supp_sec:n_3}
In Table \ \ref{tab:n_2}, we observe that for our best set of values of $K$ and $\kappa$, the multiplexed version beats direct transmission, but the non-multiplexed version fails to do so because of the loss in coherence of the memory. An intuition to make the waiting time smaller is to reduce the length of an elementary link. We do so by increasing the nesting level to 3. Table \ref{tab:n_3} lists the fidelities, rates, and cross-over points for the two sets of values of $K$ and $\kappa$.  

The additional set of local operations cost us the advantage gained in coherence of the cavity for $K/\kappa=10^4$, and brought down the highest fidelity possible to approx. $60 \%$. Therefore, if one were to design a repeater with $K/\kappa=10^4$, then the optimum number of nesting levels is 2. For our best set of values of $K$ and $\kappa$, the difference in fidelities of local operations is not too significant. Though we can comfortably beat direct transmission with the multiplexed version without needing additional storage cavities, we could still not beat direction transmission without needing those longer lived cavities for $m=1$. We can, in principle go for higher $n$ to make the waiting time shorter, such that the repeater beats direct transmission with a useful fidelity for $m=1$ without needing to transfer the photons to additional longer lived cavities (e.g.\ for $m=1$ and $n=4$, the final state fidelity at the cross-over point is approx. $57\%$ without requiring additional cavities). But the fidelity of the multiplexed version would reduce because of the additional operations needed. It's also practically more reasonable to use additional cavities for $m=1$ rather than adding another nesting level with shorter elementary links. We choose $n=3$ in the main text (see Fig.\ 6 in the main text) while comparing different schemes to distribute entanglement because both the multiplexed and non-multiplexed versions beat direct transmission with greater than $90 \%$ fidelities here. This is possible even without needing to convert from cat basis to Fock basis for storage. For the multiplexed version, we do not need additional longer lived cavities for storage, while for the non-multiplexed version, we need those. Compared to $n=2$, $n=3$ yields much higher entanglement distribution rates (see the last column in Table \ref{tab:n_2} and Table \ref{tab:n_3}). However, our scheme still outperforms the other schemes mentioned in the main text, with good fidelities, even for $n=2$.  

\begin{table*}
\centering
\includegraphics[width=0.9\linewidth]{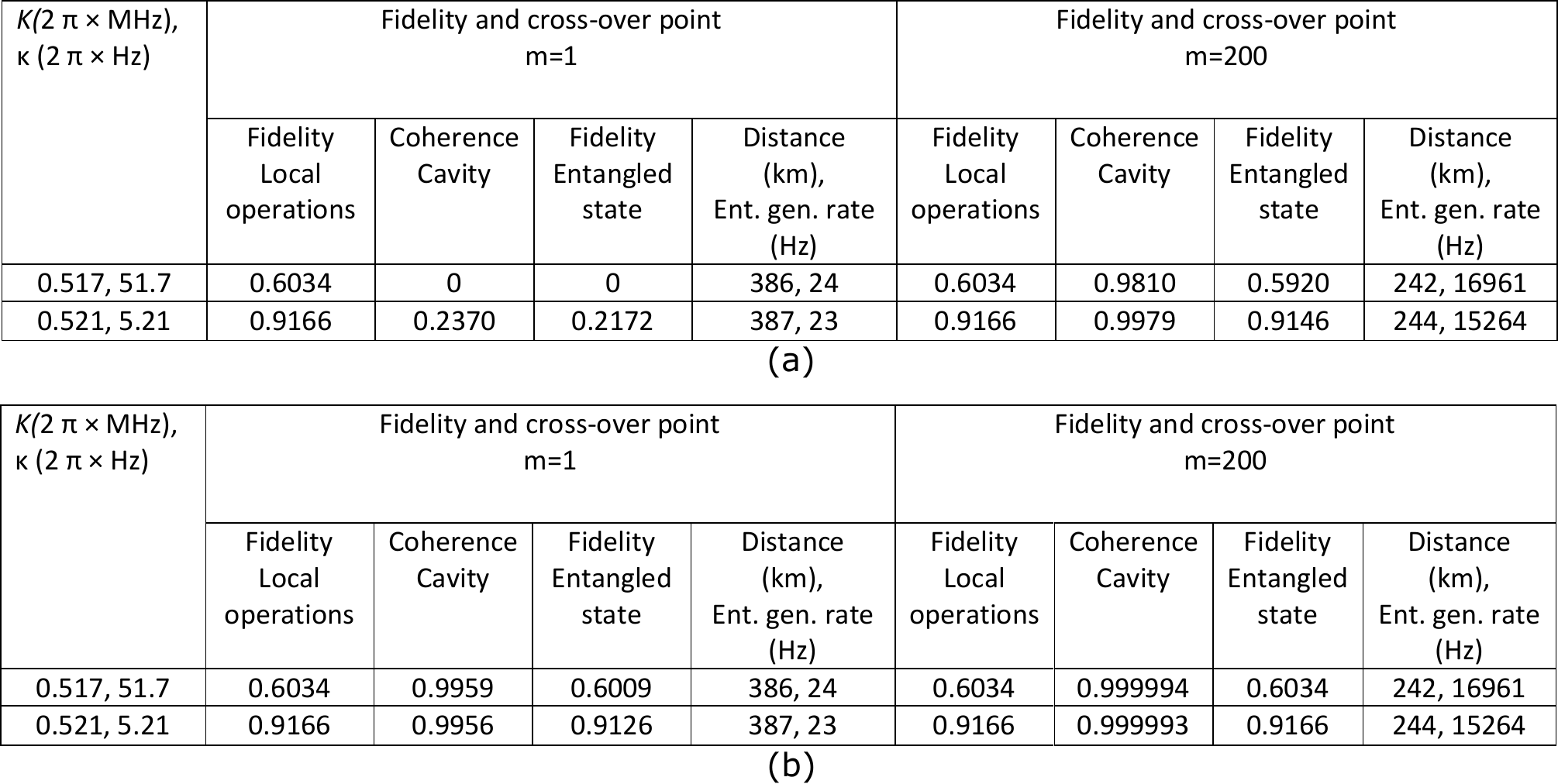}
\caption{Fidelities and entanglement distribution rates for different cross-over points for nesting level $n=3$. (a) The photonic states are stored in the same cavities used for computation. (b) Additional longer-lived cavities are used to store the photonic states during the waiting times. The cross-over distances are listed in the fifth and ninth columns for different values of $K$ and $\kappa$. One set of reading is for the non-multiplexed architecture ($m=1$), and another set is for the multiplexed architecture with multiplexing $m=200$.}
\label{tab:n_3}
\end{table*}

\subsection{Error-correction for cavity qubits}\label{sec:error-correction} 
One of the most attractive features of working with cat states is the relatively more efficient error-correction codes designed for them \cite{01_Encoding_Preskill, 13_Hardware_Mirrahimi, 14_Dynamically_Devoret,15_Confining_Devoret, 16_Extending_Schoelkopf,17_Degeneracy_Mirrahimi,18_Coherent_Devoret, 18_Fault_Schoelkopf, 18_Stabilized_Girvin}. However, the hardware-resource efficiency of the implementation of these codes depends on the way the qubits are encoded. If the qubits are encoded in cat states of the same parity, then single photon loss, which is the most prominent source of errors, is easy to keep track of and correct \cite{13_Hardware_Mirrahimi, 14_Dynamically_Devoret,16_Extending_Schoelkopf,18_Fault_Schoelkopf}. Single photon loss turns the states of even (odd) parity into odd (even) parity, and Quantum Non-Demolition (QND) measurements on the parity (of individual cavities) can track this leakage out of the qubit space.  One way to measure parity is by using an ancillary transmon qubit and a fast-decaying readout cavity coupled to the storage cavity \cite{14_Tracking_Schoelkopf,16_Extending_Schoelkopf}. A Ramsey type measurement is performed. Here, two $\pi/2$ pulses are applied on the transmon (initially prepared in its ground state), separated by a period of dispersive interaction with the storage cavity (equivalent to a Controlled-Phase gate). These operations take the transmon to either its ground or excited state, depending on the photon-number parity in the storage cavity. The state of the transmon is then read-out projectively using the other cavity.\par

On the other hand, if the qubit is encoded in states of different parity, as is true of our qubits ($\ket{C_{\alpha}^\pm}=\mathcal{N}_{\alpha}^\pm (\ket{\alpha}\pm\ket{-\alpha})$), then error-correction turns out to be more challenging. However, if there is protection against amplitude decay of the coherent states, and the qubits are encoded in cat states, then one needs to worry only about bit-flip errors, which can be corrected with joint parity measurements on several cavities \cite{15_Confining_Devoret, 17_Degeneracy_Mirrahimi, 18_Coherent_Devoret}. Joint parity measurements would be more hardware-resource consuming than parity measurements on single cavities, but a few experiments have demonstrated their feasibility, e.g.\ the one in ref.\ \cite{16_Schrodinger_Schoelkopf}. Similar challenges rise in applying error correction when encoding the qubits in the different parity Fock states, $\ket{0}$ and $\ket{1}$ \cite{16_New_Girvin, 18_Performance_Jiang}.\par

The instantaneous eigenstates of our Hamiltonian $H(t)=-K {a^{\dagger}}^2 a^2 + (\mathcal{E}_p(t) {a^{\dagger}}^2 +\mathcal{E}_p^*(t) a^2)$ are the coherent states $\ket{\alpha}$ and $\ket{-\alpha}$, where $\alpha=\sqrt{\mathcal{E}_p/K}$. It is interesting to note that a general Hamiltonian of the form $H(t)=-K {a^{\dagger}}^n a^n + (\mathcal{E}_p(t) {a^{\dagger}}^n +\mathcal{E}_p^*(t) a^n)$ would have $`n\textrm'$ different coherent states as their eigenstates, e.g.\ for $n=4$, the eigenstates are $\ket{\alpha}$, $\ket{-\alpha}$, $\ket{i \alpha}$, and $\ket{-i \alpha}$, where $\alpha=({\mathcal{E}_p/K})^{1/4}$. Different cat states, i.e.\ superposition of these coherent states can be reached by driving the Fock states, $\ket{0}-\ket{3}$ with a 4-photon drive. Using these 4 coherent states, it is possible to encode a qubit in states of the same parity, e.g.\ the logical qubits  $ \ket {\bar{0}}$ and $ \ket {\bar{1}}$ could be $\mathcal{N} (\ket{C_{\alpha}^+} + \ket{C_{i\alpha}^+})$ and $\mathcal{N} (\ket{C_{\alpha}^+} - \ket{C_{i\alpha}^+})$ respectively, where  $\mathcal{N}$ is a normalization constant. The well developed cat codes for error-correction could be directly used in such a system, however it is experimentally difficult to generate a reasonably high value for this higher order nonlinearity, $ K{a^{\dagger}}^4 a^4$. Even if one were to achieve this, it is unclear how to use the then undriven Fock states of same parity, e.g.\ $\ket{0}$ and $\ket{2}$, for robust distribution of entanglement after transduction to optical frequencies. Photon loss in the fiber and the detector inefficiencies make it difficult to envision a robust scheme similar to the one we used in our paper (for Fock states $\ket{0}$ and $\ket{1}$). Coming up with schemes to efficiently distribute entanglement between cavity states of the same parity would likely be interesting and useful. 

\putbib{}
\end{bibunit}

\begin{thebibliography}{136}%
\makeatletter
\providecommand \@ifxundefined [1]{%
 \@ifx{#1\undefined}
}%
\providecommand \@ifnum [1]{%
 \ifnum #1\expandafter \@firstoftwo
 \else \expandafter \@secondoftwo
 \fi
}%
\providecommand \@ifx [1]{%
 \ifx #1\expandafter \@firstoftwo
 \else \expandafter \@secondoftwo
 \fi
}%
\providecommand \natexlab [1]{#1}%
\providecommand \enquote  [1]{``#1''}%
\providecommand \bibnamefont  [1]{#1}%
\providecommand \bibfnamefont [1]{#1}%
\providecommand \citenamefont [1]{#1}%
\providecommand \href@noop [0]{\@secondoftwo}%
\providecommand \href [0]{\begingroup \@sanitize@url \@href}%
\providecommand \@href[1]{\@@startlink{#1}\@@href}%
\providecommand \@@href[1]{\endgroup#1\@@endlink}%
\providecommand \@sanitize@url [0]{\catcode `\\12\catcode `\$12\catcode
  `\&12\catcode `\#12\catcode `\^12\catcode `\_12\catcode `\%12\relax}%
\providecommand \@@startlink[1]{}%
\providecommand \@@endlink[0]{}%
\providecommand \url  [0]{\begingroup\@sanitize@url \@url }%
\providecommand \@url [1]{\endgroup\@href {#1}{\urlprefix }}%
\providecommand \urlprefix  [0]{URL }%
\providecommand \Eprint [0]{\href }%
\providecommand \doibase [0]{http://dx.doi.org/}%
\providecommand \selectlanguage [0]{\@gobble}%
\providecommand \bibinfo  [0]{\@secondoftwo}%
\providecommand \bibfield  [0]{\@secondoftwo}%
\providecommand \translation [1]{[#1]}%
\providecommand \BibitemOpen [0]{}%
\providecommand \bibitemStop [0]{}%
\providecommand \bibitemNoStop [0]{.\EOS\space}%
\providecommand \EOS [0]{\spacefactor3000\relax}%
\providecommand \BibitemShut  [1]{\csname bibitem#1\endcsname}%
\let\auto@bib@innerbib\@empty
\bibitem [{\citenamefont {Bennett}\ and\ \citenamefont
  {Brassard}(2014)}]{84_Cryptography_Brassard}%
  \BibitemOpen
  \bibfield  {author} {\bibinfo {author} {\bibfnamefont {C.~H.}\ \bibnamefont
  {Bennett}}\ and\ \bibinfo {author} {\bibfnamefont {G.}~\bibnamefont
  {Brassard}},\ }\href {\doibase 10.1016/j.tcs.2014.05.025} {\bibfield
  {journal} {\bibinfo  {journal} {Theoretical Computer Science}\ }\textbf
  {\bibinfo {volume} {560}},\ \bibinfo {pages} {7 } (\bibinfo {year}
  {2014})}\BibitemShut {NoStop}%
\bibitem [{\citenamefont {Gisin}\ \emph {et~al.}(2002)\citenamefont {Gisin},
  \citenamefont {Ribordy}, \citenamefont {Tittel},\ and\ \citenamefont
  {Zbinden}}]{02_Cryptography_Zbinden}%
  \BibitemOpen
  \bibfield  {author} {\bibinfo {author} {\bibfnamefont {N.}~\bibnamefont
  {Gisin}}, \bibinfo {author} {\bibfnamefont {G.}~\bibnamefont {Ribordy}},
  \bibinfo {author} {\bibfnamefont {W.}~\bibnamefont {Tittel}}, \ and\ \bibinfo
  {author} {\bibfnamefont {H.}~\bibnamefont {Zbinden}},\ }\href@noop {}
  {\bibfield  {journal} {\bibinfo  {journal} {Reviews of Modern Physics}\
  }\textbf {\bibinfo {volume} {74}},\ \bibinfo {pages} {145} (\bibinfo {year}
  {2002})}\BibitemShut {NoStop}%
\bibitem [{\citenamefont {{Broadbent}}\ \emph {et~al.}(2009)\citenamefont
  {{Broadbent}}, \citenamefont {{Fitzsimons}},\ and\ \citenamefont
  {{Kashefi}}}]{09_Universal_Kashefi}%
  \BibitemOpen
  \bibfield  {author} {\bibinfo {author} {\bibfnamefont {A.}~\bibnamefont
  {{Broadbent}}}, \bibinfo {author} {\bibfnamefont {J.}~\bibnamefont
  {{Fitzsimons}}}, \ and\ \bibinfo {author} {\bibfnamefont {E.}~\bibnamefont
  {{Kashefi}}},\ }in\ \href {\doibase 10.1109/FOCS.2009.36} {\emph {\bibinfo
  {booktitle} {2009 50th Annual IEEE Symposium on Foundations of Computer
  Science}}}\ (\bibinfo {year} {2009})\ pp.\ \bibinfo {pages}
  {517--526}\BibitemShut {NoStop}%
\bibitem [{\citenamefont {Barz}\ \emph {et~al.}(2012)\citenamefont {Barz},
  \citenamefont {Kashefi}, \citenamefont {Broadbent}, \citenamefont
  {Fitzsimons}, \citenamefont {Zeilinger},\ and\ \citenamefont
  {Walther}}]{12_Demonstration_Walther}%
  \BibitemOpen
  \bibfield  {author} {\bibinfo {author} {\bibfnamefont {S.}~\bibnamefont
  {Barz}}, \bibinfo {author} {\bibfnamefont {E.}~\bibnamefont {Kashefi}},
  \bibinfo {author} {\bibfnamefont {A.}~\bibnamefont {Broadbent}}, \bibinfo
  {author} {\bibfnamefont {J.~F.}\ \bibnamefont {Fitzsimons}}, \bibinfo
  {author} {\bibfnamefont {A.}~\bibnamefont {Zeilinger}}, \ and\ \bibinfo
  {author} {\bibfnamefont {P.}~\bibnamefont {Walther}},\ }\href@noop {}
  {\bibfield  {journal} {\bibinfo  {journal} {Science}\ }\textbf {\bibinfo
  {volume} {335}},\ \bibinfo {pages} {303} (\bibinfo {year}
  {2012})}\BibitemShut {NoStop}%
\bibitem [{\citenamefont {Jakobi}\ \emph {et~al.}(2011)\citenamefont {Jakobi},
  \citenamefont {Simon}, \citenamefont {Gisin}, \citenamefont {Bancal},
  \citenamefont {Branciard}, \citenamefont {Walenta},\ and\ \citenamefont
  {Zbinden}}]{11_Practical_Zbinden}%
  \BibitemOpen
  \bibfield  {author} {\bibinfo {author} {\bibfnamefont {M.}~\bibnamefont
  {Jakobi}}, \bibinfo {author} {\bibfnamefont {C.}~\bibnamefont {Simon}},
  \bibinfo {author} {\bibfnamefont {N.}~\bibnamefont {Gisin}}, \bibinfo
  {author} {\bibfnamefont {J.-D.}\ \bibnamefont {Bancal}}, \bibinfo {author}
  {\bibfnamefont {C.}~\bibnamefont {Branciard}}, \bibinfo {author}
  {\bibfnamefont {N.}~\bibnamefont {Walenta}}, \ and\ \bibinfo {author}
  {\bibfnamefont {H.}~\bibnamefont {Zbinden}},\ }\href@noop {} {\bibfield
  {journal} {\bibinfo  {journal} {Physical Review A}\ }\textbf {\bibinfo
  {volume} {83}},\ \bibinfo {pages} {022301} (\bibinfo {year}
  {2011})}\BibitemShut {NoStop}%
\bibitem [{\citenamefont {Komar}\ \emph {et~al.}(2014)\citenamefont {Komar},
  \citenamefont {Kessler}, \citenamefont {Bishof}, \citenamefont {Jiang},
  \citenamefont {S{\o}rensen}, \citenamefont {Ye},\ and\ \citenamefont
  {Lukin}}]{14_Quantum_Lukin}%
  \BibitemOpen
  \bibfield  {author} {\bibinfo {author} {\bibfnamefont {P.}~\bibnamefont
  {Komar}}, \bibinfo {author} {\bibfnamefont {E.~M.}\ \bibnamefont {Kessler}},
  \bibinfo {author} {\bibfnamefont {M.}~\bibnamefont {Bishof}}, \bibinfo
  {author} {\bibfnamefont {L.}~\bibnamefont {Jiang}}, \bibinfo {author}
  {\bibfnamefont {A.~S.}\ \bibnamefont {S{\o}rensen}}, \bibinfo {author}
  {\bibfnamefont {J.}~\bibnamefont {Ye}}, \ and\ \bibinfo {author}
  {\bibfnamefont {M.~D.}\ \bibnamefont {Lukin}},\ }\href@noop {} {\bibfield
  {journal} {\bibinfo  {journal} {Nature Physics}\ }\textbf {\bibinfo {volume}
  {10}},\ \bibinfo {pages} {582} (\bibinfo {year} {2014})}\BibitemShut
  {NoStop}%
\bibitem [{\citenamefont {Gottesman}\ \emph {et~al.}(2012)\citenamefont
  {Gottesman}, \citenamefont {Jennewein},\ and\ \citenamefont
  {Croke}}]{12_Longer_Croke}%
  \BibitemOpen
  \bibfield  {author} {\bibinfo {author} {\bibfnamefont {D.}~\bibnamefont
  {Gottesman}}, \bibinfo {author} {\bibfnamefont {T.}~\bibnamefont
  {Jennewein}}, \ and\ \bibinfo {author} {\bibfnamefont {S.}~\bibnamefont
  {Croke}},\ }\href@noop {} {\bibfield  {journal} {\bibinfo  {journal}
  {Physical Review Letters}\ }\textbf {\bibinfo {volume} {109}},\ \bibinfo
  {pages} {070503} (\bibinfo {year} {2012})}\BibitemShut {NoStop}%
\bibitem [{\citenamefont {Rideout}\ \emph {et~al.}(2012)\citenamefont
  {Rideout}, \citenamefont {Jennewein}, \citenamefont {Amelino-Camelia},
  \citenamefont {Demarie}, \citenamefont {Higgins}, \citenamefont {Kempf},
  \citenamefont {Kent}, \citenamefont {Laflamme}, \citenamefont {Ma},
  \citenamefont {Mann} \emph {et~al.}}]{12_Fundamental_Terno}%
  \BibitemOpen
  \bibfield  {author} {\bibinfo {author} {\bibfnamefont {D.}~\bibnamefont
  {Rideout}}, \bibinfo {author} {\bibfnamefont {T.}~\bibnamefont {Jennewein}},
  \bibinfo {author} {\bibfnamefont {G.}~\bibnamefont {Amelino-Camelia}},
  \bibinfo {author} {\bibfnamefont {T.~F.}\ \bibnamefont {Demarie}}, \bibinfo
  {author} {\bibfnamefont {B.~L.}\ \bibnamefont {Higgins}}, \bibinfo {author}
  {\bibfnamefont {A.}~\bibnamefont {Kempf}}, \bibinfo {author} {\bibfnamefont
  {A.}~\bibnamefont {Kent}}, \bibinfo {author} {\bibfnamefont {R.}~\bibnamefont
  {Laflamme}}, \bibinfo {author} {\bibfnamefont {X.}~\bibnamefont {Ma}},
  \bibinfo {author} {\bibfnamefont {R.~B.}\ \bibnamefont {Mann}},  \emph
  {et~al.},\ }\href@noop {} {\bibfield  {journal} {\bibinfo  {journal}
  {Classical and Quantum Gravity}\ }\textbf {\bibinfo {volume} {29}},\ \bibinfo
  {pages} {224011} (\bibinfo {year} {2012})}\BibitemShut {NoStop}%
\bibitem [{\citenamefont {Briegel}\ \emph {et~al.}(1998)\citenamefont
  {Briegel}, \citenamefont {D{\"u}r}, \citenamefont {Cirac},\ and\
  \citenamefont {Zoller}}]{98_Quantum_Zoller}%
  \BibitemOpen
  \bibfield  {author} {\bibinfo {author} {\bibfnamefont {H.-J.}\ \bibnamefont
  {Briegel}}, \bibinfo {author} {\bibfnamefont {W.}~\bibnamefont {D{\"u}r}},
  \bibinfo {author} {\bibfnamefont {J.~I.}\ \bibnamefont {Cirac}}, \ and\
  \bibinfo {author} {\bibfnamefont {P.}~\bibnamefont {Zoller}},\ }\href@noop {}
  {\bibfield  {journal} {\bibinfo  {journal} {Physical Review Letters}\
  }\textbf {\bibinfo {volume} {81}},\ \bibinfo {pages} {5932} (\bibinfo {year}
  {1998})}\BibitemShut {NoStop}%
\bibitem [{\citenamefont {Sangouard}\ \emph {et~al.}(2011)\citenamefont
  {Sangouard}, \citenamefont {Simon}, \citenamefont {De~Riedmatten},\ and\
  \citenamefont {Gisin}}]{11_Quantum_Gisin}%
  \BibitemOpen
  \bibfield  {author} {\bibinfo {author} {\bibfnamefont {N.}~\bibnamefont
  {Sangouard}}, \bibinfo {author} {\bibfnamefont {C.}~\bibnamefont {Simon}},
  \bibinfo {author} {\bibfnamefont {H.}~\bibnamefont {De~Riedmatten}}, \ and\
  \bibinfo {author} {\bibfnamefont {N.}~\bibnamefont {Gisin}},\ }\href@noop {}
  {\bibfield  {journal} {\bibinfo  {journal} {Reviews of Modern Physics}\
  }\textbf {\bibinfo {volume} {83}},\ \bibinfo {pages} {33} (\bibinfo {year}
  {2011})}\BibitemShut {NoStop}%
\bibitem [{\citenamefont {O'Malley}\ \emph {et~al.}(2016)\citenamefont
  {O'Malley}, \citenamefont {Babbush}, \citenamefont {Kivlichan}, \citenamefont
  {Romero}, \citenamefont {McClean}, \citenamefont {Barends}, \citenamefont
  {Kelly}, \citenamefont {Roushan}, \citenamefont {Tranter}, \citenamefont
  {Ding}, \citenamefont {Campbell}, \citenamefont {Chen}, \citenamefont {Chen},
  \citenamefont {Chiaro}, \citenamefont {Dunsworth}, \citenamefont {Fowler},
  \citenamefont {Jeffrey}, \citenamefont {Lucero}, \citenamefont {Megrant},
  \citenamefont {Mutus}, \citenamefont {Neeley}, \citenamefont {Neill},
  \citenamefont {Quintana}, \citenamefont {Sank}, \citenamefont {Vainsencher},
  \citenamefont {Wenner}, \citenamefont {White}, \citenamefont {Coveney},
  \citenamefont {Love}, \citenamefont {Neven}, \citenamefont {Aspuru-Guzik},\
  and\ \citenamefont {Martinis}}]{16_Scalable_Martinis}%
  \BibitemOpen
  \bibfield  {author} {\bibinfo {author} {\bibfnamefont {P.~J.~J.}\
  \bibnamefont {O'Malley}}, \bibinfo {author} {\bibfnamefont {R.}~\bibnamefont
  {Babbush}}, \bibinfo {author} {\bibfnamefont {I.~D.}\ \bibnamefont
  {Kivlichan}}, \bibinfo {author} {\bibfnamefont {J.}~\bibnamefont {Romero}},
  \bibinfo {author} {\bibfnamefont {J.~R.}\ \bibnamefont {McClean}}, \bibinfo
  {author} {\bibfnamefont {R.}~\bibnamefont {Barends}}, \bibinfo {author}
  {\bibfnamefont {J.}~\bibnamefont {Kelly}}, \bibinfo {author} {\bibfnamefont
  {P.}~\bibnamefont {Roushan}}, \bibinfo {author} {\bibfnamefont
  {A.}~\bibnamefont {Tranter}}, \bibinfo {author} {\bibfnamefont
  {N.}~\bibnamefont {Ding}}, \bibinfo {author} {\bibfnamefont {B.}~\bibnamefont
  {Campbell}}, \bibinfo {author} {\bibfnamefont {Y.}~\bibnamefont {Chen}},
  \bibinfo {author} {\bibfnamefont {Z.}~\bibnamefont {Chen}}, \bibinfo {author}
  {\bibfnamefont {B.}~\bibnamefont {Chiaro}}, \bibinfo {author} {\bibfnamefont
  {A.}~\bibnamefont {Dunsworth}}, \bibinfo {author} {\bibfnamefont {A.~G.}\
  \bibnamefont {Fowler}}, \bibinfo {author} {\bibfnamefont {E.}~\bibnamefont
  {Jeffrey}}, \bibinfo {author} {\bibfnamefont {E.}~\bibnamefont {Lucero}},
  \bibinfo {author} {\bibfnamefont {A.}~\bibnamefont {Megrant}}, \bibinfo
  {author} {\bibfnamefont {J.~Y.}\ \bibnamefont {Mutus}}, \bibinfo {author}
  {\bibfnamefont {M.}~\bibnamefont {Neeley}}, \bibinfo {author} {\bibfnamefont
  {C.}~\bibnamefont {Neill}}, \bibinfo {author} {\bibfnamefont
  {C.}~\bibnamefont {Quintana}}, \bibinfo {author} {\bibfnamefont
  {D.}~\bibnamefont {Sank}}, \bibinfo {author} {\bibfnamefont {A.}~\bibnamefont
  {Vainsencher}}, \bibinfo {author} {\bibfnamefont {J.}~\bibnamefont {Wenner}},
  \bibinfo {author} {\bibfnamefont {T.~C.}\ \bibnamefont {White}}, \bibinfo
  {author} {\bibfnamefont {P.~V.}\ \bibnamefont {Coveney}}, \bibinfo {author}
  {\bibfnamefont {P.~J.}\ \bibnamefont {Love}}, \bibinfo {author}
  {\bibfnamefont {H.}~\bibnamefont {Neven}}, \bibinfo {author} {\bibfnamefont
  {A.}~\bibnamefont {Aspuru-Guzik}}, \ and\ \bibinfo {author} {\bibfnamefont
  {J.~M.}\ \bibnamefont {Martinis}},\ }\href {\doibase
  10.1103/PhysRevX.6.031007} {\bibfield  {journal} {\bibinfo  {journal}
  {Physical Review X}\ }\textbf {\bibinfo {volume} {6}},\ \bibinfo {pages}
  {031007} (\bibinfo {year} {2016})}\BibitemShut {NoStop}%
\bibitem [{\citenamefont {Mott}\ \emph {et~al.}(2017)\citenamefont {Mott},
  \citenamefont {Job}, \citenamefont {Vlimant}, \citenamefont {Lidar},\ and\
  \citenamefont {Spiropulu}}]{17_Solving_Spiropulu}%
  \BibitemOpen
  \bibfield  {author} {\bibinfo {author} {\bibfnamefont {A.}~\bibnamefont
  {Mott}}, \bibinfo {author} {\bibfnamefont {J.}~\bibnamefont {Job}}, \bibinfo
  {author} {\bibfnamefont {J.-R.}\ \bibnamefont {Vlimant}}, \bibinfo {author}
  {\bibfnamefont {D.}~\bibnamefont {Lidar}}, \ and\ \bibinfo {author}
  {\bibfnamefont {M.}~\bibnamefont {Spiropulu}},\ }\href
  {http://dx.doi.org/10.1038/nature24047} {\bibfield  {journal} {\bibinfo
  {journal} {Nature}\ }\textbf {\bibinfo {volume} {550}},\ \bibinfo {pages}
  {375} (\bibinfo {year} {2017})}\BibitemShut {NoStop}%
\bibitem [{\citenamefont {Rosenblum}\ \emph
  {et~al.}(2018{\natexlab{a}})\citenamefont {Rosenblum}, \citenamefont
  {Reinhold}, \citenamefont {Mirrahimi}, \citenamefont {Jiang}, \citenamefont
  {Frunzio},\ and\ \citenamefont {Schoelkopf}}]{18_Fault_Schoelkopf}%
  \BibitemOpen
  \bibfield  {author} {\bibinfo {author} {\bibfnamefont {S.}~\bibnamefont
  {Rosenblum}}, \bibinfo {author} {\bibfnamefont {P.}~\bibnamefont {Reinhold}},
  \bibinfo {author} {\bibfnamefont {M.}~\bibnamefont {Mirrahimi}}, \bibinfo
  {author} {\bibfnamefont {L.}~\bibnamefont {Jiang}}, \bibinfo {author}
  {\bibfnamefont {L.}~\bibnamefont {Frunzio}}, \ and\ \bibinfo {author}
  {\bibfnamefont {R.~J.}\ \bibnamefont {Schoelkopf}},\ }\href {\doibase
  10.1126/science.aat3996} {\bibfield  {journal} {\bibinfo  {journal}
  {Science}\ }\textbf {\bibinfo {volume} {361}},\ \bibinfo {pages} {266}
  (\bibinfo {year} {2018}{\natexlab{a}})}\BibitemShut {NoStop}%
\bibitem [{\citenamefont {Savage}(2018)}]{18_Quantum_Savage}%
  \BibitemOpen
  \bibfield  {author} {\bibinfo {author} {\bibfnamefont {N.}~\bibnamefont
  {Savage}},\ }\href
  {https://www.scientificamerican.com/article/quantum-computers-compete-for-supremacy/}
  {\bibfield  {journal} {\bibinfo  {journal} {Scientific American}\ }\textbf
  {\bibinfo {volume} {27}},\ \bibinfo {pages} {108} (\bibinfo {year}
  {2018})}\BibitemShut {NoStop}%
\bibitem [{\citenamefont {Kimble}(2008)}]{08_Quantum_Kimble}%
  \BibitemOpen
  \bibfield  {author} {\bibinfo {author} {\bibfnamefont {H.~J.}\ \bibnamefont
  {Kimble}},\ }\href@noop {} {\bibfield  {journal} {\bibinfo  {journal}
  {Nature}\ }\textbf {\bibinfo {volume} {453}},\ \bibinfo {pages} {1023}
  (\bibinfo {year} {2008})}\BibitemShut {NoStop}%
\bibitem [{\citenamefont {Wehner}\ \emph {et~al.}(2018)\citenamefont {Wehner},
  \citenamefont {Elkouss},\ and\ \citenamefont {Hanson}}]{18_Internet_Hanson}%
  \BibitemOpen
  \bibfield  {author} {\bibinfo {author} {\bibfnamefont {S.}~\bibnamefont
  {Wehner}}, \bibinfo {author} {\bibfnamefont {D.}~\bibnamefont {Elkouss}}, \
  and\ \bibinfo {author} {\bibfnamefont {R.}~\bibnamefont {Hanson}},\ }\href
  {\doibase 10.1126/science.aam9288} {\bibfield  {journal} {\bibinfo  {journal}
  {Science}\ }\textbf {\bibinfo {volume} {362}},\ \bibinfo {pages} {eaam9288}
  (\bibinfo {year} {2018})}\BibitemShut {NoStop}%
\bibitem [{\citenamefont {Simon}(2017)}]{17_Towards_Simon}%
  \BibitemOpen
  \bibfield  {author} {\bibinfo {author} {\bibfnamefont {C.}~\bibnamefont
  {Simon}},\ }\href@noop {} {\bibfield  {journal} {\bibinfo  {journal} {Nature
  Photonics}\ }\textbf {\bibinfo {volume} {11}},\ \bibinfo {pages} {678}
  (\bibinfo {year} {2017})}\BibitemShut {NoStop}%
\bibitem [{\citenamefont {Browne}\ and\ \citenamefont
  {Rudolph}(2005)}]{05_Resource_Rudolph}%
  \BibitemOpen
  \bibfield  {author} {\bibinfo {author} {\bibfnamefont {D.~E.}\ \bibnamefont
  {Browne}}\ and\ \bibinfo {author} {\bibfnamefont {T.}~\bibnamefont
  {Rudolph}},\ }\href@noop {} {\bibfield  {journal} {\bibinfo  {journal}
  {Physical Review Letters}\ }\textbf {\bibinfo {volume} {95}},\ \bibinfo
  {pages} {010501} (\bibinfo {year} {2005})}\BibitemShut {NoStop}%
\bibitem [{\citenamefont {Kielpinski}\ \emph {et~al.}(2002)\citenamefont
  {Kielpinski}, \citenamefont {Monroe},\ and\ \citenamefont
  {Wineland}}]{02_Architecture_Wineland}%
  \BibitemOpen
  \bibfield  {author} {\bibinfo {author} {\bibfnamefont {D.}~\bibnamefont
  {Kielpinski}}, \bibinfo {author} {\bibfnamefont {C.}~\bibnamefont {Monroe}},
  \ and\ \bibinfo {author} {\bibfnamefont {D.~J.}\ \bibnamefont {Wineland}},\
  }\href@noop {} {\bibfield  {journal} {\bibinfo  {journal} {Nature}\ }\textbf
  {\bibinfo {volume} {417}},\ \bibinfo {pages} {709} (\bibinfo {year}
  {2002})}\BibitemShut {NoStop}%
\bibitem [{\citenamefont {Zagoskin}\ and\ \citenamefont
  {Blais}(2007)}]{07_SQ_Blais}%
  \BibitemOpen
  \bibfield  {author} {\bibinfo {author} {\bibfnamefont {A.}~\bibnamefont
  {Zagoskin}}\ and\ \bibinfo {author} {\bibfnamefont {A.}~\bibnamefont
  {Blais}},\ }\href
  {https://pic-pac.cap.ca/static/downloads/7431c64c8e42dfeddae5e4768467a125f938ad2f.pdf}
  {\bibfield  {journal} {\bibinfo  {journal} {Physics in Canada}\ }\textbf
  {\bibinfo {volume} {63}},\ \bibinfo {pages} {215} (\bibinfo {year}
  {2007})}\BibitemShut {NoStop}%
\bibitem [{\citenamefont {Imamoglu}\ \emph {et~al.}(1999)\citenamefont
  {Imamoglu}, \citenamefont {Awschalom}, \citenamefont {Burkard}, \citenamefont
  {DiVincenzo}, \citenamefont {Loss}, \citenamefont {Sherwin}, \citenamefont
  {Small} \emph {et~al.}}]{99_QIP_Small}%
  \BibitemOpen
  \bibfield  {author} {\bibinfo {author} {\bibfnamefont {A.}~\bibnamefont
  {Imamoglu}}, \bibinfo {author} {\bibfnamefont {D.~D.}\ \bibnamefont
  {Awschalom}}, \bibinfo {author} {\bibfnamefont {G.}~\bibnamefont {Burkard}},
  \bibinfo {author} {\bibfnamefont {D.~P.}\ \bibnamefont {DiVincenzo}},
  \bibinfo {author} {\bibfnamefont {D.}~\bibnamefont {Loss}}, \bibinfo {author}
  {\bibfnamefont {M.}~\bibnamefont {Sherwin}}, \bibinfo {author} {\bibfnamefont
  {A.}~\bibnamefont {Small}},  \emph {et~al.},\ }\href@noop {} {\bibfield
  {journal} {\bibinfo  {journal} {Physical Review Letters}\ }\textbf {\bibinfo
  {volume} {83}},\ \bibinfo {pages} {4204} (\bibinfo {year}
  {1999})}\BibitemShut {NoStop}%
\bibitem [{\citenamefont {Kane}(1998)}]{98_Silicon_Kane}%
  \BibitemOpen
  \bibfield  {author} {\bibinfo {author} {\bibfnamefont {B.~E.}\ \bibnamefont
  {Kane}},\ }\href@noop {} {\bibfield  {journal} {\bibinfo  {journal} {Nature}\
  }\textbf {\bibinfo {volume} {393}},\ \bibinfo {pages} {133} (\bibinfo {year}
  {1998})}\BibitemShut {NoStop}%
\bibitem [{\citenamefont {Wallraff}\ \emph {et~al.}(2004)\citenamefont
  {Wallraff}, \citenamefont {Schuster}, \citenamefont {Blais}, \citenamefont
  {Frunzio}, \citenamefont {Huang}, \citenamefont {Majer}, \citenamefont
  {Kumar}, \citenamefont {Girvin},\ and\ \citenamefont
  {Schoelkopf}}]{04_Strong_Schoelkopf}%
  \BibitemOpen
  \bibfield  {author} {\bibinfo {author} {\bibfnamefont {A.}~\bibnamefont
  {Wallraff}}, \bibinfo {author} {\bibfnamefont {D.~I.}\ \bibnamefont
  {Schuster}}, \bibinfo {author} {\bibfnamefont {A.}~\bibnamefont {Blais}},
  \bibinfo {author} {\bibfnamefont {L.}~\bibnamefont {Frunzio}}, \bibinfo
  {author} {\bibfnamefont {R.-S.}\ \bibnamefont {Huang}}, \bibinfo {author}
  {\bibfnamefont {J.}~\bibnamefont {Majer}}, \bibinfo {author} {\bibfnamefont
  {S.}~\bibnamefont {Kumar}}, \bibinfo {author} {\bibfnamefont {S.~M.}\
  \bibnamefont {Girvin}}, \ and\ \bibinfo {author} {\bibfnamefont {R.~J.}\
  \bibnamefont {Schoelkopf}},\ }\href@noop {} {\bibfield  {journal} {\bibinfo
  {journal} {Nature}\ }\textbf {\bibinfo {volume} {431}},\ \bibinfo {pages}
  {162} (\bibinfo {year} {2004})}\BibitemShut {NoStop}%
\bibitem [{\citenamefont {Niemczyk}\ \emph {et~al.}(2010)\citenamefont
  {Niemczyk}, \citenamefont {Deppe}, \citenamefont {Huebl}, \citenamefont
  {Menzel}, \citenamefont {Hocke}, \citenamefont {Schwarz}, \citenamefont
  {Garcia-Ripoll}, \citenamefont {Zueco}, \citenamefont {H{\"u}mmer},
  \citenamefont {Solano} \emph {et~al.}}]{10_CQED_Gross}%
  \BibitemOpen
  \bibfield  {author} {\bibinfo {author} {\bibfnamefont {T.}~\bibnamefont
  {Niemczyk}}, \bibinfo {author} {\bibfnamefont {F.}~\bibnamefont {Deppe}},
  \bibinfo {author} {\bibfnamefont {H.}~\bibnamefont {Huebl}}, \bibinfo
  {author} {\bibfnamefont {E.}~\bibnamefont {Menzel}}, \bibinfo {author}
  {\bibfnamefont {F.}~\bibnamefont {Hocke}}, \bibinfo {author} {\bibfnamefont
  {M.}~\bibnamefont {Schwarz}}, \bibinfo {author} {\bibfnamefont
  {J.}~\bibnamefont {Garcia-Ripoll}}, \bibinfo {author} {\bibfnamefont
  {D.}~\bibnamefont {Zueco}}, \bibinfo {author} {\bibfnamefont
  {T.}~\bibnamefont {H{\"u}mmer}}, \bibinfo {author} {\bibfnamefont
  {E.}~\bibnamefont {Solano}},  \emph {et~al.},\ }\href@noop {} {\bibfield
  {journal} {\bibinfo  {journal} {Nature Physics}\ }\textbf {\bibinfo {volume}
  {6}},\ \bibinfo {pages} {772} (\bibinfo {year} {2010})}\BibitemShut {NoStop}%
\bibitem [{\citenamefont {Gu}\ \emph {et~al.}(2017)\citenamefont {Gu},
  \citenamefont {Kockum}, \citenamefont {Miranowicz}, \citenamefont {Liu},\
  and\ \citenamefont {Nori}}]{17_Microwave_Nori}%
  \BibitemOpen
  \bibfield  {author} {\bibinfo {author} {\bibfnamefont {X.}~\bibnamefont
  {Gu}}, \bibinfo {author} {\bibfnamefont {A.~F.}\ \bibnamefont {Kockum}},
  \bibinfo {author} {\bibfnamefont {A.}~\bibnamefont {Miranowicz}}, \bibinfo
  {author} {\bibfnamefont {Y.-x.}\ \bibnamefont {Liu}}, \ and\ \bibinfo
  {author} {\bibfnamefont {F.}~\bibnamefont {Nori}},\ }\href@noop {} {\bibfield
   {journal} {\bibinfo  {journal} {Physics Reports}\ }\textbf {\bibinfo
  {volume} {718}},\ \bibinfo {pages} {1} (\bibinfo {year} {2017})}\BibitemShut
  {NoStop}%
\bibitem [{\citenamefont {Yoshihara}\ \emph {et~al.}(2017)\citenamefont
  {Yoshihara}, \citenamefont {Fuse}, \citenamefont {Ashhab}, \citenamefont
  {Kakuyanagi}, \citenamefont {Saito},\ and\ \citenamefont
  {Semba}}]{16_SQ_Semba}%
  \BibitemOpen
  \bibfield  {author} {\bibinfo {author} {\bibfnamefont {F.}~\bibnamefont
  {Yoshihara}}, \bibinfo {author} {\bibfnamefont {T.}~\bibnamefont {Fuse}},
  \bibinfo {author} {\bibfnamefont {S.}~\bibnamefont {Ashhab}}, \bibinfo
  {author} {\bibfnamefont {K.}~\bibnamefont {Kakuyanagi}}, \bibinfo {author}
  {\bibfnamefont {S.}~\bibnamefont {Saito}}, \ and\ \bibinfo {author}
  {\bibfnamefont {K.}~\bibnamefont {Semba}},\ }\href@noop {} {\bibfield
  {journal} {\bibinfo  {journal} {Nature Physics}\ }\textbf {\bibinfo {volume}
  {13}},\ \bibinfo {pages} {44} (\bibinfo {year} {2017})}\BibitemShut {NoStop}%
\bibitem [{\citenamefont {Rigetti}\ \emph {et~al.}(2012)\citenamefont
  {Rigetti}, \citenamefont {Gambetta}, \citenamefont {Poletto}, \citenamefont
  {Plourde}, \citenamefont {Chow}, \citenamefont {C{\'o}rcoles}, \citenamefont
  {Smolin}, \citenamefont {Merkel}, \citenamefont {Rozen}, \citenamefont
  {Keefe} \emph {et~al.}}]{12_SQ_Steffen}%
  \BibitemOpen
  \bibfield  {author} {\bibinfo {author} {\bibfnamefont {C.}~\bibnamefont
  {Rigetti}}, \bibinfo {author} {\bibfnamefont {J.~M.}\ \bibnamefont
  {Gambetta}}, \bibinfo {author} {\bibfnamefont {S.}~\bibnamefont {Poletto}},
  \bibinfo {author} {\bibfnamefont {B.}~\bibnamefont {Plourde}}, \bibinfo
  {author} {\bibfnamefont {J.~M.}\ \bibnamefont {Chow}}, \bibinfo {author}
  {\bibfnamefont {A.}~\bibnamefont {C{\'o}rcoles}}, \bibinfo {author}
  {\bibfnamefont {J.~A.}\ \bibnamefont {Smolin}}, \bibinfo {author}
  {\bibfnamefont {S.~T.}\ \bibnamefont {Merkel}}, \bibinfo {author}
  {\bibfnamefont {J.}~\bibnamefont {Rozen}}, \bibinfo {author} {\bibfnamefont
  {G.~A.}\ \bibnamefont {Keefe}},  \emph {et~al.},\ }\href@noop {} {\bibfield
  {journal} {\bibinfo  {journal} {Physical Review B}\ }\textbf {\bibinfo
  {volume} {86}},\ \bibinfo {pages} {100506} (\bibinfo {year}
  {2012})}\BibitemShut {NoStop}%
\bibitem [{\citenamefont {Leghtas}\ \emph {et~al.}(2013)\citenamefont
  {Leghtas}, \citenamefont {Kirchmair}, \citenamefont {Vlastakis},
  \citenamefont {Schoelkopf}, \citenamefont {Devoret},\ and\ \citenamefont
  {Mirrahimi}}]{13_Hardware_Mirrahimi}%
  \BibitemOpen
  \bibfield  {author} {\bibinfo {author} {\bibfnamefont {Z.}~\bibnamefont
  {Leghtas}}, \bibinfo {author} {\bibfnamefont {G.}~\bibnamefont {Kirchmair}},
  \bibinfo {author} {\bibfnamefont {B.}~\bibnamefont {Vlastakis}}, \bibinfo
  {author} {\bibfnamefont {R.~J.}\ \bibnamefont {Schoelkopf}}, \bibinfo
  {author} {\bibfnamefont {M.~H.}\ \bibnamefont {Devoret}}, \ and\ \bibinfo
  {author} {\bibfnamefont {M.}~\bibnamefont {Mirrahimi}},\ }\href@noop {}
  {\bibfield  {journal} {\bibinfo  {journal} {Physical Review Letters}\
  }\textbf {\bibinfo {volume} {111}},\ \bibinfo {pages} {120501} (\bibinfo
  {year} {2013})}\BibitemShut {NoStop}%
\bibitem [{\citenamefont {Mirrahimi}\ \emph {et~al.}(2014)\citenamefont
  {Mirrahimi}, \citenamefont {Leghtas}, \citenamefont {Albert}, \citenamefont
  {Touzard}, \citenamefont {Schoelkopf}, \citenamefont {Jiang},\ and\
  \citenamefont {Devoret}}]{14_Dynamically_Devoret}%
  \BibitemOpen
  \bibfield  {author} {\bibinfo {author} {\bibfnamefont {M.}~\bibnamefont
  {Mirrahimi}}, \bibinfo {author} {\bibfnamefont {Z.}~\bibnamefont {Leghtas}},
  \bibinfo {author} {\bibfnamefont {V.~V.}\ \bibnamefont {Albert}}, \bibinfo
  {author} {\bibfnamefont {S.}~\bibnamefont {Touzard}}, \bibinfo {author}
  {\bibfnamefont {R.~J.}\ \bibnamefont {Schoelkopf}}, \bibinfo {author}
  {\bibfnamefont {L.}~\bibnamefont {Jiang}}, \ and\ \bibinfo {author}
  {\bibfnamefont {M.~H.}\ \bibnamefont {Devoret}},\ }\href@noop {} {\bibfield
  {journal} {\bibinfo  {journal} {New Journal of Physics}\ }\textbf {\bibinfo
  {volume} {16}},\ \bibinfo {pages} {045014} (\bibinfo {year}
  {2014})}\BibitemShut {NoStop}%
\bibitem [{\citenamefont {Leghtas}\ \emph {et~al.}(2015)\citenamefont
  {Leghtas}, \citenamefont {Touzard}, \citenamefont {Pop}, \citenamefont {Kou},
  \citenamefont {Vlastakis}, \citenamefont {Petrenko}, \citenamefont {Sliwa},
  \citenamefont {Narla}, \citenamefont {Shankar}, \citenamefont {Hatridge}
  \emph {et~al.}}]{15_Confining_Devoret}%
  \BibitemOpen
  \bibfield  {author} {\bibinfo {author} {\bibfnamefont {Z.}~\bibnamefont
  {Leghtas}}, \bibinfo {author} {\bibfnamefont {S.}~\bibnamefont {Touzard}},
  \bibinfo {author} {\bibfnamefont {I.~M.}\ \bibnamefont {Pop}}, \bibinfo
  {author} {\bibfnamefont {A.}~\bibnamefont {Kou}}, \bibinfo {author}
  {\bibfnamefont {B.}~\bibnamefont {Vlastakis}}, \bibinfo {author}
  {\bibfnamefont {A.}~\bibnamefont {Petrenko}}, \bibinfo {author}
  {\bibfnamefont {K.~M.}\ \bibnamefont {Sliwa}}, \bibinfo {author}
  {\bibfnamefont {A.}~\bibnamefont {Narla}}, \bibinfo {author} {\bibfnamefont
  {S.}~\bibnamefont {Shankar}}, \bibinfo {author} {\bibfnamefont {M.~J.}\
  \bibnamefont {Hatridge}},  \emph {et~al.},\ }\href@noop {} {\bibfield
  {journal} {\bibinfo  {journal} {Science}\ }\textbf {\bibinfo {volume}
  {347}},\ \bibinfo {pages} {853} (\bibinfo {year} {2015})}\BibitemShut
  {NoStop}%
\bibitem [{\citenamefont {Ofek}\ \emph {et~al.}(2016)\citenamefont {Ofek},
  \citenamefont {Petrenko}, \citenamefont {Heeres}, \citenamefont {Reinhold},
  \citenamefont {Leghtas}, \citenamefont {Vlastakis}, \citenamefont {Liu},
  \citenamefont {Frunzio}, \citenamefont {Girvin}, \citenamefont {Jiang} \emph
  {et~al.}}]{16_Extending_Schoelkopf}%
  \BibitemOpen
  \bibfield  {author} {\bibinfo {author} {\bibfnamefont {N.}~\bibnamefont
  {Ofek}}, \bibinfo {author} {\bibfnamefont {A.}~\bibnamefont {Petrenko}},
  \bibinfo {author} {\bibfnamefont {R.}~\bibnamefont {Heeres}}, \bibinfo
  {author} {\bibfnamefont {P.}~\bibnamefont {Reinhold}}, \bibinfo {author}
  {\bibfnamefont {Z.}~\bibnamefont {Leghtas}}, \bibinfo {author} {\bibfnamefont
  {B.}~\bibnamefont {Vlastakis}}, \bibinfo {author} {\bibfnamefont
  {Y.}~\bibnamefont {Liu}}, \bibinfo {author} {\bibfnamefont {L.}~\bibnamefont
  {Frunzio}}, \bibinfo {author} {\bibfnamefont {S.}~\bibnamefont {Girvin}},
  \bibinfo {author} {\bibfnamefont {L.}~\bibnamefont {Jiang}},  \emph
  {et~al.},\ }\href@noop {} {\bibfield  {journal} {\bibinfo  {journal}
  {Nature}\ }\textbf {\bibinfo {volume} {536}},\ \bibinfo {pages} {441}
  (\bibinfo {year} {2016})}\BibitemShut {NoStop}%
\bibitem [{\citenamefont {Wang}\ \emph {et~al.}(2016)\citenamefont {Wang},
  \citenamefont {Gao}, \citenamefont {Reinhold}, \citenamefont {Heeres},
  \citenamefont {Ofek}, \citenamefont {Chou}, \citenamefont {Axline},
  \citenamefont {Reagor}, \citenamefont {Blumoff}, \citenamefont {Sliwa} \emph
  {et~al.}}]{16_Schrodinger_Schoelkopf}%
  \BibitemOpen
  \bibfield  {author} {\bibinfo {author} {\bibfnamefont {C.}~\bibnamefont
  {Wang}}, \bibinfo {author} {\bibfnamefont {Y.~Y.}\ \bibnamefont {Gao}},
  \bibinfo {author} {\bibfnamefont {P.}~\bibnamefont {Reinhold}}, \bibinfo
  {author} {\bibfnamefont {R.~W.}\ \bibnamefont {Heeres}}, \bibinfo {author}
  {\bibfnamefont {N.}~\bibnamefont {Ofek}}, \bibinfo {author} {\bibfnamefont
  {K.}~\bibnamefont {Chou}}, \bibinfo {author} {\bibfnamefont {C.}~\bibnamefont
  {Axline}}, \bibinfo {author} {\bibfnamefont {M.}~\bibnamefont {Reagor}},
  \bibinfo {author} {\bibfnamefont {J.}~\bibnamefont {Blumoff}}, \bibinfo
  {author} {\bibfnamefont {K.}~\bibnamefont {Sliwa}},  \emph {et~al.},\
  }\href@noop {} {\bibfield  {journal} {\bibinfo  {journal} {Science}\ }\textbf
  {\bibinfo {volume} {352}},\ \bibinfo {pages} {1087} (\bibinfo {year}
  {2016})}\BibitemShut {NoStop}%
\bibitem [{\citenamefont {Goto}(2016{\natexlab{a}})}]{16_Universal_Goto}%
  \BibitemOpen
  \bibfield  {author} {\bibinfo {author} {\bibfnamefont {H.}~\bibnamefont
  {Goto}},\ }\href@noop {} {\bibfield  {journal} {\bibinfo  {journal} {Physical
  Review A}\ }\textbf {\bibinfo {volume} {93}},\ \bibinfo {pages} {050301}
  (\bibinfo {year} {2016}{\natexlab{a}})}\BibitemShut {NoStop}%
\bibitem [{\citenamefont {Puri}\ \emph
  {et~al.}(2017{\natexlab{a}})\citenamefont {Puri}, \citenamefont {Boutin},\
  and\ \citenamefont {Blais}}]{17_Engineering_Blais}%
  \BibitemOpen
  \bibfield  {author} {\bibinfo {author} {\bibfnamefont {S.}~\bibnamefont
  {Puri}}, \bibinfo {author} {\bibfnamefont {S.}~\bibnamefont {Boutin}}, \ and\
  \bibinfo {author} {\bibfnamefont {A.}~\bibnamefont {Blais}},\ }\href@noop {}
  {\bibfield  {journal} {\bibinfo  {journal} {npj Quantum Information}\
  }\textbf {\bibinfo {volume} {3}},\ \bibinfo {pages} {18} (\bibinfo {year}
  {2017}{\natexlab{a}})}\BibitemShut {NoStop}%
\bibitem [{\citenamefont {Goto}(2016{\natexlab{b}})}]{16_Bifurcation_Goto}%
  \BibitemOpen
  \bibfield  {author} {\bibinfo {author} {\bibfnamefont {H.}~\bibnamefont
  {Goto}},\ }\href@noop {} {\bibfield  {journal} {\bibinfo  {journal}
  {Scientific reports}\ }\textbf {\bibinfo {volume} {6}},\ \bibinfo {pages}
  {21686} (\bibinfo {year} {2016}{\natexlab{b}})}\BibitemShut {NoStop}%
\bibitem [{\citenamefont {Nigg}\ \emph {et~al.}(2017)\citenamefont {Nigg},
  \citenamefont {L{\"o}rch},\ and\ \citenamefont {Tiwari}}]{17_Robust_Tiwari}%
  \BibitemOpen
  \bibfield  {author} {\bibinfo {author} {\bibfnamefont {S.~E.}\ \bibnamefont
  {Nigg}}, \bibinfo {author} {\bibfnamefont {N.}~\bibnamefont {L{\"o}rch}}, \
  and\ \bibinfo {author} {\bibfnamefont {R.~P.}\ \bibnamefont {Tiwari}},\
  }\href@noop {} {\bibfield  {journal} {\bibinfo  {journal} {Science Advances}\
  }\textbf {\bibinfo {volume} {3}},\ \bibinfo {pages} {e1602273} (\bibinfo
  {year} {2017})}\BibitemShut {NoStop}%
\bibitem [{\citenamefont {Puri}\ \emph
  {et~al.}(2017{\natexlab{b}})\citenamefont {Puri}, \citenamefont {Andersen},
  \citenamefont {Grimsmo},\ and\ \citenamefont {Blais}}]{17_Quantum_Blais}%
  \BibitemOpen
  \bibfield  {author} {\bibinfo {author} {\bibfnamefont {S.}~\bibnamefont
  {Puri}}, \bibinfo {author} {\bibfnamefont {C.~K.}\ \bibnamefont {Andersen}},
  \bibinfo {author} {\bibfnamefont {A.~L.}\ \bibnamefont {Grimsmo}}, \ and\
  \bibinfo {author} {\bibfnamefont {A.}~\bibnamefont {Blais}},\ }\href@noop {}
  {\bibfield  {journal} {\bibinfo  {journal} {Nature Communications}\ }\textbf
  {\bibinfo {volume} {8}},\ \bibinfo {pages} {15785} (\bibinfo {year}
  {2017}{\natexlab{b}})}\BibitemShut {NoStop}%
\bibitem [{\citenamefont {Rosenblum}\ \emph
  {et~al.}(2018{\natexlab{b}})\citenamefont {Rosenblum}, \citenamefont {Gao},
  \citenamefont {Reinhold}, \citenamefont {Wang}, \citenamefont {Axline},
  \citenamefont {Frunzio}, \citenamefont {Girvin}, \citenamefont {Jiang},
  \citenamefont {Mirrahimi}, \citenamefont {Devoret} \emph
  {et~al.}}]{18_CNOT_Schoelkopf}%
  \BibitemOpen
  \bibfield  {author} {\bibinfo {author} {\bibfnamefont {S.}~\bibnamefont
  {Rosenblum}}, \bibinfo {author} {\bibfnamefont {Y.~Y.}\ \bibnamefont {Gao}},
  \bibinfo {author} {\bibfnamefont {P.}~\bibnamefont {Reinhold}}, \bibinfo
  {author} {\bibfnamefont {C.}~\bibnamefont {Wang}}, \bibinfo {author}
  {\bibfnamefont {C.~J.}\ \bibnamefont {Axline}}, \bibinfo {author}
  {\bibfnamefont {L.}~\bibnamefont {Frunzio}}, \bibinfo {author} {\bibfnamefont
  {S.~M.}\ \bibnamefont {Girvin}}, \bibinfo {author} {\bibfnamefont
  {L.}~\bibnamefont {Jiang}}, \bibinfo {author} {\bibfnamefont
  {M.}~\bibnamefont {Mirrahimi}}, \bibinfo {author} {\bibfnamefont {M.~H.}\
  \bibnamefont {Devoret}},  \emph {et~al.},\ }\href@noop {} {\bibfield
  {journal} {\bibinfo  {journal} {Nature Communications}\ }\textbf {\bibinfo
  {volume} {9}},\ \bibinfo {pages} {652} (\bibinfo {year}
  {2018}{\natexlab{b}})}\BibitemShut {NoStop}%
\bibitem [{\citenamefont {Kuhr}\ \emph {et~al.}(2007)\citenamefont {Kuhr},
  \citenamefont {Gleyzes}, \citenamefont {Guerlin}, \citenamefont {Bernu},
  \citenamefont {Hoff}, \citenamefont {Del{\'e}glise}, \citenamefont {Osnaghi},
  \citenamefont {Brune}, \citenamefont {Raimond}, \citenamefont {Haroche} \emph
  {et~al.}}]{07_Ultrahigh_Visentin}%
  \BibitemOpen
  \bibfield  {author} {\bibinfo {author} {\bibfnamefont {S.}~\bibnamefont
  {Kuhr}}, \bibinfo {author} {\bibfnamefont {S.}~\bibnamefont {Gleyzes}},
  \bibinfo {author} {\bibfnamefont {C.}~\bibnamefont {Guerlin}}, \bibinfo
  {author} {\bibfnamefont {J.}~\bibnamefont {Bernu}}, \bibinfo {author}
  {\bibfnamefont {U.~B.}\ \bibnamefont {Hoff}}, \bibinfo {author}
  {\bibfnamefont {S.}~\bibnamefont {Del{\'e}glise}}, \bibinfo {author}
  {\bibfnamefont {S.}~\bibnamefont {Osnaghi}}, \bibinfo {author} {\bibfnamefont
  {M.}~\bibnamefont {Brune}}, \bibinfo {author} {\bibfnamefont {J.-M.}\
  \bibnamefont {Raimond}}, \bibinfo {author} {\bibfnamefont {S.}~\bibnamefont
  {Haroche}},  \emph {et~al.},\ }\href@noop {} {\bibfield  {journal} {\bibinfo
  {journal} {Applied Physics Letters}\ }\textbf {\bibinfo {volume} {90}},\
  \bibinfo {pages} {164101} (\bibinfo {year} {2007})}\BibitemShut {NoStop}%
\bibitem [{\citenamefont {Reagor}\ \emph {et~al.}(2013)\citenamefont {Reagor},
  \citenamefont {Paik}, \citenamefont {Catelani}, \citenamefont {Sun},
  \citenamefont {Axline}, \citenamefont {Holland}, \citenamefont {Pop},
  \citenamefont {Masluk}, \citenamefont {Brecht}, \citenamefont {Frunzio} \emph
  {et~al.}}]{13_Reaching_Schoelkopf}%
  \BibitemOpen
  \bibfield  {author} {\bibinfo {author} {\bibfnamefont {M.}~\bibnamefont
  {Reagor}}, \bibinfo {author} {\bibfnamefont {H.}~\bibnamefont {Paik}},
  \bibinfo {author} {\bibfnamefont {G.}~\bibnamefont {Catelani}}, \bibinfo
  {author} {\bibfnamefont {L.}~\bibnamefont {Sun}}, \bibinfo {author}
  {\bibfnamefont {C.}~\bibnamefont {Axline}}, \bibinfo {author} {\bibfnamefont
  {E.}~\bibnamefont {Holland}}, \bibinfo {author} {\bibfnamefont {I.~M.}\
  \bibnamefont {Pop}}, \bibinfo {author} {\bibfnamefont {N.~A.}\ \bibnamefont
  {Masluk}}, \bibinfo {author} {\bibfnamefont {T.}~\bibnamefont {Brecht}},
  \bibinfo {author} {\bibfnamefont {L.}~\bibnamefont {Frunzio}},  \emph
  {et~al.},\ }\href@noop {} {\bibfield  {journal} {\bibinfo  {journal} {Applied
  Physics Letters}\ }\textbf {\bibinfo {volume} {102}},\ \bibinfo {pages}
  {192604} (\bibinfo {year} {2013})}\BibitemShut {NoStop}%
\bibitem [{\citenamefont {Romanenko}\ \emph {et~al.}(2018)\citenamefont
  {Romanenko}, \citenamefont {Pilipenko}, \citenamefont {Zorzetti},
  \citenamefont {Frolov}, \citenamefont {Awida}, \citenamefont {Posen},\ and\
  \citenamefont {Grassellino}}]{18_3D_Grassellino}%
  \BibitemOpen
  \bibfield  {author} {\bibinfo {author} {\bibfnamefont {A.}~\bibnamefont
  {Romanenko}}, \bibinfo {author} {\bibfnamefont {R.}~\bibnamefont
  {Pilipenko}}, \bibinfo {author} {\bibfnamefont {S.}~\bibnamefont {Zorzetti}},
  \bibinfo {author} {\bibfnamefont {D.}~\bibnamefont {Frolov}}, \bibinfo
  {author} {\bibfnamefont {M.}~\bibnamefont {Awida}}, \bibinfo {author}
  {\bibfnamefont {S.}~\bibnamefont {Posen}}, \ and\ \bibinfo {author}
  {\bibfnamefont {A.}~\bibnamefont {Grassellino}},\ }\href@noop {} {\bibfield
  {journal} {\bibinfo  {journal} {arXiv:1810.03703}\ } (\bibinfo {year}
  {2018})}\BibitemShut {NoStop}%
\bibitem [{\citenamefont {Gottesman}\ \emph {et~al.}(2001)\citenamefont
  {Gottesman}, \citenamefont {Kitaev},\ and\ \citenamefont
  {Preskill}}]{01_Encoding_Preskill}%
  \BibitemOpen
  \bibfield  {author} {\bibinfo {author} {\bibfnamefont {D.}~\bibnamefont
  {Gottesman}}, \bibinfo {author} {\bibfnamefont {A.}~\bibnamefont {Kitaev}}, \
  and\ \bibinfo {author} {\bibfnamefont {J.}~\bibnamefont {Preskill}},\
  }\href@noop {} {\bibfield  {journal} {\bibinfo  {journal} {Physical Review
  A}\ }\textbf {\bibinfo {volume} {64}},\ \bibinfo {pages} {012310} (\bibinfo
  {year} {2001})}\BibitemShut {NoStop}%
\bibitem [{\citenamefont {Cohen}\ \emph {et~al.}(2017)\citenamefont {Cohen},
  \citenamefont {Smith}, \citenamefont {Devoret},\ and\ \citenamefont
  {Mirrahimi}}]{17_Degeneracy_Mirrahimi}%
  \BibitemOpen
  \bibfield  {author} {\bibinfo {author} {\bibfnamefont {J.}~\bibnamefont
  {Cohen}}, \bibinfo {author} {\bibfnamefont {W.~C.}\ \bibnamefont {Smith}},
  \bibinfo {author} {\bibfnamefont {M.~H.}\ \bibnamefont {Devoret}}, \ and\
  \bibinfo {author} {\bibfnamefont {M.}~\bibnamefont {Mirrahimi}},\ }\href
  {\doibase 10.1103/PhysRevLett.119.060503} {\bibfield  {journal} {\bibinfo
  {journal} {Physical Review Letters}\ }\textbf {\bibinfo {volume} {119}},\
  \bibinfo {pages} {060503} (\bibinfo {year} {2017})}\BibitemShut {NoStop}%
\bibitem [{\citenamefont {Puri}\ \emph {et~al.}(2018)\citenamefont {Puri},
  \citenamefont {Grimm}, \citenamefont {Campagne-Ibarcq}, \citenamefont
  {Eickbusch}, \citenamefont {Noh}, \citenamefont {Roberts}, \citenamefont
  {Jiang}, \citenamefont {Mirrahimi}, \citenamefont {Devoret},\ and\
  \citenamefont {Girvin}}]{18_Stabilized_Girvin}%
  \BibitemOpen
  \bibfield  {author} {\bibinfo {author} {\bibfnamefont {S.}~\bibnamefont
  {Puri}}, \bibinfo {author} {\bibfnamefont {A.}~\bibnamefont {Grimm}},
  \bibinfo {author} {\bibfnamefont {P.}~\bibnamefont {Campagne-Ibarcq}},
  \bibinfo {author} {\bibfnamefont {A.}~\bibnamefont {Eickbusch}}, \bibinfo
  {author} {\bibfnamefont {K.}~\bibnamefont {Noh}}, \bibinfo {author}
  {\bibfnamefont {G.}~\bibnamefont {Roberts}}, \bibinfo {author} {\bibfnamefont
  {L.}~\bibnamefont {Jiang}}, \bibinfo {author} {\bibfnamefont
  {M.}~\bibnamefont {Mirrahimi}}, \bibinfo {author} {\bibfnamefont {M.~H.}\
  \bibnamefont {Devoret}}, \ and\ \bibinfo {author} {\bibfnamefont {S.~M.}\
  \bibnamefont {Girvin}},\ }\href@noop {} {\bibfield  {journal} {\bibinfo
  {journal} {arXiv:1807.09334}\ } (\bibinfo {year} {2018})}\BibitemShut
  {NoStop}%
\bibitem [{\citenamefont {Gao}\ \emph {et~al.}(2019)\citenamefont {Gao},
  \citenamefont {Lester}, \citenamefont {Chou}, \citenamefont {Frunzio},
  \citenamefont {Devoret}, \citenamefont {Jiang}, \citenamefont {Girvin},\ and\
  \citenamefont {Schoelkopf}}]{19_Entanglement_Schoelkopf}%
  \BibitemOpen
  \bibfield  {author} {\bibinfo {author} {\bibfnamefont {Y.~Y.}\ \bibnamefont
  {Gao}}, \bibinfo {author} {\bibfnamefont {B.~J.}\ \bibnamefont {Lester}},
  \bibinfo {author} {\bibfnamefont {K.~S.}\ \bibnamefont {Chou}}, \bibinfo
  {author} {\bibfnamefont {L.}~\bibnamefont {Frunzio}}, \bibinfo {author}
  {\bibfnamefont {M.~H.}\ \bibnamefont {Devoret}}, \bibinfo {author}
  {\bibfnamefont {L.}~\bibnamefont {Jiang}}, \bibinfo {author} {\bibfnamefont
  {S.}~\bibnamefont {Girvin}}, \ and\ \bibinfo {author} {\bibfnamefont {R.~J.}\
  \bibnamefont {Schoelkopf}},\ }\href@noop {} {\bibfield  {journal} {\bibinfo
  {journal} {Nature}\ }\textbf {\bibinfo {volume} {566}},\ \bibinfo {pages}
  {509} (\bibinfo {year} {2019})}\BibitemShut {NoStop}%
\bibitem [{\citenamefont {Kubo}\ \emph {et~al.}(2010)\citenamefont {Kubo},
  \citenamefont {Ong}, \citenamefont {Bertet}, \citenamefont {Vion},
  \citenamefont {Jacques}, \citenamefont {Zheng}, \citenamefont {Dr{\'e}au},
  \citenamefont {Roch}, \citenamefont {Auff{\`e}ves}, \citenamefont {Jelezko}
  \emph {et~al.}}]{10_Strong_Esteve}%
  \BibitemOpen
  \bibfield  {author} {\bibinfo {author} {\bibfnamefont {Y.}~\bibnamefont
  {Kubo}}, \bibinfo {author} {\bibfnamefont {F.}~\bibnamefont {Ong}}, \bibinfo
  {author} {\bibfnamefont {P.}~\bibnamefont {Bertet}}, \bibinfo {author}
  {\bibfnamefont {D.}~\bibnamefont {Vion}}, \bibinfo {author} {\bibfnamefont
  {V.}~\bibnamefont {Jacques}}, \bibinfo {author} {\bibfnamefont
  {D.}~\bibnamefont {Zheng}}, \bibinfo {author} {\bibfnamefont
  {A.}~\bibnamefont {Dr{\'e}au}}, \bibinfo {author} {\bibfnamefont {J.-F.}\
  \bibnamefont {Roch}}, \bibinfo {author} {\bibfnamefont {A.}~\bibnamefont
  {Auff{\`e}ves}}, \bibinfo {author} {\bibfnamefont {F.}~\bibnamefont
  {Jelezko}},  \emph {et~al.},\ }\href@noop {} {\bibfield  {journal} {\bibinfo
  {journal} {Physical Review Letters}\ }\textbf {\bibinfo {volume} {105}},\
  \bibinfo {pages} {140502} (\bibinfo {year} {2010})}\BibitemShut {NoStop}%
\bibitem [{\citenamefont {Verd{\'u}}\ \emph {et~al.}(2009)\citenamefont
  {Verd{\'u}}, \citenamefont {Zoubi}, \citenamefont {Koller}, \citenamefont
  {Majer}, \citenamefont {Ritsch},\ and\ \citenamefont
  {Schmiedmayer}}]{09_Strong_Schmiedmayer}%
  \BibitemOpen
  \bibfield  {author} {\bibinfo {author} {\bibfnamefont {J.}~\bibnamefont
  {Verd{\'u}}}, \bibinfo {author} {\bibfnamefont {H.}~\bibnamefont {Zoubi}},
  \bibinfo {author} {\bibfnamefont {C.}~\bibnamefont {Koller}}, \bibinfo
  {author} {\bibfnamefont {J.}~\bibnamefont {Majer}}, \bibinfo {author}
  {\bibfnamefont {H.}~\bibnamefont {Ritsch}}, \ and\ \bibinfo {author}
  {\bibfnamefont {J.}~\bibnamefont {Schmiedmayer}},\ }\href@noop {} {\bibfield
  {journal} {\bibinfo  {journal} {Physical Review Letters}\ }\textbf {\bibinfo
  {volume} {103}},\ \bibinfo {pages} {043603} (\bibinfo {year}
  {2009})}\BibitemShut {NoStop}%
\bibitem [{\citenamefont {Staudt}\ \emph {et~al.}(2012)\citenamefont {Staudt},
  \citenamefont {Hoi}, \citenamefont {Krantz}, \citenamefont {Sandberg},
  \citenamefont {Simoen}, \citenamefont {Bushev}, \citenamefont {Sangouard},
  \citenamefont {Afzelius}, \citenamefont {Shumeiko}, \citenamefont {Johansson}
  \emph {et~al.}}]{12_Coupling_Johansson}%
  \BibitemOpen
  \bibfield  {author} {\bibinfo {author} {\bibfnamefont {M.~U.}\ \bibnamefont
  {Staudt}}, \bibinfo {author} {\bibfnamefont {I.-C.}\ \bibnamefont {Hoi}},
  \bibinfo {author} {\bibfnamefont {P.}~\bibnamefont {Krantz}}, \bibinfo
  {author} {\bibfnamefont {M.}~\bibnamefont {Sandberg}}, \bibinfo {author}
  {\bibfnamefont {M.}~\bibnamefont {Simoen}}, \bibinfo {author} {\bibfnamefont
  {P.}~\bibnamefont {Bushev}}, \bibinfo {author} {\bibfnamefont
  {N.}~\bibnamefont {Sangouard}}, \bibinfo {author} {\bibfnamefont
  {M.}~\bibnamefont {Afzelius}}, \bibinfo {author} {\bibfnamefont {V.~S.}\
  \bibnamefont {Shumeiko}}, \bibinfo {author} {\bibfnamefont {G.}~\bibnamefont
  {Johansson}},  \emph {et~al.},\ }\href@noop {} {\bibfield  {journal}
  {\bibinfo  {journal} {Journal of Physics B: Atomic, Molecular and Optical
  Physics}\ }\textbf {\bibinfo {volume} {45}},\ \bibinfo {pages} {124019}
  (\bibinfo {year} {2012})}\BibitemShut {NoStop}%
\bibitem [{\citenamefont {O’Brien}\ \emph {et~al.}(2014)\citenamefont
  {O’Brien}, \citenamefont {Lauk}, \citenamefont {Blum}, \citenamefont
  {Morigi},\ and\ \citenamefont {Fleischhauer}}]{14_Interfacing_Fleischhauer}%
  \BibitemOpen
  \bibfield  {author} {\bibinfo {author} {\bibfnamefont {C.}~\bibnamefont
  {O’Brien}}, \bibinfo {author} {\bibfnamefont {N.}~\bibnamefont {Lauk}},
  \bibinfo {author} {\bibfnamefont {S.}~\bibnamefont {Blum}}, \bibinfo {author}
  {\bibfnamefont {G.}~\bibnamefont {Morigi}}, \ and\ \bibinfo {author}
  {\bibfnamefont {M.}~\bibnamefont {Fleischhauer}},\ }\href@noop {} {\bibfield
  {journal} {\bibinfo  {journal} {Physical Review Letters}\ }\textbf {\bibinfo
  {volume} {113}},\ \bibinfo {pages} {063603} (\bibinfo {year}
  {2014})}\BibitemShut {NoStop}%
\bibitem [{\citenamefont {Williamson}\ \emph {et~al.}(2014)\citenamefont
  {Williamson}, \citenamefont {Chen},\ and\ \citenamefont
  {Longdell}}]{14_Magneto_Longdell}%
  \BibitemOpen
  \bibfield  {author} {\bibinfo {author} {\bibfnamefont {L.~A.}\ \bibnamefont
  {Williamson}}, \bibinfo {author} {\bibfnamefont {Y.-H.}\ \bibnamefont
  {Chen}}, \ and\ \bibinfo {author} {\bibfnamefont {J.~J.}\ \bibnamefont
  {Longdell}},\ }\href@noop {} {\bibfield  {journal} {\bibinfo  {journal}
  {Physical Review Letters}\ }\textbf {\bibinfo {volume} {113}},\ \bibinfo
  {pages} {203601} (\bibinfo {year} {2014})}\BibitemShut {NoStop}%
\bibitem [{\citenamefont {Bochmann}\ \emph {et~al.}(2013)\citenamefont
  {Bochmann}, \citenamefont {Vainsencher}, \citenamefont {Awschalom},\ and\
  \citenamefont {Cleland}}]{13_Nanomechanical_Cleland}%
  \BibitemOpen
  \bibfield  {author} {\bibinfo {author} {\bibfnamefont {J.}~\bibnamefont
  {Bochmann}}, \bibinfo {author} {\bibfnamefont {A.}~\bibnamefont
  {Vainsencher}}, \bibinfo {author} {\bibfnamefont {D.~D.}\ \bibnamefont
  {Awschalom}}, \ and\ \bibinfo {author} {\bibfnamefont {A.~N.}\ \bibnamefont
  {Cleland}},\ }\href@noop {} {\bibfield  {journal} {\bibinfo  {journal}
  {Nature Physics}\ }\textbf {\bibinfo {volume} {9}},\ \bibinfo {pages} {712}
  (\bibinfo {year} {2013})}\BibitemShut {NoStop}%
\bibitem [{\citenamefont {Andrews}\ \emph {et~al.}(2014)\citenamefont
  {Andrews}, \citenamefont {Peterson}, \citenamefont {Purdy}, \citenamefont
  {Cicak}, \citenamefont {Simmonds}, \citenamefont {Regal},\ and\ \citenamefont
  {Lehnert}}]{14_Bidirectional_Lehnert}%
  \BibitemOpen
  \bibfield  {author} {\bibinfo {author} {\bibfnamefont {R.~W.}\ \bibnamefont
  {Andrews}}, \bibinfo {author} {\bibfnamefont {R.~W.}\ \bibnamefont
  {Peterson}}, \bibinfo {author} {\bibfnamefont {T.~P.}\ \bibnamefont {Purdy}},
  \bibinfo {author} {\bibfnamefont {K.}~\bibnamefont {Cicak}}, \bibinfo
  {author} {\bibfnamefont {R.~W.}\ \bibnamefont {Simmonds}}, \bibinfo {author}
  {\bibfnamefont {C.~A.}\ \bibnamefont {Regal}}, \ and\ \bibinfo {author}
  {\bibfnamefont {K.~W.}\ \bibnamefont {Lehnert}},\ }\href@noop {} {\bibfield
  {journal} {\bibinfo  {journal} {Nature Physics}\ }\textbf {\bibinfo {volume}
  {10}},\ \bibinfo {pages} {321} (\bibinfo {year} {2014})}\BibitemShut
  {NoStop}%
\bibitem [{\citenamefont {Higginbotham}\ \emph {et~al.}(2018)\citenamefont
  {Higginbotham}, \citenamefont {Burns}, \citenamefont {Urmey}, \citenamefont
  {Peterson}, \citenamefont {Kampel}, \citenamefont {Brubaker}, \citenamefont
  {Smith}, \citenamefont {Lehnert},\ and\ \citenamefont
  {Regal}}]{18_Harnessing_Regal}%
  \BibitemOpen
  \bibfield  {author} {\bibinfo {author} {\bibfnamefont {A.}~\bibnamefont
  {Higginbotham}}, \bibinfo {author} {\bibfnamefont {P.}~\bibnamefont {Burns}},
  \bibinfo {author} {\bibfnamefont {M.}~\bibnamefont {Urmey}}, \bibinfo
  {author} {\bibfnamefont {R.}~\bibnamefont {Peterson}}, \bibinfo {author}
  {\bibfnamefont {N.}~\bibnamefont {Kampel}}, \bibinfo {author} {\bibfnamefont
  {B.}~\bibnamefont {Brubaker}}, \bibinfo {author} {\bibfnamefont
  {G.}~\bibnamefont {Smith}}, \bibinfo {author} {\bibfnamefont
  {K.}~\bibnamefont {Lehnert}}, \ and\ \bibinfo {author} {\bibfnamefont
  {C.}~\bibnamefont {Regal}},\ }\href@noop {} {\bibfield  {journal} {\bibinfo
  {journal} {Nature Physics}\ }\textbf {\bibinfo {volume} {14}},\ \bibinfo
  {pages} {1038} (\bibinfo {year} {2018})}\BibitemShut {NoStop}%
\bibitem [{\citenamefont {Forsch}\ \emph {et~al.}(2018)\citenamefont {Forsch},
  \citenamefont {Stockill}, \citenamefont {Wallucks}, \citenamefont
  {Marinkovic}, \citenamefont {G{\"a}rtner}, \citenamefont {Norte},
  \citenamefont {van Otten}, \citenamefont {Fiore}, \citenamefont
  {Srinivasan},\ and\ \citenamefont
  {Gr{\"o}blacher}}]{18_Microwave_Groblacher}%
  \BibitemOpen
  \bibfield  {author} {\bibinfo {author} {\bibfnamefont {M.}~\bibnamefont
  {Forsch}}, \bibinfo {author} {\bibfnamefont {R.}~\bibnamefont {Stockill}},
  \bibinfo {author} {\bibfnamefont {A.}~\bibnamefont {Wallucks}}, \bibinfo
  {author} {\bibfnamefont {I.}~\bibnamefont {Marinkovic}}, \bibinfo {author}
  {\bibfnamefont {C.}~\bibnamefont {G{\"a}rtner}}, \bibinfo {author}
  {\bibfnamefont {R.~A.}\ \bibnamefont {Norte}}, \bibinfo {author}
  {\bibfnamefont {F.}~\bibnamefont {van Otten}}, \bibinfo {author}
  {\bibfnamefont {A.}~\bibnamefont {Fiore}}, \bibinfo {author} {\bibfnamefont
  {K.}~\bibnamefont {Srinivasan}}, \ and\ \bibinfo {author} {\bibfnamefont
  {S.}~\bibnamefont {Gr{\"o}blacher}},\ }\href@noop {} {\bibfield  {journal}
  {\bibinfo  {journal} {arXiv:1812.07588}\ } (\bibinfo {year}
  {2018})}\BibitemShut {NoStop}%
\bibitem [{\citenamefont {Welinski}\ \emph {et~al.}(2019)\citenamefont
  {Welinski}, \citenamefont {Woodburn}, \citenamefont {Lauk}, \citenamefont
  {Cone}, \citenamefont {Simon}, \citenamefont {Goldner},\ and\ \citenamefont
  {Thiel}}]{18_Electron_Thiel}%
  \BibitemOpen
  \bibfield  {author} {\bibinfo {author} {\bibfnamefont {S.}~\bibnamefont
  {Welinski}}, \bibinfo {author} {\bibfnamefont {P.~J.}\ \bibnamefont
  {Woodburn}}, \bibinfo {author} {\bibfnamefont {N.}~\bibnamefont {Lauk}},
  \bibinfo {author} {\bibfnamefont {R.~L.}\ \bibnamefont {Cone}}, \bibinfo
  {author} {\bibfnamefont {C.}~\bibnamefont {Simon}}, \bibinfo {author}
  {\bibfnamefont {P.}~\bibnamefont {Goldner}}, \ and\ \bibinfo {author}
  {\bibfnamefont {C.~W.}\ \bibnamefont {Thiel}},\ }\href@noop {} {\bibfield
  {journal} {\bibinfo  {journal} {Physical Review Letters}\ }\textbf {\bibinfo
  {volume} {122}},\ \bibinfo {pages} {247401} (\bibinfo {year}
  {2019})}\BibitemShut {NoStop}%
\bibitem [{\citenamefont {Duan}\ \emph {et~al.}(2001)\citenamefont {Duan},
  \citenamefont {Lukin}, \citenamefont {Cirac},\ and\ \citenamefont
  {Zoller}}]{01_Long_Zoller}%
  \BibitemOpen
  \bibfield  {author} {\bibinfo {author} {\bibfnamefont {L.-M.}\ \bibnamefont
  {Duan}}, \bibinfo {author} {\bibfnamefont {M.}~\bibnamefont {Lukin}},
  \bibinfo {author} {\bibfnamefont {J.~I.}\ \bibnamefont {Cirac}}, \ and\
  \bibinfo {author} {\bibfnamefont {P.}~\bibnamefont {Zoller}},\ }\href@noop {}
  {\bibfield  {journal} {\bibinfo  {journal} {Nature}\ }\textbf {\bibinfo
  {volume} {414}},\ \bibinfo {pages} {413} (\bibinfo {year}
  {2001})}\BibitemShut {NoStop}%
\bibitem [{\citenamefont {L{\"u}tkenhaus}\ \emph {et~al.}(1999)\citenamefont
  {L{\"u}tkenhaus}, \citenamefont {Calsamiglia},\ and\ \citenamefont
  {Suominen}}]{99_Bell_Suominem}%
  \BibitemOpen
  \bibfield  {author} {\bibinfo {author} {\bibfnamefont {N.}~\bibnamefont
  {L{\"u}tkenhaus}}, \bibinfo {author} {\bibfnamefont {J.}~\bibnamefont
  {Calsamiglia}}, \ and\ \bibinfo {author} {\bibfnamefont {K.-A.}\ \bibnamefont
  {Suominen}},\ }\href@noop {} {\bibfield  {journal} {\bibinfo  {journal}
  {Physical Review A}\ }\textbf {\bibinfo {volume} {59}},\ \bibinfo {pages}
  {3295} (\bibinfo {year} {1999})}\BibitemShut {NoStop}%
\bibitem [{\citenamefont {Grice}(2011)}]{11_Arbitrarily_Grice}%
  \BibitemOpen
  \bibfield  {author} {\bibinfo {author} {\bibfnamefont {W.~P.}\ \bibnamefont
  {Grice}},\ }\href@noop {} {\bibfield  {journal} {\bibinfo  {journal}
  {Physical Review A}\ }\textbf {\bibinfo {volume} {84}},\ \bibinfo {pages}
  {042331} (\bibinfo {year} {2011})}\BibitemShut {NoStop}%
\bibitem [{\citenamefont {Wein}\ \emph {et~al.}(2016)\citenamefont {Wein},
  \citenamefont {Heshami}, \citenamefont {Fuchs}, \citenamefont {Krovi},
  \citenamefont {Dutton}, \citenamefont {Tittel},\ and\ \citenamefont
  {Simon}}]{16_Efficiency_Simon}%
  \BibitemOpen
  \bibfield  {author} {\bibinfo {author} {\bibfnamefont {S.}~\bibnamefont
  {Wein}}, \bibinfo {author} {\bibfnamefont {K.}~\bibnamefont {Heshami}},
  \bibinfo {author} {\bibfnamefont {C.~A.}\ \bibnamefont {Fuchs}}, \bibinfo
  {author} {\bibfnamefont {H.}~\bibnamefont {Krovi}}, \bibinfo {author}
  {\bibfnamefont {Z.}~\bibnamefont {Dutton}}, \bibinfo {author} {\bibfnamefont
  {W.}~\bibnamefont {Tittel}}, \ and\ \bibinfo {author} {\bibfnamefont
  {C.}~\bibnamefont {Simon}},\ }\href@noop {} {\bibfield  {journal} {\bibinfo
  {journal} {Physical Review A}\ }\textbf {\bibinfo {volume} {94}},\ \bibinfo
  {pages} {032332} (\bibinfo {year} {2016})}\BibitemShut {NoStop}%
\bibitem [{\citenamefont {Sangouard}\ \emph {et~al.}(2009)\citenamefont
  {Sangouard}, \citenamefont {Dubessy},\ and\ \citenamefont
  {Simon}}]{09_Quantum_Simon}%
  \BibitemOpen
  \bibfield  {author} {\bibinfo {author} {\bibfnamefont {N.}~\bibnamefont
  {Sangouard}}, \bibinfo {author} {\bibfnamefont {R.}~\bibnamefont {Dubessy}},
  \ and\ \bibinfo {author} {\bibfnamefont {C.}~\bibnamefont {Simon}},\
  }\href@noop {} {\bibfield  {journal} {\bibinfo  {journal} {Physical Review
  A}\ }\textbf {\bibinfo {volume} {79}},\ \bibinfo {pages} {042340} (\bibinfo
  {year} {2009})}\BibitemShut {NoStop}%
\bibitem [{\citenamefont {Zhao}\ \emph {et~al.}(2010)\citenamefont {Zhao},
  \citenamefont {M{\"u}ller}, \citenamefont {Hammerer},\ and\ \citenamefont
  {Zoller}}]{10_Efficient_Zoller}%
  \BibitemOpen
  \bibfield  {author} {\bibinfo {author} {\bibfnamefont {B.}~\bibnamefont
  {Zhao}}, \bibinfo {author} {\bibfnamefont {M.}~\bibnamefont {M{\"u}ller}},
  \bibinfo {author} {\bibfnamefont {K.}~\bibnamefont {Hammerer}}, \ and\
  \bibinfo {author} {\bibfnamefont {P.}~\bibnamefont {Zoller}},\ }\href@noop {}
  {\bibfield  {journal} {\bibinfo  {journal} {Physical Review A}\ }\textbf
  {\bibinfo {volume} {81}},\ \bibinfo {pages} {052329} (\bibinfo {year}
  {2010})}\BibitemShut {NoStop}%
\bibitem [{\citenamefont {Han}\ \emph {et~al.}(2010)\citenamefont {Han},
  \citenamefont {He}, \citenamefont {Heshami}, \citenamefont {Li},\ and\
  \citenamefont {Simon}}]{10_Quantumrepeaters_Simon}%
  \BibitemOpen
  \bibfield  {author} {\bibinfo {author} {\bibfnamefont {Y.}~\bibnamefont
  {Han}}, \bibinfo {author} {\bibfnamefont {B.}~\bibnamefont {He}}, \bibinfo
  {author} {\bibfnamefont {K.}~\bibnamefont {Heshami}}, \bibinfo {author}
  {\bibfnamefont {C.-Z.}\ \bibnamefont {Li}}, \ and\ \bibinfo {author}
  {\bibfnamefont {C.}~\bibnamefont {Simon}},\ }\href@noop {} {\bibfield
  {journal} {\bibinfo  {journal} {Physical Review A}\ }\textbf {\bibinfo
  {volume} {81}},\ \bibinfo {pages} {052311} (\bibinfo {year}
  {2010})}\BibitemShut {NoStop}%
\bibitem [{\citenamefont {Reiserer}\ and\ \citenamefont
  {Rempe}(2015)}]{15_Cavity-based_Rempe}%
  \BibitemOpen
  \bibfield  {author} {\bibinfo {author} {\bibfnamefont {A.}~\bibnamefont
  {Reiserer}}\ and\ \bibinfo {author} {\bibfnamefont {G.}~\bibnamefont
  {Rempe}},\ }\href@noop {} {\bibfield  {journal} {\bibinfo  {journal} {Reviews
  of Modern Physics}\ }\textbf {\bibinfo {volume} {87}},\ \bibinfo {pages}
  {1379} (\bibinfo {year} {2015})}\BibitemShut {NoStop}%
\bibitem [{\citenamefont {Covey}\ \emph {et~al.}(2019)\citenamefont {Covey},
  \citenamefont {Sipahigil}, \citenamefont {Szoke}, \citenamefont {Sinclair},
  \citenamefont {Endres},\ and\ \citenamefont {Painter}}]{19_Telecom_Painter}%
  \BibitemOpen
  \bibfield  {author} {\bibinfo {author} {\bibfnamefont {J.~P.}\ \bibnamefont
  {Covey}}, \bibinfo {author} {\bibfnamefont {A.}~\bibnamefont {Sipahigil}},
  \bibinfo {author} {\bibfnamefont {S.}~\bibnamefont {Szoke}}, \bibinfo
  {author} {\bibfnamefont {N.}~\bibnamefont {Sinclair}}, \bibinfo {author}
  {\bibfnamefont {M.}~\bibnamefont {Endres}}, \ and\ \bibinfo {author}
  {\bibfnamefont {O.}~\bibnamefont {Painter}},\ }\href@noop {} {\bibfield
  {journal} {\bibinfo  {journal} {Physical Review Applied}\ }\textbf {\bibinfo
  {volume} {11}},\ \bibinfo {pages} {034044} (\bibinfo {year}
  {2019})}\BibitemShut {NoStop}%
\bibitem [{\citenamefont {Asadi}\ \emph {et~al.}(2018)\citenamefont {Asadi},
  \citenamefont {Lauk}, \citenamefont {Wein}, \citenamefont {Sinclair},
  \citenamefont {O{\textquotesingle}Brien},\ and\ \citenamefont
  {Simon}}]{18_Quantum_Simon}%
  \BibitemOpen
  \bibfield  {author} {\bibinfo {author} {\bibfnamefont {F.~K.}\ \bibnamefont
  {Asadi}}, \bibinfo {author} {\bibfnamefont {N.}~\bibnamefont {Lauk}},
  \bibinfo {author} {\bibfnamefont {S.}~\bibnamefont {Wein}}, \bibinfo {author}
  {\bibfnamefont {N.}~\bibnamefont {Sinclair}}, \bibinfo {author}
  {\bibfnamefont {C.}~\bibnamefont {O{\textquotesingle}Brien}}, \ and\ \bibinfo
  {author} {\bibfnamefont {C.}~\bibnamefont {Simon}},\ }\href {\doibase
  10.22331/q-2018-09-13-93} {\bibfield  {journal} {\bibinfo  {journal}
  {Quantum}\ }\textbf {\bibinfo {volume} {2}},\ \bibinfo {pages} {93} (\bibinfo
  {year} {2018})}\BibitemShut {NoStop}%
\bibitem [{\citenamefont {Hensen}\ \emph {et~al.}(2015)\citenamefont {Hensen},
  \citenamefont {Bernien}, \citenamefont {Dr{\'e}au}, \citenamefont {Reiserer},
  \citenamefont {Kalb}, \citenamefont {Blok}, \citenamefont {Ruitenberg},
  \citenamefont {Vermeulen}, \citenamefont {Schouten}, \citenamefont
  {Abell{\'a}n} \emph {et~al.}}]{15_Loophole_Hanson}%
  \BibitemOpen
  \bibfield  {author} {\bibinfo {author} {\bibfnamefont {B.}~\bibnamefont
  {Hensen}}, \bibinfo {author} {\bibfnamefont {H.}~\bibnamefont {Bernien}},
  \bibinfo {author} {\bibfnamefont {A.~E.}\ \bibnamefont {Dr{\'e}au}}, \bibinfo
  {author} {\bibfnamefont {A.}~\bibnamefont {Reiserer}}, \bibinfo {author}
  {\bibfnamefont {N.}~\bibnamefont {Kalb}}, \bibinfo {author} {\bibfnamefont
  {M.~S.}\ \bibnamefont {Blok}}, \bibinfo {author} {\bibfnamefont
  {J.}~\bibnamefont {Ruitenberg}}, \bibinfo {author} {\bibfnamefont {R.~F.}\
  \bibnamefont {Vermeulen}}, \bibinfo {author} {\bibfnamefont {R.~N.}\
  \bibnamefont {Schouten}}, \bibinfo {author} {\bibfnamefont {C.}~\bibnamefont
  {Abell{\'a}n}},  \emph {et~al.},\ }\href@noop {} {\bibfield  {journal}
  {\bibinfo  {journal} {Nature}\ }\textbf {\bibinfo {volume} {526}},\ \bibinfo
  {pages} {682} (\bibinfo {year} {2015})}\BibitemShut {NoStop}%
\bibitem [{\citenamefont {Moehring}\ \emph {et~al.}(2007)\citenamefont
  {Moehring}, \citenamefont {Maunz}, \citenamefont {Olmschenk}, \citenamefont
  {Younge}, \citenamefont {Matsukevich}, \citenamefont {Duan},\ and\
  \citenamefont {Monroe}}]{07_Entanglement_Monroe}%
  \BibitemOpen
  \bibfield  {author} {\bibinfo {author} {\bibfnamefont {D.}~\bibnamefont
  {Moehring}}, \bibinfo {author} {\bibfnamefont {P.}~\bibnamefont {Maunz}},
  \bibinfo {author} {\bibfnamefont {S.}~\bibnamefont {Olmschenk}}, \bibinfo
  {author} {\bibfnamefont {K.}~\bibnamefont {Younge}}, \bibinfo {author}
  {\bibfnamefont {D.}~\bibnamefont {Matsukevich}}, \bibinfo {author}
  {\bibfnamefont {L.-M.}\ \bibnamefont {Duan}}, \ and\ \bibinfo {author}
  {\bibfnamefont {C.}~\bibnamefont {Monroe}},\ }\href@noop {} {\bibfield
  {journal} {\bibinfo  {journal} {Nature}\ }\textbf {\bibinfo {volume} {449}},\
  \bibinfo {pages} {68} (\bibinfo {year} {2007})}\BibitemShut {NoStop}%
\bibitem [{\citenamefont {Delteil}\ \emph {et~al.}(2015)\citenamefont
  {Delteil}, \citenamefont {Sun}, \citenamefont {Gao}, \citenamefont {Togan},
  \citenamefont {Faelt},\ and\ \citenamefont
  {Imamo{\u{g}}lu}}]{16_Generation_Imamoglu}%
  \BibitemOpen
  \bibfield  {author} {\bibinfo {author} {\bibfnamefont {A.}~\bibnamefont
  {Delteil}}, \bibinfo {author} {\bibfnamefont {Z.}~\bibnamefont {Sun}},
  \bibinfo {author} {\bibfnamefont {W.}~\bibnamefont {Gao}}, \bibinfo {author}
  {\bibfnamefont {E.}~\bibnamefont {Togan}}, \bibinfo {author} {\bibfnamefont
  {S.}~\bibnamefont {Faelt}}, \ and\ \bibinfo {author} {\bibfnamefont
  {A.}~\bibnamefont {Imamo{\u{g}}lu}},\ }\href {\doibase 10.1038/nphys3605}
  {\bibfield  {journal} {\bibinfo  {journal} {Nature Physics}\ }\textbf
  {\bibinfo {volume} {12}},\ \bibinfo {pages} {218} (\bibinfo {year}
  {2015})}\BibitemShut {NoStop}%
\bibitem [{\citenamefont {Barrett}\ and\ \citenamefont
  {Kok}(2005)}]{05_Efficient_Kok}%
  \BibitemOpen
  \bibfield  {author} {\bibinfo {author} {\bibfnamefont {S.~D.}\ \bibnamefont
  {Barrett}}\ and\ \bibinfo {author} {\bibfnamefont {P.}~\bibnamefont {Kok}},\
  }\href@noop {} {\bibfield  {journal} {\bibinfo  {journal} {Physical Review
  A}\ }\textbf {\bibinfo {volume} {71}},\ \bibinfo {pages} {060310} (\bibinfo
  {year} {2005})}\BibitemShut {NoStop}%
\bibitem [{\citenamefont {Berry}(2009)}]{09_Transitionless_Berry}%
  \BibitemOpen
  \bibfield  {author} {\bibinfo {author} {\bibfnamefont {M.}~\bibnamefont
  {Berry}},\ }\href@noop {} {\bibfield  {journal} {\bibinfo  {journal} {Journal
  of Physics A: Mathematical and Theoretical}\ }\textbf {\bibinfo {volume}
  {42}},\ \bibinfo {pages} {365303} (\bibinfo {year} {2009})}\BibitemShut
  {NoStop}%
\bibitem [{\citenamefont {Bernien}\ \emph {et~al.}(2013)\citenamefont
  {Bernien}, \citenamefont {Hensen}, \citenamefont {Pfaff}, \citenamefont
  {Koolstra}, \citenamefont {Blok}, \citenamefont {Robledo}, \citenamefont
  {Taminiau}, \citenamefont {Markham}, \citenamefont {Twitchen}, \citenamefont
  {Childress} \emph {et~al.}}]{13_Heralded_Hanson}%
  \BibitemOpen
  \bibfield  {author} {\bibinfo {author} {\bibfnamefont {H.}~\bibnamefont
  {Bernien}}, \bibinfo {author} {\bibfnamefont {B.}~\bibnamefont {Hensen}},
  \bibinfo {author} {\bibfnamefont {W.}~\bibnamefont {Pfaff}}, \bibinfo
  {author} {\bibfnamefont {G.}~\bibnamefont {Koolstra}}, \bibinfo {author}
  {\bibfnamefont {M.}~\bibnamefont {Blok}}, \bibinfo {author} {\bibfnamefont
  {L.}~\bibnamefont {Robledo}}, \bibinfo {author} {\bibfnamefont
  {T.}~\bibnamefont {Taminiau}}, \bibinfo {author} {\bibfnamefont
  {M.}~\bibnamefont {Markham}}, \bibinfo {author} {\bibfnamefont
  {D.}~\bibnamefont {Twitchen}}, \bibinfo {author} {\bibfnamefont
  {L.}~\bibnamefont {Childress}},  \emph {et~al.},\ }\href@noop {} {\bibfield
  {journal} {\bibinfo  {journal} {Nature}\ }\textbf {\bibinfo {volume} {497}},\
  \bibinfo {pages} {86} (\bibinfo {year} {2013})}\BibitemShut {NoStop}%
\bibitem [{\citenamefont {Narla}\ \emph {et~al.}(2016)\citenamefont {Narla},
  \citenamefont {Shankar}, \citenamefont {Hatridge}, \citenamefont {Leghtas},
  \citenamefont {Sliwa}, \citenamefont {Zalys-Geller}, \citenamefont
  {Mundhada}, \citenamefont {Pfaff}, \citenamefont {Frunzio}, \citenamefont
  {Schoelkopf},\ and\ \citenamefont {Devoret}}]{16_Robust_Devoret}%
  \BibitemOpen
  \bibfield  {author} {\bibinfo {author} {\bibfnamefont {A.}~\bibnamefont
  {Narla}}, \bibinfo {author} {\bibfnamefont {S.}~\bibnamefont {Shankar}},
  \bibinfo {author} {\bibfnamefont {M.}~\bibnamefont {Hatridge}}, \bibinfo
  {author} {\bibfnamefont {Z.}~\bibnamefont {Leghtas}}, \bibinfo {author}
  {\bibfnamefont {K.~M.}\ \bibnamefont {Sliwa}}, \bibinfo {author}
  {\bibfnamefont {E.}~\bibnamefont {Zalys-Geller}}, \bibinfo {author}
  {\bibfnamefont {S.~O.}\ \bibnamefont {Mundhada}}, \bibinfo {author}
  {\bibfnamefont {W.}~\bibnamefont {Pfaff}}, \bibinfo {author} {\bibfnamefont
  {L.}~\bibnamefont {Frunzio}}, \bibinfo {author} {\bibfnamefont {R.~J.}\
  \bibnamefont {Schoelkopf}}, \ and\ \bibinfo {author} {\bibfnamefont {M.~H.}\
  \bibnamefont {Devoret}},\ }\href {\doibase 10.1103/PhysRevX.6.031036}
  {\bibfield  {journal} {\bibinfo  {journal} {Physical Review X}\ }\textbf
  {\bibinfo {volume} {6}},\ \bibinfo {pages} {031036} (\bibinfo {year}
  {2016})}\BibitemShut {NoStop}%
\bibitem [{\citenamefont {Dutt}\ \emph {et~al.}(2007)\citenamefont {Dutt},
  \citenamefont {Childress}, \citenamefont {Jiang}, \citenamefont {Togan},
  \citenamefont {Maze}, \citenamefont {Jelezko}, \citenamefont {Zibrov},
  \citenamefont {Hemmer},\ and\ \citenamefont
  {Lukin}}]{07_Quantum_register_Lukin}%
  \BibitemOpen
  \bibfield  {author} {\bibinfo {author} {\bibfnamefont {M.~G.}\ \bibnamefont
  {Dutt}}, \bibinfo {author} {\bibfnamefont {L.}~\bibnamefont {Childress}},
  \bibinfo {author} {\bibfnamefont {L.}~\bibnamefont {Jiang}}, \bibinfo
  {author} {\bibfnamefont {E.}~\bibnamefont {Togan}}, \bibinfo {author}
  {\bibfnamefont {J.}~\bibnamefont {Maze}}, \bibinfo {author} {\bibfnamefont
  {F.}~\bibnamefont {Jelezko}}, \bibinfo {author} {\bibfnamefont
  {A.}~\bibnamefont {Zibrov}}, \bibinfo {author} {\bibfnamefont
  {P.}~\bibnamefont {Hemmer}}, \ and\ \bibinfo {author} {\bibfnamefont
  {M.}~\bibnamefont {Lukin}},\ }\href@noop {} {\bibfield  {journal} {\bibinfo
  {journal} {Science}\ }\textbf {\bibinfo {volume} {316}},\ \bibinfo {pages}
  {1312} (\bibinfo {year} {2007})}\BibitemShut {NoStop}%
\bibitem [{\citenamefont {Reiserer}\ \emph {et~al.}(2016)\citenamefont
  {Reiserer}, \citenamefont {Kalb}, \citenamefont {Blok}, \citenamefont {van
  Bemmelen}, \citenamefont {Taminiau}, \citenamefont {Hanson}, \citenamefont
  {Twitchen},\ and\ \citenamefont {Markham}}]{16_Robust_Markham}%
  \BibitemOpen
  \bibfield  {author} {\bibinfo {author} {\bibfnamefont {A.}~\bibnamefont
  {Reiserer}}, \bibinfo {author} {\bibfnamefont {N.}~\bibnamefont {Kalb}},
  \bibinfo {author} {\bibfnamefont {M.~S.}\ \bibnamefont {Blok}}, \bibinfo
  {author} {\bibfnamefont {K.~J.~M.}\ \bibnamefont {van Bemmelen}}, \bibinfo
  {author} {\bibfnamefont {T.~H.}\ \bibnamefont {Taminiau}}, \bibinfo {author}
  {\bibfnamefont {R.}~\bibnamefont {Hanson}}, \bibinfo {author} {\bibfnamefont
  {D.~J.}\ \bibnamefont {Twitchen}}, \ and\ \bibinfo {author} {\bibfnamefont
  {M.}~\bibnamefont {Markham}},\ }\href {\doibase 10.1103/PhysRevX.6.021040}
  {\bibfield  {journal} {\bibinfo  {journal} {Physical Review X}\ }\textbf
  {\bibinfo {volume} {6}},\ \bibinfo {pages} {021040} (\bibinfo {year}
  {2016})}\BibitemShut {NoStop}%
\bibitem [{\citenamefont {Bennett}\ \emph {et~al.}(1993)\citenamefont
  {Bennett}, \citenamefont {Brassard}, \citenamefont {Cr{\'e}peau},
  \citenamefont {Jozsa}, \citenamefont {Peres},\ and\ \citenamefont
  {Wootters}}]{93_Teleporting_Wootters}%
  \BibitemOpen
  \bibfield  {author} {\bibinfo {author} {\bibfnamefont {C.~H.}\ \bibnamefont
  {Bennett}}, \bibinfo {author} {\bibfnamefont {G.}~\bibnamefont {Brassard}},
  \bibinfo {author} {\bibfnamefont {C.}~\bibnamefont {Cr{\'e}peau}}, \bibinfo
  {author} {\bibfnamefont {R.}~\bibnamefont {Jozsa}}, \bibinfo {author}
  {\bibfnamefont {A.}~\bibnamefont {Peres}}, \ and\ \bibinfo {author}
  {\bibfnamefont {W.~K.}\ \bibnamefont {Wootters}},\ }\href@noop {} {\bibfield
  {journal} {\bibinfo  {journal} {Physical Review Letters}\ }\textbf {\bibinfo
  {volume} {70}},\ \bibinfo {pages} {1895} (\bibinfo {year}
  {1993})}\BibitemShut {NoStop}%
\bibitem [{\citenamefont {Li}\ \emph {et~al.}(2017)\citenamefont {Li},
  \citenamefont {Zou}, \citenamefont {Albert}, \citenamefont {Muralidharan},
  \citenamefont {Girvin},\ and\ \citenamefont {Jiang}}]{17_Cat_Jiang}%
  \BibitemOpen
  \bibfield  {author} {\bibinfo {author} {\bibfnamefont {L.}~\bibnamefont
  {Li}}, \bibinfo {author} {\bibfnamefont {C.-L.}\ \bibnamefont {Zou}},
  \bibinfo {author} {\bibfnamefont {V.~V.}\ \bibnamefont {Albert}}, \bibinfo
  {author} {\bibfnamefont {S.}~\bibnamefont {Muralidharan}}, \bibinfo {author}
  {\bibfnamefont {S.}~\bibnamefont {Girvin}}, \ and\ \bibinfo {author}
  {\bibfnamefont {L.}~\bibnamefont {Jiang}},\ }\href@noop {} {\bibfield
  {journal} {\bibinfo  {journal} {Physical Review Letters}\ }\textbf {\bibinfo
  {volume} {119}},\ \bibinfo {pages} {030502} (\bibinfo {year}
  {2017})}\BibitemShut {NoStop}%
\bibitem [{\citenamefont {Roy}\ \emph {et~al.}(2016)\citenamefont {Roy},
  \citenamefont {Stone},\ and\ \citenamefont {Jiang}}]{16_Concurrent_Jiang}%
  \BibitemOpen
  \bibfield  {author} {\bibinfo {author} {\bibfnamefont {A.}~\bibnamefont
  {Roy}}, \bibinfo {author} {\bibfnamefont {A.~D.}\ \bibnamefont {Stone}}, \
  and\ \bibinfo {author} {\bibfnamefont {L.}~\bibnamefont {Jiang}},\
  }\href@noop {} {\bibfield  {journal} {\bibinfo  {journal} {Physical Review
  A}\ }\textbf {\bibinfo {volume} {94}},\ \bibinfo {pages} {032333} (\bibinfo
  {year} {2016})}\BibitemShut {NoStop}%
\bibitem [{\citenamefont {Xiang}\ \emph {et~al.}(2013)\citenamefont {Xiang},
  \citenamefont {Ashhab}, \citenamefont {You},\ and\ \citenamefont
  {Nori}}]{13_Hybrid_Nori}%
  \BibitemOpen
  \bibfield  {author} {\bibinfo {author} {\bibfnamefont {Z.-L.}\ \bibnamefont
  {Xiang}}, \bibinfo {author} {\bibfnamefont {S.}~\bibnamefont {Ashhab}},
  \bibinfo {author} {\bibfnamefont {J.}~\bibnamefont {You}}, \ and\ \bibinfo
  {author} {\bibfnamefont {F.}~\bibnamefont {Nori}},\ }\href@noop {} {\bibfield
   {journal} {\bibinfo  {journal} {Reviews of Modern Physics}\ }\textbf
  {\bibinfo {volume} {85}},\ \bibinfo {pages} {623} (\bibinfo {year}
  {2013})}\BibitemShut {NoStop}%
\bibitem [{\citenamefont {Probst}\ \emph {et~al.}(2013)\citenamefont {Probst},
  \citenamefont {Rotzinger}, \citenamefont {W{\"u}nsch}, \citenamefont {Jung},
  \citenamefont {Jerger}, \citenamefont {Siegel}, \citenamefont {Ustinov},\
  and\ \citenamefont {Bushev}}]{13_Anisotropic_Bushev}%
  \BibitemOpen
  \bibfield  {author} {\bibinfo {author} {\bibfnamefont {S.}~\bibnamefont
  {Probst}}, \bibinfo {author} {\bibfnamefont {H.}~\bibnamefont {Rotzinger}},
  \bibinfo {author} {\bibfnamefont {S.}~\bibnamefont {W{\"u}nsch}}, \bibinfo
  {author} {\bibfnamefont {P.}~\bibnamefont {Jung}}, \bibinfo {author}
  {\bibfnamefont {M.}~\bibnamefont {Jerger}}, \bibinfo {author} {\bibfnamefont
  {M.}~\bibnamefont {Siegel}}, \bibinfo {author} {\bibfnamefont
  {A.}~\bibnamefont {Ustinov}}, \ and\ \bibinfo {author} {\bibfnamefont
  {P.}~\bibnamefont {Bushev}},\ }\href@noop {} {\bibfield  {journal} {\bibinfo
  {journal} {Physical Review Letters}\ }\textbf {\bibinfo {volume} {110}},\
  \bibinfo {pages} {157001} (\bibinfo {year} {2013})}\BibitemShut {NoStop}%
\bibitem [{\citenamefont {Blum}\ \emph {et~al.}(2015)\citenamefont {Blum},
  \citenamefont {O'Brien}, \citenamefont {Lauk}, \citenamefont {Bushev},
  \citenamefont {Fleischhauer},\ and\ \citenamefont
  {Morigi}}]{15_Interfacing_Morigi}%
  \BibitemOpen
  \bibfield  {author} {\bibinfo {author} {\bibfnamefont {S.}~\bibnamefont
  {Blum}}, \bibinfo {author} {\bibfnamefont {C.}~\bibnamefont {O'Brien}},
  \bibinfo {author} {\bibfnamefont {N.}~\bibnamefont {Lauk}}, \bibinfo {author}
  {\bibfnamefont {P.}~\bibnamefont {Bushev}}, \bibinfo {author} {\bibfnamefont
  {M.}~\bibnamefont {Fleischhauer}}, \ and\ \bibinfo {author} {\bibfnamefont
  {G.}~\bibnamefont {Morigi}},\ }\href@noop {} {\bibfield  {journal} {\bibinfo
  {journal} {Physical Review A}\ }\textbf {\bibinfo {volume} {91}},\ \bibinfo
  {pages} {033834} (\bibinfo {year} {2015})}\BibitemShut {NoStop}%
\bibitem [{\citenamefont {Moiseev}\ and\ \citenamefont
  {Kr{\"o}ll}(2001)}]{01_Complete_Kroll}%
  \BibitemOpen
  \bibfield  {author} {\bibinfo {author} {\bibfnamefont {S.}~\bibnamefont
  {Moiseev}}\ and\ \bibinfo {author} {\bibfnamefont {S.}~\bibnamefont
  {Kr{\"o}ll}},\ }\href@noop {} {\bibfield  {journal} {\bibinfo  {journal}
  {Physical Review Letters}\ }\textbf {\bibinfo {volume} {87}},\ \bibinfo
  {pages} {173601} (\bibinfo {year} {2001})}\BibitemShut {NoStop}%
\bibitem [{\citenamefont {Sangouard}\ \emph {et~al.}(2007)\citenamefont
  {Sangouard}, \citenamefont {Simon}, \citenamefont {Afzelius},\ and\
  \citenamefont {Gisin}}]{07_Analysis_Gisin}%
  \BibitemOpen
  \bibfield  {author} {\bibinfo {author} {\bibfnamefont {N.}~\bibnamefont
  {Sangouard}}, \bibinfo {author} {\bibfnamefont {C.}~\bibnamefont {Simon}},
  \bibinfo {author} {\bibfnamefont {M.}~\bibnamefont {Afzelius}}, \ and\
  \bibinfo {author} {\bibfnamefont {N.}~\bibnamefont {Gisin}},\ }\href@noop {}
  {\bibfield  {journal} {\bibinfo  {journal} {Physical Review A}\ }\textbf
  {\bibinfo {volume} {75}},\ \bibinfo {pages} {032327} (\bibinfo {year}
  {2007})}\BibitemShut {NoStop}%
\bibitem [{\citenamefont {Staudt}\ \emph {et~al.}(2007)\citenamefont {Staudt},
  \citenamefont {Afzelius}, \citenamefont {De~Riedmatten}, \citenamefont
  {Hastings-Simon}, \citenamefont {Simon}, \citenamefont {Ricken},
  \citenamefont {Suche}, \citenamefont {Sohler},\ and\ \citenamefont
  {Gisin}}]{07_Interference_Gisin}%
  \BibitemOpen
  \bibfield  {author} {\bibinfo {author} {\bibfnamefont {M.~U.}\ \bibnamefont
  {Staudt}}, \bibinfo {author} {\bibfnamefont {M.}~\bibnamefont {Afzelius}},
  \bibinfo {author} {\bibfnamefont {H.}~\bibnamefont {De~Riedmatten}}, \bibinfo
  {author} {\bibfnamefont {S.~R.}\ \bibnamefont {Hastings-Simon}}, \bibinfo
  {author} {\bibfnamefont {C.}~\bibnamefont {Simon}}, \bibinfo {author}
  {\bibfnamefont {R.}~\bibnamefont {Ricken}}, \bibinfo {author} {\bibfnamefont
  {H.}~\bibnamefont {Suche}}, \bibinfo {author} {\bibfnamefont
  {W.}~\bibnamefont {Sohler}}, \ and\ \bibinfo {author} {\bibfnamefont
  {N.}~\bibnamefont {Gisin}},\ }\href@noop {} {\bibfield  {journal} {\bibinfo
  {journal} {Physical Review Letters}\ }\textbf {\bibinfo {volume} {99}},\
  \bibinfo {pages} {173602} (\bibinfo {year} {2007})}\BibitemShut {NoStop}%
\bibitem [{\citenamefont {G{\"u}ndo{\u{g}}an}\ \emph
  {et~al.}(2015)\citenamefont {G{\"u}ndo{\u{g}}an}, \citenamefont {Ledingham},
  \citenamefont {Kutluer}, \citenamefont {Mazzera},\ and\ \citenamefont
  {De~Riedmatten}}]{15_Solid_Riedmatten}%
  \BibitemOpen
  \bibfield  {author} {\bibinfo {author} {\bibfnamefont {M.}~\bibnamefont
  {G{\"u}ndo{\u{g}}an}}, \bibinfo {author} {\bibfnamefont {P.~M.}\ \bibnamefont
  {Ledingham}}, \bibinfo {author} {\bibfnamefont {K.}~\bibnamefont {Kutluer}},
  \bibinfo {author} {\bibfnamefont {M.}~\bibnamefont {Mazzera}}, \ and\
  \bibinfo {author} {\bibfnamefont {H.}~\bibnamefont {De~Riedmatten}},\
  }\href@noop {} {\bibfield  {journal} {\bibinfo  {journal} {Physical Review
  Letters}\ }\textbf {\bibinfo {volume} {114}},\ \bibinfo {pages} {230501}
  (\bibinfo {year} {2015})}\BibitemShut {NoStop}%
\bibitem [{\citenamefont {Beavan}\ \emph {et~al.}(2013)\citenamefont {Beavan},
  \citenamefont {Goldschmidt},\ and\ \citenamefont
  {Sellars}}]{13_Demonstration_Sellars}%
  \BibitemOpen
  \bibfield  {author} {\bibinfo {author} {\bibfnamefont {S.~E.}\ \bibnamefont
  {Beavan}}, \bibinfo {author} {\bibfnamefont {E.~A.}\ \bibnamefont
  {Goldschmidt}}, \ and\ \bibinfo {author} {\bibfnamefont {M.~J.}\ \bibnamefont
  {Sellars}},\ }\href@noop {} {\bibfield  {journal} {\bibinfo  {journal} {JOSA
  B}\ }\textbf {\bibinfo {volume} {30}},\ \bibinfo {pages} {1173} (\bibinfo
  {year} {2013})}\BibitemShut {NoStop}%
\bibitem [{\citenamefont {Simon}\ \emph {et~al.}(2010)\citenamefont {Simon},
  \citenamefont {Afzelius}, \citenamefont {Appel}, \citenamefont
  {de~La~Giroday}, \citenamefont {Dewhurst}, \citenamefont {Gisin},
  \citenamefont {Hu}, \citenamefont {Jelezko}, \citenamefont {Kr{\"o}ll},
  \citenamefont {M{\"u}ller} \emph {et~al.}}]{10_Quantum_Young}%
  \BibitemOpen
  \bibfield  {author} {\bibinfo {author} {\bibfnamefont {C.}~\bibnamefont
  {Simon}}, \bibinfo {author} {\bibfnamefont {M.}~\bibnamefont {Afzelius}},
  \bibinfo {author} {\bibfnamefont {J.}~\bibnamefont {Appel}}, \bibinfo
  {author} {\bibfnamefont {A.~B.}\ \bibnamefont {de~La~Giroday}}, \bibinfo
  {author} {\bibfnamefont {S.}~\bibnamefont {Dewhurst}}, \bibinfo {author}
  {\bibfnamefont {N.}~\bibnamefont {Gisin}}, \bibinfo {author} {\bibfnamefont
  {C.}~\bibnamefont {Hu}}, \bibinfo {author} {\bibfnamefont {F.}~\bibnamefont
  {Jelezko}}, \bibinfo {author} {\bibfnamefont {S.}~\bibnamefont {Kr{\"o}ll}},
  \bibinfo {author} {\bibfnamefont {J.}~\bibnamefont {M{\"u}ller}},  \emph
  {et~al.},\ }\href@noop {} {\bibfield  {journal} {\bibinfo  {journal} {The
  European Physical Journal D}\ }\textbf {\bibinfo {volume} {58}},\ \bibinfo
  {pages} {1} (\bibinfo {year} {2010})}\BibitemShut {NoStop}%
\bibitem [{\citenamefont {Lvovsky}\ \emph {et~al.}(2009)\citenamefont
  {Lvovsky}, \citenamefont {Sanders},\ and\ \citenamefont
  {Tittel}}]{09_Optical_Tittel}%
  \BibitemOpen
  \bibfield  {author} {\bibinfo {author} {\bibfnamefont {A.~I.}\ \bibnamefont
  {Lvovsky}}, \bibinfo {author} {\bibfnamefont {B.~C.}\ \bibnamefont
  {Sanders}}, \ and\ \bibinfo {author} {\bibfnamefont {W.}~\bibnamefont
  {Tittel}},\ }\href@noop {} {\bibfield  {journal} {\bibinfo  {journal} {Nature
  Photonics}\ }\textbf {\bibinfo {volume} {3}},\ \bibinfo {pages} {706}
  (\bibinfo {year} {2009})}\BibitemShut {NoStop}%
\bibitem [{\citenamefont {Bao}\ \emph {et~al.}(2012)\citenamefont {Bao},
  \citenamefont {Reingruber}, \citenamefont {Dietrich}, \citenamefont {Rui},
  \citenamefont {D{\"u}ck}, \citenamefont {Strassel}, \citenamefont {Li},
  \citenamefont {Liu}, \citenamefont {Zhao},\ and\ \citenamefont
  {Pan}}]{12_Efficient_Pan}%
  \BibitemOpen
  \bibfield  {author} {\bibinfo {author} {\bibfnamefont {X.-H.}\ \bibnamefont
  {Bao}}, \bibinfo {author} {\bibfnamefont {A.}~\bibnamefont {Reingruber}},
  \bibinfo {author} {\bibfnamefont {P.}~\bibnamefont {Dietrich}}, \bibinfo
  {author} {\bibfnamefont {J.}~\bibnamefont {Rui}}, \bibinfo {author}
  {\bibfnamefont {A.}~\bibnamefont {D{\"u}ck}}, \bibinfo {author}
  {\bibfnamefont {T.}~\bibnamefont {Strassel}}, \bibinfo {author}
  {\bibfnamefont {L.}~\bibnamefont {Li}}, \bibinfo {author} {\bibfnamefont
  {N.-L.}\ \bibnamefont {Liu}}, \bibinfo {author} {\bibfnamefont
  {B.}~\bibnamefont {Zhao}}, \ and\ \bibinfo {author} {\bibfnamefont {J.-W.}\
  \bibnamefont {Pan}},\ }\href@noop {} {\bibfield  {journal} {\bibinfo
  {journal} {Nature Physics}\ }\textbf {\bibinfo {volume} {8}},\ \bibinfo
  {pages} {517} (\bibinfo {year} {2012})}\BibitemShut {NoStop}%
\bibitem [{\citenamefont {Yang}\ \emph {et~al.}(2016)\citenamefont {Yang},
  \citenamefont {Wang}, \citenamefont {Bao},\ and\ \citenamefont
  {Pan}}]{16_Efficient_Pan}%
  \BibitemOpen
  \bibfield  {author} {\bibinfo {author} {\bibfnamefont {S.-J.}\ \bibnamefont
  {Yang}}, \bibinfo {author} {\bibfnamefont {X.-J.}\ \bibnamefont {Wang}},
  \bibinfo {author} {\bibfnamefont {X.-H.}\ \bibnamefont {Bao}}, \ and\
  \bibinfo {author} {\bibfnamefont {J.-W.}\ \bibnamefont {Pan}},\ }\href@noop
  {} {\bibfield  {journal} {\bibinfo  {journal} {Nature Photonics}\ }\textbf
  {\bibinfo {volume} {10}},\ \bibinfo {pages} {381} (\bibinfo {year}
  {2016})}\BibitemShut {NoStop}%
\bibitem [{\citenamefont {Collins}\ \emph {et~al.}(2007)\citenamefont
  {Collins}, \citenamefont {Jenkins}, \citenamefont {Kuzmich},\ and\
  \citenamefont {Kennedy}}]{07_Multiplexed_Kennedy}%
  \BibitemOpen
  \bibfield  {author} {\bibinfo {author} {\bibfnamefont {O.}~\bibnamefont
  {Collins}}, \bibinfo {author} {\bibfnamefont {S.}~\bibnamefont {Jenkins}},
  \bibinfo {author} {\bibfnamefont {A.}~\bibnamefont {Kuzmich}}, \ and\
  \bibinfo {author} {\bibfnamefont {T.}~\bibnamefont {Kennedy}},\ }\href@noop
  {} {\bibfield  {journal} {\bibinfo  {journal} {Physical Review Letters}\
  }\textbf {\bibinfo {volume} {98}},\ \bibinfo {pages} {060502} (\bibinfo
  {year} {2007})}\BibitemShut {NoStop}%
\bibitem [{\citenamefont {Sinclair}\ \emph {et~al.}(2014)\citenamefont
  {Sinclair}, \citenamefont {Saglamyurek}, \citenamefont {Mallahzadeh},
  \citenamefont {Slater}, \citenamefont {George}, \citenamefont {Ricken},
  \citenamefont {Hedges}, \citenamefont {Oblak}, \citenamefont {Simon},
  \citenamefont {Sohler} \emph {et~al.}}]{14_Spectral_Tittel}%
  \BibitemOpen
  \bibfield  {author} {\bibinfo {author} {\bibfnamefont {N.}~\bibnamefont
  {Sinclair}}, \bibinfo {author} {\bibfnamefont {E.}~\bibnamefont
  {Saglamyurek}}, \bibinfo {author} {\bibfnamefont {H.}~\bibnamefont
  {Mallahzadeh}}, \bibinfo {author} {\bibfnamefont {J.~A.}\ \bibnamefont
  {Slater}}, \bibinfo {author} {\bibfnamefont {M.}~\bibnamefont {George}},
  \bibinfo {author} {\bibfnamefont {R.}~\bibnamefont {Ricken}}, \bibinfo
  {author} {\bibfnamefont {M.~P.}\ \bibnamefont {Hedges}}, \bibinfo {author}
  {\bibfnamefont {D.}~\bibnamefont {Oblak}}, \bibinfo {author} {\bibfnamefont
  {C.}~\bibnamefont {Simon}}, \bibinfo {author} {\bibfnamefont
  {W.}~\bibnamefont {Sohler}},  \emph {et~al.},\ }\href@noop {} {\bibfield
  {journal} {\bibinfo  {journal} {Physical Review Letters}\ }\textbf {\bibinfo
  {volume} {113}},\ \bibinfo {pages} {053603} (\bibinfo {year}
  {2014})}\BibitemShut {NoStop}%
\bibitem [{\citenamefont {Puigibert}\ \emph {et~al.}(2017)\citenamefont
  {Puigibert}, \citenamefont {Aguilar}, \citenamefont {Zhou}, \citenamefont
  {Marsili}, \citenamefont {Shaw}, \citenamefont {Verma}, \citenamefont {Nam},
  \citenamefont {Oblak},\ and\ \citenamefont {Tittel}}]{17_Heralded_Tittel}%
  \BibitemOpen
  \bibfield  {author} {\bibinfo {author} {\bibfnamefont {M.~G.}\ \bibnamefont
  {Puigibert}}, \bibinfo {author} {\bibfnamefont {G.}~\bibnamefont {Aguilar}},
  \bibinfo {author} {\bibfnamefont {Q.}~\bibnamefont {Zhou}}, \bibinfo {author}
  {\bibfnamefont {F.}~\bibnamefont {Marsili}}, \bibinfo {author} {\bibfnamefont
  {M.}~\bibnamefont {Shaw}}, \bibinfo {author} {\bibfnamefont {V.}~\bibnamefont
  {Verma}}, \bibinfo {author} {\bibfnamefont {S.}~\bibnamefont {Nam}}, \bibinfo
  {author} {\bibfnamefont {D.}~\bibnamefont {Oblak}}, \ and\ \bibinfo {author}
  {\bibfnamefont {W.}~\bibnamefont {Tittel}},\ }\href@noop {} {\bibfield
  {journal} {\bibinfo  {journal} {Physical Review Letters}\ }\textbf {\bibinfo
  {volume} {119}},\ \bibinfo {pages} {083601} (\bibinfo {year}
  {2017})}\BibitemShut {NoStop}%
\bibitem [{\citenamefont {Yang}\ \emph {et~al.}(2018)\citenamefont {Yang},
  \citenamefont {Oh}, \citenamefont {Lee}, \citenamefont {Yang}, \citenamefont
  {Yi}, \citenamefont {Shen}, \citenamefont {Wang},\ and\ \citenamefont
  {Vahala}}]{18_Bridging_Vahala}%
  \BibitemOpen
  \bibfield  {author} {\bibinfo {author} {\bibfnamefont {K.~Y.}\ \bibnamefont
  {Yang}}, \bibinfo {author} {\bibfnamefont {D.~Y.}\ \bibnamefont {Oh}},
  \bibinfo {author} {\bibfnamefont {S.~H.}\ \bibnamefont {Lee}}, \bibinfo
  {author} {\bibfnamefont {Q.-F.}\ \bibnamefont {Yang}}, \bibinfo {author}
  {\bibfnamefont {X.}~\bibnamefont {Yi}}, \bibinfo {author} {\bibfnamefont
  {B.}~\bibnamefont {Shen}}, \bibinfo {author} {\bibfnamefont {H.}~\bibnamefont
  {Wang}}, \ and\ \bibinfo {author} {\bibfnamefont {K.}~\bibnamefont
  {Vahala}},\ }\href@noop {} {\bibfield  {journal} {\bibinfo  {journal} {Nature
  Photonics}\ }\textbf {\bibinfo {volume} {12}},\ \bibinfo {pages} {297}
  (\bibinfo {year} {2018})}\BibitemShut {NoStop}%
\bibitem [{\citenamefont {Yamamoto}\ \emph {et~al.}(2008)\citenamefont
  {Yamamoto}, \citenamefont {Inomata}, \citenamefont {Watanabe}, \citenamefont
  {Matsuba}, \citenamefont {Miyazaki}, \citenamefont {Oliver}, \citenamefont
  {Nakamura},\ and\ \citenamefont {Tsai}}]{08_Flux_Tsai}%
  \BibitemOpen
  \bibfield  {author} {\bibinfo {author} {\bibfnamefont {T.}~\bibnamefont
  {Yamamoto}}, \bibinfo {author} {\bibfnamefont {K.}~\bibnamefont {Inomata}},
  \bibinfo {author} {\bibfnamefont {M.}~\bibnamefont {Watanabe}}, \bibinfo
  {author} {\bibfnamefont {K.}~\bibnamefont {Matsuba}}, \bibinfo {author}
  {\bibfnamefont {T.}~\bibnamefont {Miyazaki}}, \bibinfo {author}
  {\bibfnamefont {W.}~\bibnamefont {Oliver}}, \bibinfo {author} {\bibfnamefont
  {Y.}~\bibnamefont {Nakamura}}, \ and\ \bibinfo {author} {\bibfnamefont
  {J.}~\bibnamefont {Tsai}},\ }\href@noop {} {\bibfield  {journal} {\bibinfo
  {journal} {Applied Physics Letters}\ }\textbf {\bibinfo {volume} {93}},\
  \bibinfo {pages} {042510} (\bibinfo {year} {2008})}\BibitemShut {NoStop}%
\bibitem [{\citenamefont {Johansson}\ \emph {et~al.}(2012)\citenamefont
  {Johansson}, \citenamefont {Nation},\ and\ \citenamefont
  {Nori}}]{12_Qutip_Nori}%
  \BibitemOpen
  \bibfield  {author} {\bibinfo {author} {\bibfnamefont {J.}~\bibnamefont
  {Johansson}}, \bibinfo {author} {\bibfnamefont {P.}~\bibnamefont {Nation}}, \
  and\ \bibinfo {author} {\bibfnamefont {F.}~\bibnamefont {Nori}},\ }\href@noop
  {} {\bibfield  {journal} {\bibinfo  {journal} {Computer Physics
  Communications}\ }\textbf {\bibinfo {volume} {183}},\ \bibinfo {pages} {1760}
  (\bibinfo {year} {2012})}\BibitemShut {NoStop}%
\bibitem [{\citenamefont {Bourassa}\ \emph {et~al.}(2012)\citenamefont
  {Bourassa}, \citenamefont {Beaudoin}, \citenamefont {Gambetta},\ and\
  \citenamefont {Blais}}]{12_Josephson_Blais}%
  \BibitemOpen
  \bibfield  {author} {\bibinfo {author} {\bibfnamefont {J.}~\bibnamefont
  {Bourassa}}, \bibinfo {author} {\bibfnamefont {F.}~\bibnamefont {Beaudoin}},
  \bibinfo {author} {\bibfnamefont {J.~M.}\ \bibnamefont {Gambetta}}, \ and\
  \bibinfo {author} {\bibfnamefont {A.}~\bibnamefont {Blais}},\ }\href@noop {}
  {\bibfield  {journal} {\bibinfo  {journal} {Physical Review A}\ }\textbf
  {\bibinfo {volume} {86}},\ \bibinfo {pages} {013814} (\bibinfo {year}
  {2012})}\BibitemShut {NoStop}%
\bibitem [{\citenamefont {Khaneja}\ \emph {et~al.}(2005)\citenamefont
  {Khaneja}, \citenamefont {Reiss}, \citenamefont {Kehlet}, \citenamefont
  {Schulte-Herbr{\"u}ggen},\ and\ \citenamefont {Glaser}}]{05_Optimal_Glaser}%
  \BibitemOpen
  \bibfield  {author} {\bibinfo {author} {\bibfnamefont {N.}~\bibnamefont
  {Khaneja}}, \bibinfo {author} {\bibfnamefont {T.}~\bibnamefont {Reiss}},
  \bibinfo {author} {\bibfnamefont {C.}~\bibnamefont {Kehlet}}, \bibinfo
  {author} {\bibfnamefont {T.}~\bibnamefont {Schulte-Herbr{\"u}ggen}}, \ and\
  \bibinfo {author} {\bibfnamefont {S.~J.}\ \bibnamefont {Glaser}},\
  }\href@noop {} {\bibfield  {journal} {\bibinfo  {journal} {Journal of
  Magnetic Resonance}\ }\textbf {\bibinfo {volume} {172}},\ \bibinfo {pages}
  {296} (\bibinfo {year} {2005})}\BibitemShut {NoStop}%
\bibitem [{\citenamefont {Machnes}\ \emph {et~al.}(2011)\citenamefont
  {Machnes}, \citenamefont {Sander}, \citenamefont {Glaser}, \citenamefont
  {de~Fouquieres}, \citenamefont {Gruslys}, \citenamefont {Schirmer},\ and\
  \citenamefont {Schulte-Herbr{\"u}ggen}}]{11_Comparing_Herbruggen}%
  \BibitemOpen
  \bibfield  {author} {\bibinfo {author} {\bibfnamefont {S.}~\bibnamefont
  {Machnes}}, \bibinfo {author} {\bibfnamefont {U.}~\bibnamefont {Sander}},
  \bibinfo {author} {\bibfnamefont {S.~J.}\ \bibnamefont {Glaser}}, \bibinfo
  {author} {\bibfnamefont {P.}~\bibnamefont {de~Fouquieres}}, \bibinfo {author}
  {\bibfnamefont {A.}~\bibnamefont {Gruslys}}, \bibinfo {author} {\bibfnamefont
  {S.}~\bibnamefont {Schirmer}}, \ and\ \bibinfo {author} {\bibfnamefont
  {T.}~\bibnamefont {Schulte-Herbr{\"u}ggen}},\ }\href@noop {} {\bibfield
  {journal} {\bibinfo  {journal} {Physical Review A}\ }\textbf {\bibinfo
  {volume} {84}},\ \bibinfo {pages} {022305} (\bibinfo {year}
  {2011})}\BibitemShut {NoStop}%
\bibitem [{\citenamefont {Kirchmair}\ \emph {et~al.}(2013)\citenamefont
  {Kirchmair}, \citenamefont {Vlastakis}, \citenamefont {Leghtas},
  \citenamefont {Nigg}, \citenamefont {Paik}, \citenamefont {Ginossar},
  \citenamefont {Mirrahimi}, \citenamefont {Frunzio}, \citenamefont {Girvin},\
  and\ \citenamefont {Schoelkopf}}]{13_Observation_Schoelkopf}%
  \BibitemOpen
  \bibfield  {author} {\bibinfo {author} {\bibfnamefont {G.}~\bibnamefont
  {Kirchmair}}, \bibinfo {author} {\bibfnamefont {B.}~\bibnamefont
  {Vlastakis}}, \bibinfo {author} {\bibfnamefont {Z.}~\bibnamefont {Leghtas}},
  \bibinfo {author} {\bibfnamefont {S.~E.}\ \bibnamefont {Nigg}}, \bibinfo
  {author} {\bibfnamefont {H.}~\bibnamefont {Paik}}, \bibinfo {author}
  {\bibfnamefont {E.}~\bibnamefont {Ginossar}}, \bibinfo {author}
  {\bibfnamefont {M.}~\bibnamefont {Mirrahimi}}, \bibinfo {author}
  {\bibfnamefont {L.}~\bibnamefont {Frunzio}}, \bibinfo {author} {\bibfnamefont
  {S.~M.}\ \bibnamefont {Girvin}}, \ and\ \bibinfo {author} {\bibfnamefont
  {R.~J.}\ \bibnamefont {Schoelkopf}},\ }\href@noop {} {\bibfield  {journal}
  {\bibinfo  {journal} {Nature}\ }\textbf {\bibinfo {volume} {495}},\ \bibinfo
  {pages} {205} (\bibinfo {year} {2013})}\BibitemShut {NoStop}%
\bibitem [{\citenamefont {Houck}\ \emph {et~al.}(2007)\citenamefont {Houck},
  \citenamefont {Schuster}, \citenamefont {Gambetta}, \citenamefont {Schreier},
  \citenamefont {Johnson}, \citenamefont {Chow}, \citenamefont {Frunzio},
  \citenamefont {Majer}, \citenamefont {Devoret}, \citenamefont {Girvin} \emph
  {et~al.}}]{07_Generating_Schoelkopf}%
  \BibitemOpen
  \bibfield  {author} {\bibinfo {author} {\bibfnamefont {A.}~\bibnamefont
  {Houck}}, \bibinfo {author} {\bibfnamefont {D.}~\bibnamefont {Schuster}},
  \bibinfo {author} {\bibfnamefont {J.}~\bibnamefont {Gambetta}}, \bibinfo
  {author} {\bibfnamefont {J.}~\bibnamefont {Schreier}}, \bibinfo {author}
  {\bibfnamefont {B.}~\bibnamefont {Johnson}}, \bibinfo {author} {\bibfnamefont
  {J.}~\bibnamefont {Chow}}, \bibinfo {author} {\bibfnamefont {L.}~\bibnamefont
  {Frunzio}}, \bibinfo {author} {\bibfnamefont {J.}~\bibnamefont {Majer}},
  \bibinfo {author} {\bibfnamefont {M.}~\bibnamefont {Devoret}}, \bibinfo
  {author} {\bibfnamefont {S.}~\bibnamefont {Girvin}},  \emph {et~al.},\
  }\href@noop {} {\bibfield  {journal} {\bibinfo  {journal} {Nature}\ }\textbf
  {\bibinfo {volume} {449}},\ \bibinfo {pages} {328} (\bibinfo {year}
  {2007})}\BibitemShut {NoStop}%
\bibitem [{\citenamefont {Reagor}\ \emph {et~al.}(2016)\citenamefont {Reagor},
  \citenamefont {Pfaff}, \citenamefont {Axline}, \citenamefont {Heeres},
  \citenamefont {Ofek}, \citenamefont {Sliwa}, \citenamefont {Holland},
  \citenamefont {Wang}, \citenamefont {Blumoff}, \citenamefont {Chou} \emph
  {et~al.}}]{16_Quantum_Schoelkopf}%
  \BibitemOpen
  \bibfield  {author} {\bibinfo {author} {\bibfnamefont {M.}~\bibnamefont
  {Reagor}}, \bibinfo {author} {\bibfnamefont {W.}~\bibnamefont {Pfaff}},
  \bibinfo {author} {\bibfnamefont {C.}~\bibnamefont {Axline}}, \bibinfo
  {author} {\bibfnamefont {R.~W.}\ \bibnamefont {Heeres}}, \bibinfo {author}
  {\bibfnamefont {N.}~\bibnamefont {Ofek}}, \bibinfo {author} {\bibfnamefont
  {K.}~\bibnamefont {Sliwa}}, \bibinfo {author} {\bibfnamefont
  {E.}~\bibnamefont {Holland}}, \bibinfo {author} {\bibfnamefont
  {C.}~\bibnamefont {Wang}}, \bibinfo {author} {\bibfnamefont {J.}~\bibnamefont
  {Blumoff}}, \bibinfo {author} {\bibfnamefont {K.}~\bibnamefont {Chou}},
  \emph {et~al.},\ }\href@noop {} {\bibfield  {journal} {\bibinfo  {journal}
  {Physical Review B}\ }\textbf {\bibinfo {volume} {94}},\ \bibinfo {pages}
  {014506} (\bibinfo {year} {2016})}\BibitemShut {NoStop}%
\bibitem [{\citenamefont {Boissonneault}\ \emph {et~al.}(2009)\citenamefont
  {Boissonneault}, \citenamefont {Gambetta},\ and\ \citenamefont
  {Blais}}]{09_Dispersive_Blais}%
  \BibitemOpen
  \bibfield  {author} {\bibinfo {author} {\bibfnamefont {M.}~\bibnamefont
  {Boissonneault}}, \bibinfo {author} {\bibfnamefont {J.~M.}\ \bibnamefont
  {Gambetta}}, \ and\ \bibinfo {author} {\bibfnamefont {A.}~\bibnamefont
  {Blais}},\ }\href@noop {} {\bibfield  {journal} {\bibinfo  {journal}
  {Physical Review A}\ }\textbf {\bibinfo {volume} {79}},\ \bibinfo {pages}
  {013819} (\bibinfo {year} {2009})}\BibitemShut {NoStop}%
\bibitem [{\citenamefont {Kjaergaard}\ \emph {et~al.}(2019)\citenamefont
  {Kjaergaard}, \citenamefont {Schwartz}, \citenamefont {Braum{\"u}ller},
  \citenamefont {Krantz}, \citenamefont {Wang}, \citenamefont {Gustavsson},\
  and\ \citenamefont {Oliver}}]{19_SQs_Oliver}%
  \BibitemOpen
  \bibfield  {author} {\bibinfo {author} {\bibfnamefont {M.}~\bibnamefont
  {Kjaergaard}}, \bibinfo {author} {\bibfnamefont {M.~E.}\ \bibnamefont
  {Schwartz}}, \bibinfo {author} {\bibfnamefont {J.}~\bibnamefont
  {Braum{\"u}ller}}, \bibinfo {author} {\bibfnamefont {P.}~\bibnamefont
  {Krantz}}, \bibinfo {author} {\bibfnamefont {J.~I.-J.}\ \bibnamefont {Wang}},
  \bibinfo {author} {\bibfnamefont {S.}~\bibnamefont {Gustavsson}}, \ and\
  \bibinfo {author} {\bibfnamefont {W.~D.}\ \bibnamefont {Oliver}},\
  }\href@noop {} {\bibfield  {journal} {\bibinfo  {journal} {arXiv:1905.13641}\
  } (\bibinfo {year} {2019})}\BibitemShut {NoStop}%
\bibitem [{\citenamefont {Devoret}\ \emph {et~al.}(2007)\citenamefont
  {Devoret}, \citenamefont {Girvin},\ and\ \citenamefont
  {Schoelkopf}}]{07_cQED_Schoelkopf}%
  \BibitemOpen
  \bibfield  {author} {\bibinfo {author} {\bibfnamefont {M.}~\bibnamefont
  {Devoret}}, \bibinfo {author} {\bibfnamefont {S.}~\bibnamefont {Girvin}}, \
  and\ \bibinfo {author} {\bibfnamefont {R.}~\bibnamefont {Schoelkopf}},\
  }\href@noop {} {\bibfield  {journal} {\bibinfo  {journal} {Annalen der
  Physik}\ }\textbf {\bibinfo {volume} {16}},\ \bibinfo {pages} {767} (\bibinfo
  {year} {2007})}\BibitemShut {NoStop}%
\bibitem [{\citenamefont {Nigg}\ \emph {et~al.}(2012)\citenamefont {Nigg},
  \citenamefont {Paik}, \citenamefont {Vlastakis}, \citenamefont {Kirchmair},
  \citenamefont {Shankar}, \citenamefont {Frunzio}, \citenamefont {Devoret},
  \citenamefont {Schoelkopf},\ and\ \citenamefont {Girvin}}]{12_Black_Girvin}%
  \BibitemOpen
  \bibfield  {author} {\bibinfo {author} {\bibfnamefont {S.~E.}\ \bibnamefont
  {Nigg}}, \bibinfo {author} {\bibfnamefont {H.}~\bibnamefont {Paik}}, \bibinfo
  {author} {\bibfnamefont {B.}~\bibnamefont {Vlastakis}}, \bibinfo {author}
  {\bibfnamefont {G.}~\bibnamefont {Kirchmair}}, \bibinfo {author}
  {\bibfnamefont {S.}~\bibnamefont {Shankar}}, \bibinfo {author} {\bibfnamefont
  {L.}~\bibnamefont {Frunzio}}, \bibinfo {author} {\bibfnamefont
  {M.}~\bibnamefont {Devoret}}, \bibinfo {author} {\bibfnamefont
  {R.}~\bibnamefont {Schoelkopf}}, \ and\ \bibinfo {author} {\bibfnamefont
  {S.}~\bibnamefont {Girvin}},\ }\href@noop {} {\bibfield  {journal} {\bibinfo
  {journal} {Physical Review Letters}\ }\textbf {\bibinfo {volume} {108}},\
  \bibinfo {pages} {240502} (\bibinfo {year} {2012})}\BibitemShut {NoStop}%
\bibitem [{\citenamefont {Shankar}\ \emph {et~al.}(2013)\citenamefont
  {Shankar}, \citenamefont {Hatridge}, \citenamefont {Leghtas}, \citenamefont
  {Sliwa}, \citenamefont {Narla}, \citenamefont {Vool}, \citenamefont {Girvin},
  \citenamefont {Frunzio}, \citenamefont {Mirrahimi},\ and\ \citenamefont
  {Devoret}}]{13_Autonomously_Devoret}%
  \BibitemOpen
  \bibfield  {author} {\bibinfo {author} {\bibfnamefont {S.}~\bibnamefont
  {Shankar}}, \bibinfo {author} {\bibfnamefont {M.}~\bibnamefont {Hatridge}},
  \bibinfo {author} {\bibfnamefont {Z.}~\bibnamefont {Leghtas}}, \bibinfo
  {author} {\bibfnamefont {K.}~\bibnamefont {Sliwa}}, \bibinfo {author}
  {\bibfnamefont {A.}~\bibnamefont {Narla}}, \bibinfo {author} {\bibfnamefont
  {U.}~\bibnamefont {Vool}}, \bibinfo {author} {\bibfnamefont {S.~M.}\
  \bibnamefont {Girvin}}, \bibinfo {author} {\bibfnamefont {L.}~\bibnamefont
  {Frunzio}}, \bibinfo {author} {\bibfnamefont {M.}~\bibnamefont {Mirrahimi}},
  \ and\ \bibinfo {author} {\bibfnamefont {M.~H.}\ \bibnamefont {Devoret}},\
  }\href@noop {} {\bibfield  {journal} {\bibinfo  {journal} {Nature}\ }\textbf
  {\bibinfo {volume} {504}},\ \bibinfo {pages} {419} (\bibinfo {year}
  {2013})}\BibitemShut {NoStop}%
\bibitem [{\citenamefont {Koch}\ \emph {et~al.}(2007)\citenamefont {Koch},
  \citenamefont {Terri}, \citenamefont {Gambetta}, \citenamefont {Houck},
  \citenamefont {Schuster}, \citenamefont {Majer}, \citenamefont {Blais},
  \citenamefont {Devoret}, \citenamefont {Girvin},\ and\ \citenamefont
  {Schoelkopf}}]{07_Charge_Schoelkopf}%
  \BibitemOpen
  \bibfield  {author} {\bibinfo {author} {\bibfnamefont {J.}~\bibnamefont
  {Koch}}, \bibinfo {author} {\bibfnamefont {M.~Y.}\ \bibnamefont {Terri}},
  \bibinfo {author} {\bibfnamefont {J.}~\bibnamefont {Gambetta}}, \bibinfo
  {author} {\bibfnamefont {A.~A.}\ \bibnamefont {Houck}}, \bibinfo {author}
  {\bibfnamefont {D.}~\bibnamefont {Schuster}}, \bibinfo {author}
  {\bibfnamefont {J.}~\bibnamefont {Majer}}, \bibinfo {author} {\bibfnamefont
  {A.}~\bibnamefont {Blais}}, \bibinfo {author} {\bibfnamefont {M.~H.}\
  \bibnamefont {Devoret}}, \bibinfo {author} {\bibfnamefont {S.~M.}\
  \bibnamefont {Girvin}}, \ and\ \bibinfo {author} {\bibfnamefont {R.~J.}\
  \bibnamefont {Schoelkopf}},\ }\href@noop {} {\bibfield  {journal} {\bibinfo
  {journal} {Physical Review A}\ }\textbf {\bibinfo {volume} {76}},\ \bibinfo
  {pages} {042319} (\bibinfo {year} {2007})}\BibitemShut {NoStop}%
\bibitem [{\citenamefont {Reshitnyk}\ \emph {et~al.}(2016)\citenamefont
  {Reshitnyk}, \citenamefont {Jerger},\ and\ \citenamefont
  {Fedorov}}]{16_3D_Fedorov}%
  \BibitemOpen
  \bibfield  {author} {\bibinfo {author} {\bibfnamefont {Y.}~\bibnamefont
  {Reshitnyk}}, \bibinfo {author} {\bibfnamefont {M.}~\bibnamefont {Jerger}}, \
  and\ \bibinfo {author} {\bibfnamefont {A.}~\bibnamefont {Fedorov}},\
  }\href@noop {} {\bibfield  {journal} {\bibinfo  {journal} {EPJ Quantum
  Technology}\ }\textbf {\bibinfo {volume} {3}},\ \bibinfo {pages} {13}
  (\bibinfo {year} {2016})}\BibitemShut {NoStop}%
\bibitem [{\citenamefont {Gargiulo}\ \emph {et~al.}(2018)\citenamefont
  {Gargiulo}, \citenamefont {Oleschko}, \citenamefont {Prat-Camps},
  \citenamefont {Zanner},\ and\ \citenamefont {Kirchmair}}]{18_Fast_Kirchmair}%
  \BibitemOpen
  \bibfield  {author} {\bibinfo {author} {\bibfnamefont {O.}~\bibnamefont
  {Gargiulo}}, \bibinfo {author} {\bibfnamefont {S.}~\bibnamefont {Oleschko}},
  \bibinfo {author} {\bibfnamefont {J.}~\bibnamefont {Prat-Camps}}, \bibinfo
  {author} {\bibfnamefont {M.}~\bibnamefont {Zanner}}, \ and\ \bibinfo {author}
  {\bibfnamefont {G.}~\bibnamefont {Kirchmair}},\ }\href@noop {} {\bibfield
  {journal} {\bibinfo  {journal} {arXiv:1811.10875}\ } (\bibinfo {year}
  {2018})}\BibitemShut {NoStop}%
\bibitem [{\citenamefont {Stammeier}\ \emph {et~al.}(2018)\citenamefont
  {Stammeier}, \citenamefont {Garcia},\ and\ \citenamefont
  {Wallraff}}]{18_Applying_Wallraff}%
  \BibitemOpen
  \bibfield  {author} {\bibinfo {author} {\bibfnamefont {M.}~\bibnamefont
  {Stammeier}}, \bibinfo {author} {\bibfnamefont {S.}~\bibnamefont {Garcia}}, \
  and\ \bibinfo {author} {\bibfnamefont {A.}~\bibnamefont {Wallraff}},\
  }\href@noop {} {\bibfield  {journal} {\bibinfo  {journal} {Quantum Science
  and Technology}\ }\textbf {\bibinfo {volume} {3}},\ \bibinfo {pages} {045007}
  (\bibinfo {year} {2018})}\BibitemShut {NoStop}%
\bibitem [{\citenamefont {Ball}\ \emph {et~al.}(2018)\citenamefont {Ball},
  \citenamefont {Yamashiro}, \citenamefont {Sumiya}, \citenamefont {Onoda},
  \citenamefont {Ohshima}, \citenamefont {Isoya}, \citenamefont
  {Konstantinov},\ and\ \citenamefont {Kubo}}]{18_Loop-gap_Kubo}%
  \BibitemOpen
  \bibfield  {author} {\bibinfo {author} {\bibfnamefont {J.~R.}\ \bibnamefont
  {Ball}}, \bibinfo {author} {\bibfnamefont {Y.}~\bibnamefont {Yamashiro}},
  \bibinfo {author} {\bibfnamefont {H.}~\bibnamefont {Sumiya}}, \bibinfo
  {author} {\bibfnamefont {S.}~\bibnamefont {Onoda}}, \bibinfo {author}
  {\bibfnamefont {T.}~\bibnamefont {Ohshima}}, \bibinfo {author} {\bibfnamefont
  {J.}~\bibnamefont {Isoya}}, \bibinfo {author} {\bibfnamefont
  {D.}~\bibnamefont {Konstantinov}}, \ and\ \bibinfo {author} {\bibfnamefont
  {Y.}~\bibnamefont {Kubo}},\ }\href@noop {} {\bibfield  {journal} {\bibinfo
  {journal} {Applied Physics Letters}\ }\textbf {\bibinfo {volume} {112}},\
  \bibinfo {pages} {204102} (\bibinfo {year} {2018})}\BibitemShut {NoStop}%
\bibitem [{\citenamefont {Eisenstein}(1954)}]{54_Superconducting_Eisenstein}%
  \BibitemOpen
  \bibfield  {author} {\bibinfo {author} {\bibfnamefont {J.}~\bibnamefont
  {Eisenstein}},\ }\href@noop {} {\bibfield  {journal} {\bibinfo  {journal}
  {Reviews of Modern Physics}\ }\textbf {\bibinfo {volume} {26}},\ \bibinfo
  {pages} {277} (\bibinfo {year} {1954})}\BibitemShut {NoStop}%
\bibitem [{\citenamefont {Saito}\ \emph {et~al.}(2001)\citenamefont {Saito}
  \emph {et~al.}}]{01_Critical_Saito}%
  \BibitemOpen
  \bibfield  {author} {\bibinfo {author} {\bibfnamefont {K.}~\bibnamefont
  {Saito}} \emph {et~al.},\ }in\ \href
  {http://conference.kek.jp/SRF2001/pdf/PH003.pdf} {\emph {\bibinfo {booktitle}
  {Proceedings of the 10th International Conference on RF Superconductivity,
  Tsukuba, Japan}}}\ (\bibinfo {year} {2001})\BibitemShut {NoStop}%
\bibitem [{\citenamefont {Yin}\ \emph {et~al.}(2013)\citenamefont {Yin},
  \citenamefont {Chen}, \citenamefont {Sank}, \citenamefont {O'Malley},
  \citenamefont {White}, \citenamefont {Barends}, \citenamefont {Kelly},
  \citenamefont {Lucero}, \citenamefont {Mariantoni}, \citenamefont {Megrant}
  \emph {et~al.}}]{13_Catch_Martinis}%
  \BibitemOpen
  \bibfield  {author} {\bibinfo {author} {\bibfnamefont {Y.}~\bibnamefont
  {Yin}}, \bibinfo {author} {\bibfnamefont {Y.}~\bibnamefont {Chen}}, \bibinfo
  {author} {\bibfnamefont {D.}~\bibnamefont {Sank}}, \bibinfo {author}
  {\bibfnamefont {P.~J.~J.}\ \bibnamefont {O'Malley}}, \bibinfo {author}
  {\bibfnamefont {T.~C.}\ \bibnamefont {White}}, \bibinfo {author}
  {\bibfnamefont {R.}~\bibnamefont {Barends}}, \bibinfo {author} {\bibfnamefont
  {J.}~\bibnamefont {Kelly}}, \bibinfo {author} {\bibfnamefont
  {E.}~\bibnamefont {Lucero}}, \bibinfo {author} {\bibfnamefont
  {M.}~\bibnamefont {Mariantoni}}, \bibinfo {author} {\bibfnamefont
  {A.}~\bibnamefont {Megrant}},  \emph {et~al.},\ }\href@noop {} {\bibfield
  {journal} {\bibinfo  {journal} {Physical Review Letters}\ }\textbf {\bibinfo
  {volume} {110}},\ \bibinfo {pages} {107001} (\bibinfo {year}
  {2013})}\BibitemShut {NoStop}%
\bibitem [{\citenamefont {Axline}\ \emph {et~al.}(2018)\citenamefont {Axline},
  \citenamefont {Burkhart}, \citenamefont {Pfaff}, \citenamefont {Zhang},
  \citenamefont {Chou}, \citenamefont {Campagne-Ibarcq}, \citenamefont
  {Reinhold}, \citenamefont {Frunzio}, \citenamefont {Girvin}, \citenamefont
  {Jiang}, \citenamefont {Devoret},\ and\ \citenamefont
  {Schoelkopf}}]{18_On-demand_Schoelkopf}%
  \BibitemOpen
  \bibfield  {author} {\bibinfo {author} {\bibfnamefont {C.~J.}\ \bibnamefont
  {Axline}}, \bibinfo {author} {\bibfnamefont {L.~D.}\ \bibnamefont
  {Burkhart}}, \bibinfo {author} {\bibfnamefont {W.}~\bibnamefont {Pfaff}},
  \bibinfo {author} {\bibfnamefont {M.}~\bibnamefont {Zhang}}, \bibinfo
  {author} {\bibfnamefont {K.}~\bibnamefont {Chou}}, \bibinfo {author}
  {\bibfnamefont {P.}~\bibnamefont {Campagne-Ibarcq}}, \bibinfo {author}
  {\bibfnamefont {P.}~\bibnamefont {Reinhold}}, \bibinfo {author}
  {\bibfnamefont {L.}~\bibnamefont {Frunzio}}, \bibinfo {author} {\bibfnamefont
  {S.~M.}\ \bibnamefont {Girvin}}, \bibinfo {author} {\bibfnamefont
  {L.}~\bibnamefont {Jiang}}, \bibinfo {author} {\bibfnamefont {M.~H.}\
  \bibnamefont {Devoret}}, \ and\ \bibinfo {author} {\bibfnamefont {R.~J.}\
  \bibnamefont {Schoelkopf}},\ }\href {\doibase 10.1038/s41567-018-0115-y}
  {\bibfield  {journal} {\bibinfo  {journal} {Nature Physics}\ }\textbf
  {\bibinfo {volume} {14}},\ \bibinfo {pages} {705} (\bibinfo {year}
  {2018})}\BibitemShut {NoStop}%
\bibitem [{\citenamefont {Simon}\ \emph {et~al.}(2007)\citenamefont {Simon},
  \citenamefont {De~Riedmatten}, \citenamefont {Afzelius}, \citenamefont
  {Sangouard}, \citenamefont {Zbinden},\ and\ \citenamefont
  {Gisin}}]{07_Quantum_repeaters_Gisin}%
  \BibitemOpen
  \bibfield  {author} {\bibinfo {author} {\bibfnamefont {C.}~\bibnamefont
  {Simon}}, \bibinfo {author} {\bibfnamefont {H.}~\bibnamefont
  {De~Riedmatten}}, \bibinfo {author} {\bibfnamefont {M.}~\bibnamefont
  {Afzelius}}, \bibinfo {author} {\bibfnamefont {N.}~\bibnamefont {Sangouard}},
  \bibinfo {author} {\bibfnamefont {H.}~\bibnamefont {Zbinden}}, \ and\
  \bibinfo {author} {\bibfnamefont {N.}~\bibnamefont {Gisin}},\ }\href@noop {}
  {\bibfield  {journal} {\bibinfo  {journal} {Physical Review Letters}\
  }\textbf {\bibinfo {volume} {98}},\ \bibinfo {pages} {190503} (\bibinfo
  {year} {2007})}\BibitemShut {NoStop}%
\bibitem [{\citenamefont {Sun}\ \emph {et~al.}(2014)\citenamefont {Sun},
  \citenamefont {Petrenko}, \citenamefont {Leghtas}, \citenamefont {Vlastakis},
  \citenamefont {Kirchmair}, \citenamefont {Sliwa}, \citenamefont {Narla},
  \citenamefont {Hatridge}, \citenamefont {Shankar}, \citenamefont {Blumoff}
  \emph {et~al.}}]{14_Tracking_Schoelkopf}%
  \BibitemOpen
  \bibfield  {author} {\bibinfo {author} {\bibfnamefont {L.}~\bibnamefont
  {Sun}}, \bibinfo {author} {\bibfnamefont {A.}~\bibnamefont {Petrenko}},
  \bibinfo {author} {\bibfnamefont {Z.}~\bibnamefont {Leghtas}}, \bibinfo
  {author} {\bibfnamefont {B.}~\bibnamefont {Vlastakis}}, \bibinfo {author}
  {\bibfnamefont {G.}~\bibnamefont {Kirchmair}}, \bibinfo {author}
  {\bibfnamefont {K.}~\bibnamefont {Sliwa}}, \bibinfo {author} {\bibfnamefont
  {A.}~\bibnamefont {Narla}}, \bibinfo {author} {\bibfnamefont
  {M.}~\bibnamefont {Hatridge}}, \bibinfo {author} {\bibfnamefont
  {S.}~\bibnamefont {Shankar}}, \bibinfo {author} {\bibfnamefont
  {J.}~\bibnamefont {Blumoff}},  \emph {et~al.},\ }\href@noop {} {\bibfield
  {journal} {\bibinfo  {journal} {Nature}\ }\textbf {\bibinfo {volume} {511}},\
  \bibinfo {pages} {444} (\bibinfo {year} {2014})}\BibitemShut {NoStop}%
\bibitem [{\citenamefont {Johnson}\ \emph {et~al.}(2010)\citenamefont
  {Johnson}, \citenamefont {Reed}, \citenamefont {Houck}, \citenamefont
  {Schuster}, \citenamefont {Bishop}, \citenamefont {Ginossar}, \citenamefont
  {Gambetta}, \citenamefont {DiCarlo}, \citenamefont {Frunzio}, \citenamefont
  {Girvin} \emph {et~al.}}]{10_QND_Schoelkopf}%
  \BibitemOpen
  \bibfield  {author} {\bibinfo {author} {\bibfnamefont {B.}~\bibnamefont
  {Johnson}}, \bibinfo {author} {\bibfnamefont {M.}~\bibnamefont {Reed}},
  \bibinfo {author} {\bibfnamefont {A.}~\bibnamefont {Houck}}, \bibinfo
  {author} {\bibfnamefont {D.}~\bibnamefont {Schuster}}, \bibinfo {author}
  {\bibfnamefont {L.~S.}\ \bibnamefont {Bishop}}, \bibinfo {author}
  {\bibfnamefont {E.}~\bibnamefont {Ginossar}}, \bibinfo {author}
  {\bibfnamefont {J.}~\bibnamefont {Gambetta}}, \bibinfo {author}
  {\bibfnamefont {L.}~\bibnamefont {DiCarlo}}, \bibinfo {author} {\bibfnamefont
  {L.}~\bibnamefont {Frunzio}}, \bibinfo {author} {\bibfnamefont
  {S.}~\bibnamefont {Girvin}},  \emph {et~al.},\ }\href@noop {} {\bibfield
  {journal} {\bibinfo  {journal} {Nature Physics}\ }\textbf {\bibinfo {volume}
  {6}},\ \bibinfo {pages} {663} (\bibinfo {year} {2010})}\BibitemShut {NoStop}%
\bibitem [{\citenamefont {Chen}\ \emph {et~al.}(2018)\citenamefont {Chen},
  \citenamefont {Zopf}, \citenamefont {Keil}, \citenamefont {Ding},\ and\
  \citenamefont {Schmidt}}]{18_Highly_Schmidt}%
  \BibitemOpen
  \bibfield  {author} {\bibinfo {author} {\bibfnamefont {Y.}~\bibnamefont
  {Chen}}, \bibinfo {author} {\bibfnamefont {M.}~\bibnamefont {Zopf}}, \bibinfo
  {author} {\bibfnamefont {R.}~\bibnamefont {Keil}}, \bibinfo {author}
  {\bibfnamefont {F.}~\bibnamefont {Ding}}, \ and\ \bibinfo {author}
  {\bibfnamefont {O.~G.}\ \bibnamefont {Schmidt}},\ }\href@noop {} {\bibfield
  {journal} {\bibinfo  {journal} {Nature Communications}\ }\textbf {\bibinfo
  {volume} {9}},\ \bibinfo {pages} {2994} (\bibinfo {year} {2018})}\BibitemShut
  {NoStop}%
\bibitem [{\citenamefont {Chou}\ \emph {et~al.}(2005)\citenamefont {Chou},
  \citenamefont {De~Riedmatten}, \citenamefont {Felinto}, \citenamefont
  {Polyakov}, \citenamefont {Van~Enk},\ and\ \citenamefont
  {Kimble}}]{05_Measurement_Kimble}%
  \BibitemOpen
  \bibfield  {author} {\bibinfo {author} {\bibfnamefont {C.-W.}\ \bibnamefont
  {Chou}}, \bibinfo {author} {\bibfnamefont {H.}~\bibnamefont {De~Riedmatten}},
  \bibinfo {author} {\bibfnamefont {D.}~\bibnamefont {Felinto}}, \bibinfo
  {author} {\bibfnamefont {S.}~\bibnamefont {Polyakov}}, \bibinfo {author}
  {\bibfnamefont {S.}~\bibnamefont {Van~Enk}}, \ and\ \bibinfo {author}
  {\bibfnamefont {H.~J.}\ \bibnamefont {Kimble}},\ }\href@noop {} {\bibfield
  {journal} {\bibinfo  {journal} {Nature}\ }\textbf {\bibinfo {volume} {438}},\
  \bibinfo {pages} {828} (\bibinfo {year} {2005})}\BibitemShut {NoStop}%
\bibitem [{\citenamefont {Kurpiers}\ \emph {et~al.}(2018)\citenamefont
  {Kurpiers}, \citenamefont {Magnard}, \citenamefont {Walter}, \citenamefont
  {Royer}, \citenamefont {Pechal}, \citenamefont {Heinsoo}, \citenamefont
  {Salath{\'e}}, \citenamefont {Akin}, \citenamefont {Storz}, \citenamefont
  {Besse} \emph {et~al.}}]{18_Deterministic_Wallraff}%
  \BibitemOpen
  \bibfield  {author} {\bibinfo {author} {\bibfnamefont {P.}~\bibnamefont
  {Kurpiers}}, \bibinfo {author} {\bibfnamefont {P.}~\bibnamefont {Magnard}},
  \bibinfo {author} {\bibfnamefont {T.}~\bibnamefont {Walter}}, \bibinfo
  {author} {\bibfnamefont {B.}~\bibnamefont {Royer}}, \bibinfo {author}
  {\bibfnamefont {M.}~\bibnamefont {Pechal}}, \bibinfo {author} {\bibfnamefont
  {J.}~\bibnamefont {Heinsoo}}, \bibinfo {author} {\bibfnamefont
  {Y.}~\bibnamefont {Salath{\'e}}}, \bibinfo {author} {\bibfnamefont
  {A.}~\bibnamefont {Akin}}, \bibinfo {author} {\bibfnamefont {S.}~\bibnamefont
  {Storz}}, \bibinfo {author} {\bibfnamefont {J.-C.}\ \bibnamefont {Besse}},
  \emph {et~al.},\ }\href@noop {} {\bibfield  {journal} {\bibinfo  {journal}
  {Nature}\ }\textbf {\bibinfo {volume} {558}},\ \bibinfo {pages} {264}
  (\bibinfo {year} {2018})}\BibitemShut {NoStop}%
\bibitem [{\citenamefont {Li}\ \emph {et~al.}(2018)\citenamefont {Li},
  \citenamefont {Ma}, \citenamefont {Han}, \citenamefont {Chen}, \citenamefont
  {Xu}, \citenamefont {Cai}, \citenamefont {Wang}, \citenamefont {Song},
  \citenamefont {Xue}, \citenamefont {Yin} \emph {et~al.}}]{18_Perfect_Sun}%
  \BibitemOpen
  \bibfield  {author} {\bibinfo {author} {\bibfnamefont {X.}~\bibnamefont
  {Li}}, \bibinfo {author} {\bibfnamefont {Y.}~\bibnamefont {Ma}}, \bibinfo
  {author} {\bibfnamefont {J.}~\bibnamefont {Han}}, \bibinfo {author}
  {\bibfnamefont {T.}~\bibnamefont {Chen}}, \bibinfo {author} {\bibfnamefont
  {Y.}~\bibnamefont {Xu}}, \bibinfo {author} {\bibfnamefont {W.}~\bibnamefont
  {Cai}}, \bibinfo {author} {\bibfnamefont {H.}~\bibnamefont {Wang}}, \bibinfo
  {author} {\bibfnamefont {Y.}~\bibnamefont {Song}}, \bibinfo {author}
  {\bibfnamefont {Z.-Y.}\ \bibnamefont {Xue}}, \bibinfo {author} {\bibfnamefont
  {Z.-q.}\ \bibnamefont {Yin}},  \emph {et~al.},\ }\href@noop {} {\bibfield
  {journal} {\bibinfo  {journal} {Physical Review Applied}\ }\textbf {\bibinfo
  {volume} {10}},\ \bibinfo {pages} {054009} (\bibinfo {year}
  {2018})}\BibitemShut {NoStop}%
\bibitem [{\citenamefont {Touzard}\ \emph {et~al.}(2018)\citenamefont
  {Touzard}, \citenamefont {Grimm}, \citenamefont {Leghtas}, \citenamefont
  {Mundhada}, \citenamefont {Reinhold}, \citenamefont {Axline}, \citenamefont
  {Reagor}, \citenamefont {Chou}, \citenamefont {Blumoff}, \citenamefont
  {Sliwa}, \citenamefont {Shankar}, \citenamefont {Frunzio}, \citenamefont
  {Schoelkopf}, \citenamefont {Mirrahimi},\ and\ \citenamefont
  {Devoret}}]{18_Coherent_Devoret}%
  \BibitemOpen
  \bibfield  {author} {\bibinfo {author} {\bibfnamefont {S.}~\bibnamefont
  {Touzard}}, \bibinfo {author} {\bibfnamefont {A.}~\bibnamefont {Grimm}},
  \bibinfo {author} {\bibfnamefont {Z.}~\bibnamefont {Leghtas}}, \bibinfo
  {author} {\bibfnamefont {S.~O.}\ \bibnamefont {Mundhada}}, \bibinfo {author}
  {\bibfnamefont {P.}~\bibnamefont {Reinhold}}, \bibinfo {author}
  {\bibfnamefont {C.}~\bibnamefont {Axline}}, \bibinfo {author} {\bibfnamefont
  {M.}~\bibnamefont {Reagor}}, \bibinfo {author} {\bibfnamefont
  {K.}~\bibnamefont {Chou}}, \bibinfo {author} {\bibfnamefont {J.}~\bibnamefont
  {Blumoff}}, \bibinfo {author} {\bibfnamefont {K.~M.}\ \bibnamefont {Sliwa}},
  \bibinfo {author} {\bibfnamefont {S.}~\bibnamefont {Shankar}}, \bibinfo
  {author} {\bibfnamefont {L.}~\bibnamefont {Frunzio}}, \bibinfo {author}
  {\bibfnamefont {R.~J.}\ \bibnamefont {Schoelkopf}}, \bibinfo {author}
  {\bibfnamefont {M.}~\bibnamefont {Mirrahimi}}, \ and\ \bibinfo {author}
  {\bibfnamefont {M.~H.}\ \bibnamefont {Devoret}},\ }\href {\doibase
  10.1103/PhysRevX.8.021005} {\bibfield  {journal} {\bibinfo  {journal}
  {Physical Review X}\ }\textbf {\bibinfo {volume} {8}},\ \bibinfo {pages}
  {021005} (\bibinfo {year} {2018})}\BibitemShut {NoStop}%
\bibitem [{\citenamefont {D{\"u}r}\ and\ \citenamefont
  {Briegel}(2007)}]{07_Entanglement_Briegel}%
  \BibitemOpen
  \bibfield  {author} {\bibinfo {author} {\bibfnamefont {W.}~\bibnamefont
  {D{\"u}r}}\ and\ \bibinfo {author} {\bibfnamefont {H.~J.}\ \bibnamefont
  {Briegel}},\ }\href@noop {} {\bibfield  {journal} {\bibinfo  {journal}
  {Reports on Progress in Physics}\ }\textbf {\bibinfo {volume} {70}},\
  \bibinfo {pages} {1381} (\bibinfo {year} {2007})}\BibitemShut {NoStop}%
\bibitem [{\citenamefont {Dassonneville}\ \emph {et~al.}(2019)\citenamefont
  {Dassonneville}, \citenamefont {Ramos}, \citenamefont {Milchakov},
  \citenamefont {Planat}, \citenamefont {Dumur}, \citenamefont {Foroughi},
  \citenamefont {Puertas}, \citenamefont {Leger}, \citenamefont {Bharadwaj},
  \citenamefont {Delaforce} \emph {et~al.}}]{19_Fast_Buisson}%
  \BibitemOpen
  \bibfield  {author} {\bibinfo {author} {\bibfnamefont {R.}~\bibnamefont
  {Dassonneville}}, \bibinfo {author} {\bibfnamefont {T.}~\bibnamefont
  {Ramos}}, \bibinfo {author} {\bibfnamefont {V.}~\bibnamefont {Milchakov}},
  \bibinfo {author} {\bibfnamefont {L.}~\bibnamefont {Planat}}, \bibinfo
  {author} {\bibfnamefont {{\'E}.}~\bibnamefont {Dumur}}, \bibinfo {author}
  {\bibfnamefont {F.}~\bibnamefont {Foroughi}}, \bibinfo {author}
  {\bibfnamefont {J.}~\bibnamefont {Puertas}}, \bibinfo {author} {\bibfnamefont
  {S.}~\bibnamefont {Leger}}, \bibinfo {author} {\bibfnamefont
  {K.}~\bibnamefont {Bharadwaj}}, \bibinfo {author} {\bibfnamefont
  {J.}~\bibnamefont {Delaforce}},  \emph {et~al.},\ }\href@noop {} {\bibfield
  {journal} {\bibinfo  {journal} {arXiv:1905.00271}\ } (\bibinfo {year}
  {2019})}\BibitemShut {NoStop}%
\bibitem [{\citenamefont {Krantz}\ \emph {et~al.}(2019)\citenamefont {Krantz},
  \citenamefont {Kjaergaard}, \citenamefont {Yan}, \citenamefont {Orlando},
  \citenamefont {Gustavsson},\ and\ \citenamefont
  {Oliver}}]{19_Quantum_Oliver}%
  \BibitemOpen
  \bibfield  {author} {\bibinfo {author} {\bibfnamefont {P.}~\bibnamefont
  {Krantz}}, \bibinfo {author} {\bibfnamefont {M.}~\bibnamefont {Kjaergaard}},
  \bibinfo {author} {\bibfnamefont {F.}~\bibnamefont {Yan}}, \bibinfo {author}
  {\bibfnamefont {T.~P.}\ \bibnamefont {Orlando}}, \bibinfo {author}
  {\bibfnamefont {S.}~\bibnamefont {Gustavsson}}, \ and\ \bibinfo {author}
  {\bibfnamefont {W.~D.}\ \bibnamefont {Oliver}},\ }\href {\doibase
  10.1063/1.5089550} {\bibfield  {journal} {\bibinfo  {journal} {Applied
  Physics Reviews}\ }\textbf {\bibinfo {volume} {6}},\ \bibinfo {pages}
  {021318} (\bibinfo {year} {2019})}\BibitemShut {NoStop}%
\bibitem [{\citenamefont {Michael}\ \emph {et~al.}(2016)\citenamefont
  {Michael}, \citenamefont {Silveri}, \citenamefont {Brierley}, \citenamefont
  {Albert}, \citenamefont {Salmilehto}, \citenamefont {Jiang},\ and\
  \citenamefont {Girvin}}]{16_New_Girvin}%
  \BibitemOpen
  \bibfield  {author} {\bibinfo {author} {\bibfnamefont {M.~H.}\ \bibnamefont
  {Michael}}, \bibinfo {author} {\bibfnamefont {M.}~\bibnamefont {Silveri}},
  \bibinfo {author} {\bibfnamefont {R.~T.}\ \bibnamefont {Brierley}}, \bibinfo
  {author} {\bibfnamefont {V.~V.}\ \bibnamefont {Albert}}, \bibinfo {author}
  {\bibfnamefont {J.}~\bibnamefont {Salmilehto}}, \bibinfo {author}
  {\bibfnamefont {L.}~\bibnamefont {Jiang}}, \ and\ \bibinfo {author}
  {\bibfnamefont {S.~M.}\ \bibnamefont {Girvin}},\ }\href {\doibase
  10.1103/PhysRevX.6.031006} {\bibfield  {journal} {\bibinfo  {journal}
  {Physical Review X}\ }\textbf {\bibinfo {volume} {6}},\ \bibinfo {pages}
  {031006} (\bibinfo {year} {2016})}\BibitemShut {NoStop}%
\bibitem [{\citenamefont {Albert}\ \emph {et~al.}(2018)\citenamefont {Albert},
  \citenamefont {Noh}, \citenamefont {Duivenvoorden}, \citenamefont {Young},
  \citenamefont {Brierley}, \citenamefont {Reinhold}, \citenamefont {Vuillot},
  \citenamefont {Li}, \citenamefont {Shen}, \citenamefont {Girvin} \emph
  {et~al.}}]{18_Performance_Jiang}%
  \BibitemOpen
  \bibfield  {author} {\bibinfo {author} {\bibfnamefont {V.~V.}\ \bibnamefont
  {Albert}}, \bibinfo {author} {\bibfnamefont {K.}~\bibnamefont {Noh}},
  \bibinfo {author} {\bibfnamefont {K.}~\bibnamefont {Duivenvoorden}}, \bibinfo
  {author} {\bibfnamefont {D.~J.}\ \bibnamefont {Young}}, \bibinfo {author}
  {\bibfnamefont {R.}~\bibnamefont {Brierley}}, \bibinfo {author}
  {\bibfnamefont {P.}~\bibnamefont {Reinhold}}, \bibinfo {author}
  {\bibfnamefont {C.}~\bibnamefont {Vuillot}}, \bibinfo {author} {\bibfnamefont
  {L.}~\bibnamefont {Li}}, \bibinfo {author} {\bibfnamefont {C.}~\bibnamefont
  {Shen}}, \bibinfo {author} {\bibfnamefont {S.}~\bibnamefont {Girvin}},  \emph
  {et~al.},\ }\href@noop {} {\bibfield  {journal} {\bibinfo  {journal}
  {Physical Review A}\ }\textbf {\bibinfo {volume} {97}},\ \bibinfo {pages}
  {032346} (\bibinfo {year} {2018})}\BibitemShut {NoStop}%
\bibitem [{\citenamefont {Hu}\ \emph {et~al.}(2019)\citenamefont {Hu},
  \citenamefont {Ma}, \citenamefont {Cai}, \citenamefont {Mu}, \citenamefont
  {Xu}, \citenamefont {Wang}, \citenamefont {Wu}, \citenamefont {Wang},
  \citenamefont {Song}, \citenamefont {Zou}, \citenamefont {Girvin},
  \citenamefont {Duan},\ and\ \citenamefont {Sun}}]{18_Demonstration_Sun}%
  \BibitemOpen
  \bibfield  {author} {\bibinfo {author} {\bibfnamefont {L.}~\bibnamefont
  {Hu}}, \bibinfo {author} {\bibfnamefont {Y.}~\bibnamefont {Ma}}, \bibinfo
  {author} {\bibfnamefont {W.}~\bibnamefont {Cai}}, \bibinfo {author}
  {\bibfnamefont {X.}~\bibnamefont {Mu}}, \bibinfo {author} {\bibfnamefont
  {Y.}~\bibnamefont {Xu}}, \bibinfo {author} {\bibfnamefont {W.}~\bibnamefont
  {Wang}}, \bibinfo {author} {\bibfnamefont {Y.}~\bibnamefont {Wu}}, \bibinfo
  {author} {\bibfnamefont {H.}~\bibnamefont {Wang}}, \bibinfo {author}
  {\bibfnamefont {Y.~P.}\ \bibnamefont {Song}}, \bibinfo {author}
  {\bibfnamefont {C.-L.}\ \bibnamefont {Zou}}, \bibinfo {author} {\bibfnamefont
  {S.~M.}\ \bibnamefont {Girvin}}, \bibinfo {author} {\bibfnamefont {L.-M.}\
  \bibnamefont {Duan}}, \ and\ \bibinfo {author} {\bibfnamefont
  {L.}~\bibnamefont {Sun}},\ }\href {\doibase 10.1038/s41567-018-0414-3}
  {\bibfield  {journal} {\bibinfo  {journal} {Nature Physics}\ }\textbf
  {\bibinfo {volume} {15}},\ \bibinfo {pages} {503} (\bibinfo {year}
  {2019})}\BibitemShut {NoStop}%
\bibitem [{\citenamefont {Beals}\ \emph {et~al.}(2013)\citenamefont {Beals},
  \citenamefont {Brierley}, \citenamefont {Gray}, \citenamefont {Harrow},
  \citenamefont {Kutin}, \citenamefont {Linden}, \citenamefont {Shepherd},\
  and\ \citenamefont {Stather}}]{13_Efficient_Stather}%
  \BibitemOpen
  \bibfield  {author} {\bibinfo {author} {\bibfnamefont {R.}~\bibnamefont
  {Beals}}, \bibinfo {author} {\bibfnamefont {S.}~\bibnamefont {Brierley}},
  \bibinfo {author} {\bibfnamefont {O.}~\bibnamefont {Gray}}, \bibinfo {author}
  {\bibfnamefont {A.~W.}\ \bibnamefont {Harrow}}, \bibinfo {author}
  {\bibfnamefont {S.}~\bibnamefont {Kutin}}, \bibinfo {author} {\bibfnamefont
  {N.}~\bibnamefont {Linden}}, \bibinfo {author} {\bibfnamefont
  {D.}~\bibnamefont {Shepherd}}, \ and\ \bibinfo {author} {\bibfnamefont
  {M.}~\bibnamefont {Stather}},\ }\href@noop {} {\bibfield  {journal} {\bibinfo
   {journal} {Proc. R. Soc. A}\ }\textbf {\bibinfo {volume} {469}},\ \bibinfo
  {pages} {20120686} (\bibinfo {year} {2013})}\BibitemShut {NoStop}%
\bibitem [{\citenamefont {Brierley}(2017)}]{15_Efficient_Brierley}%
  \BibitemOpen
  \bibfield  {author} {\bibinfo {author} {\bibfnamefont {S.}~\bibnamefont
  {Brierley}},\ }\href {http://dl.acm.org/citation.cfm?id=3179575.3179577}
  {\bibfield  {journal} {\bibinfo  {journal} {Quantum Info. Comput.}\ }\textbf
  {\bibinfo {volume} {17}},\ \bibinfo {pages} {1096} (\bibinfo {year}
  {2017})}\BibitemShut {NoStop}%
\bibitem [{\citenamefont {Nickerson}\ \emph {et~al.}(2014)\citenamefont
  {Nickerson}, \citenamefont {Fitzsimons},\ and\ \citenamefont
  {Benjamin}}]{14_Freely_Simon}%
  \BibitemOpen
  \bibfield  {author} {\bibinfo {author} {\bibfnamefont {N.~H.}\ \bibnamefont
  {Nickerson}}, \bibinfo {author} {\bibfnamefont {J.~F.}\ \bibnamefont
  {Fitzsimons}}, \ and\ \bibinfo {author} {\bibfnamefont {S.~C.}\ \bibnamefont
  {Benjamin}},\ }\href {\doibase 10.1103/PhysRevX.4.041041} {\bibfield
  {journal} {\bibinfo  {journal} {Physical Review X}\ }\textbf {\bibinfo
  {volume} {4}},\ \bibinfo {pages} {041041} (\bibinfo {year}
  {2014})}\BibitemShut {NoStop}%
\bibitem [{\citenamefont {Monroe}\ \emph {et~al.}(2014)\citenamefont {Monroe},
  \citenamefont {Raussendorf}, \citenamefont {Ruthven}, \citenamefont {Brown},
  \citenamefont {Maunz}, \citenamefont {Duan},\ and\ \citenamefont
  {Kim}}]{14_Large-scale_Monroe}%
  \BibitemOpen
  \bibfield  {author} {\bibinfo {author} {\bibfnamefont {C.}~\bibnamefont
  {Monroe}}, \bibinfo {author} {\bibfnamefont {R.}~\bibnamefont {Raussendorf}},
  \bibinfo {author} {\bibfnamefont {A.}~\bibnamefont {Ruthven}}, \bibinfo
  {author} {\bibfnamefont {K.}~\bibnamefont {Brown}}, \bibinfo {author}
  {\bibfnamefont {P.}~\bibnamefont {Maunz}}, \bibinfo {author} {\bibfnamefont
  {L.-M.}\ \bibnamefont {Duan}}, \ and\ \bibinfo {author} {\bibfnamefont
  {J.}~\bibnamefont {Kim}},\ }\href@noop {} {\bibfield  {journal} {\bibinfo
  {journal} {Physical Review A}\ }\textbf {\bibinfo {volume} {89}},\ \bibinfo
  {pages} {022317} (\bibinfo {year} {2014})}\BibitemShut {NoStop}%
\bibitem [{\citenamefont {Peng}\ \emph {et~al.}(2005)\citenamefont {Peng},
  \citenamefont {Yang}, \citenamefont {Bao}, \citenamefont {Zhang},
  \citenamefont {Jin}, \citenamefont {Feng}, \citenamefont {Yang},
  \citenamefont {Yang}, \citenamefont {Yin}, \citenamefont {Zhang} \emph
  {et~al.}}]{05_Experimental_Pan}%
  \BibitemOpen
  \bibfield  {author} {\bibinfo {author} {\bibfnamefont {C.-Z.}\ \bibnamefont
  {Peng}}, \bibinfo {author} {\bibfnamefont {T.}~\bibnamefont {Yang}}, \bibinfo
  {author} {\bibfnamefont {X.-H.}\ \bibnamefont {Bao}}, \bibinfo {author}
  {\bibfnamefont {J.}~\bibnamefont {Zhang}}, \bibinfo {author} {\bibfnamefont
  {X.-M.}\ \bibnamefont {Jin}}, \bibinfo {author} {\bibfnamefont {F.-Y.}\
  \bibnamefont {Feng}}, \bibinfo {author} {\bibfnamefont {B.}~\bibnamefont
  {Yang}}, \bibinfo {author} {\bibfnamefont {J.}~\bibnamefont {Yang}}, \bibinfo
  {author} {\bibfnamefont {J.}~\bibnamefont {Yin}}, \bibinfo {author}
  {\bibfnamefont {Q.}~\bibnamefont {Zhang}},  \emph {et~al.},\ }\href@noop {}
  {\bibfield  {journal} {\bibinfo  {journal} {Physical Review Letters}\
  }\textbf {\bibinfo {volume} {94}},\ \bibinfo {pages} {150501} (\bibinfo
  {year} {2005})}\BibitemShut {NoStop}%
\bibitem [{\citenamefont {Boone}\ \emph {et~al.}(2015)\citenamefont {Boone},
  \citenamefont {Bourgoin}, \citenamefont {Meyer-Scott}, \citenamefont
  {Heshami}, \citenamefont {Jennewein},\ and\ \citenamefont
  {Simon}}]{15_Entanglement_Simon}%
  \BibitemOpen
  \bibfield  {author} {\bibinfo {author} {\bibfnamefont {K.}~\bibnamefont
  {Boone}}, \bibinfo {author} {\bibfnamefont {J.-P.}\ \bibnamefont {Bourgoin}},
  \bibinfo {author} {\bibfnamefont {E.}~\bibnamefont {Meyer-Scott}}, \bibinfo
  {author} {\bibfnamefont {K.}~\bibnamefont {Heshami}}, \bibinfo {author}
  {\bibfnamefont {T.}~\bibnamefont {Jennewein}}, \ and\ \bibinfo {author}
  {\bibfnamefont {C.}~\bibnamefont {Simon}},\ }\href@noop {} {\bibfield
  {journal} {\bibinfo  {journal} {Physical Review A}\ }\textbf {\bibinfo
  {volume} {91}},\ \bibinfo {pages} {052325} (\bibinfo {year}
  {2015})}\BibitemShut {NoStop}%
\bibitem [{\citenamefont {Yin}\ \emph {et~al.}(2017)\citenamefont {Yin},
  \citenamefont {Cao}, \citenamefont {Li}, \citenamefont {Liao}, \citenamefont
  {Zhang}, \citenamefont {Ren}, \citenamefont {Cai}, \citenamefont {Liu},
  \citenamefont {Li}, \citenamefont {Dai} \emph {et~al.}}]{17_Satellite_Pan}%
  \BibitemOpen
  \bibfield  {author} {\bibinfo {author} {\bibfnamefont {J.}~\bibnamefont
  {Yin}}, \bibinfo {author} {\bibfnamefont {Y.}~\bibnamefont {Cao}}, \bibinfo
  {author} {\bibfnamefont {Y.-H.}\ \bibnamefont {Li}}, \bibinfo {author}
  {\bibfnamefont {S.-K.}\ \bibnamefont {Liao}}, \bibinfo {author}
  {\bibfnamefont {L.}~\bibnamefont {Zhang}}, \bibinfo {author} {\bibfnamefont
  {J.-G.}\ \bibnamefont {Ren}}, \bibinfo {author} {\bibfnamefont {W.-Q.}\
  \bibnamefont {Cai}}, \bibinfo {author} {\bibfnamefont {W.-Y.}\ \bibnamefont
  {Liu}}, \bibinfo {author} {\bibfnamefont {B.}~\bibnamefont {Li}}, \bibinfo
  {author} {\bibfnamefont {H.}~\bibnamefont {Dai}},  \emph {et~al.},\
  }\href@noop {} {\bibfield  {journal} {\bibinfo  {journal} {Science}\ }\textbf
  {\bibinfo {volume} {356}},\ \bibinfo {pages} {1140} (\bibinfo {year}
  {2017})}\BibitemShut {NoStop}%
\end{thebibliography}%


\begin{thebibliography}{23}%
\makeatletter
\providecommand \@ifxundefined [1]{%
 \@ifx{#1\undefined}
}%
\providecommand \@ifnum [1]{%
 \ifnum #1\expandafter \@firstoftwo
 \else \expandafter \@secondoftwo
 \fi
}%
\providecommand \@ifx [1]{%
 \ifx #1\expandafter \@firstoftwo
 \else \expandafter \@secondoftwo
 \fi
}%
\providecommand \natexlab [1]{#1}%
\providecommand \enquote  [1]{``#1''}%
\providecommand \bibnamefont  [1]{#1}%
\providecommand \bibfnamefont [1]{#1}%
\providecommand \citenamefont [1]{#1}%
\providecommand \href@noop [0]{\@secondoftwo}%
\providecommand \href [0]{\begingroup \@sanitize@url \@href}%
\providecommand \@href[1]{\@@startlink{#1}\@@href}%
\providecommand \@@href[1]{\endgroup#1\@@endlink}%
\providecommand \@sanitize@url [0]{\catcode `\\12\catcode `\$12\catcode
  `\&12\catcode `\#12\catcode `\^12\catcode `\_12\catcode `\%12\relax}%
\providecommand \@@startlink[1]{}%
\providecommand \@@endlink[0]{}%
\providecommand \url  [0]{\begingroup\@sanitize@url \@url }%
\providecommand \@url [1]{\endgroup\@href {#1}{\urlprefix }}%
\providecommand \urlprefix  [0]{URL }%
\providecommand \Eprint [0]{\href }%
\providecommand \doibase [0]{http://dx.doi.org/}%
\providecommand \selectlanguage [0]{\@gobble}%
\providecommand \bibinfo  [0]{\@secondoftwo}%
\providecommand \bibfield  [0]{\@secondoftwo}%
\providecommand \translation [1]{[#1]}%
\providecommand \BibitemOpen [0]{}%
\providecommand \bibitemStop [0]{}%
\providecommand \bibitemNoStop [0]{.\EOS\space}%
\providecommand \EOS [0]{\spacefactor3000\relax}%
\providecommand \BibitemShut  [1]{\csname bibitem#1\endcsname}%
\let\auto@bib@innerbib\@empty
\bibitem [{\citenamefont {Sangouard}\ \emph {et~al.}(2011)\citenamefont
  {Sangouard}, \citenamefont {Simon}, \citenamefont {De~Riedmatten},\ and\
  \citenamefont {Gisin}}]{11_Quantum_Gisin}%
  \BibitemOpen
  \bibfield  {author} {\bibinfo {author} {\bibfnamefont {N.}~\bibnamefont
  {Sangouard}}, \bibinfo {author} {\bibfnamefont {C.}~\bibnamefont {Simon}},
  \bibinfo {author} {\bibfnamefont {H.}~\bibnamefont {De~Riedmatten}}, \ and\
  \bibinfo {author} {\bibfnamefont {N.}~\bibnamefont {Gisin}},\ }\href@noop {}
  {\bibfield  {journal} {\bibinfo  {journal} {Reviews of Modern Physics}\
  }\textbf {\bibinfo {volume} {83}},\ \bibinfo {pages} {33} (\bibinfo {year}
  {2011})}\BibitemShut {NoStop}%
\bibitem [{\citenamefont {Yin}\ \emph {et~al.}(2013)\citenamefont {Yin},
  \citenamefont {Chen}, \citenamefont {Sank}, \citenamefont {O'Malley},
  \citenamefont {White}, \citenamefont {Barends}, \citenamefont {Kelly},
  \citenamefont {Lucero}, \citenamefont {Mariantoni}, \citenamefont {Megrant}
  \emph {et~al.}}]{13_Catch_Martinis}%
  \BibitemOpen
  \bibfield  {author} {\bibinfo {author} {\bibfnamefont {Y.}~\bibnamefont
  {Yin}}, \bibinfo {author} {\bibfnamefont {Y.}~\bibnamefont {Chen}}, \bibinfo
  {author} {\bibfnamefont {D.}~\bibnamefont {Sank}}, \bibinfo {author}
  {\bibfnamefont {P.~J.~J.}\ \bibnamefont {O'Malley}}, \bibinfo {author}
  {\bibfnamefont {T.~C.}\ \bibnamefont {White}}, \bibinfo {author}
  {\bibfnamefont {R.}~\bibnamefont {Barends}}, \bibinfo {author} {\bibfnamefont
  {J.}~\bibnamefont {Kelly}}, \bibinfo {author} {\bibfnamefont
  {E.}~\bibnamefont {Lucero}}, \bibinfo {author} {\bibfnamefont
  {M.}~\bibnamefont {Mariantoni}}, \bibinfo {author} {\bibfnamefont
  {A.}~\bibnamefont {Megrant}},  \emph {et~al.},\ }\href@noop {} {\bibfield
  {journal} {\bibinfo  {journal} {Physical Review Letters}\ }\textbf {\bibinfo
  {volume} {110}},\ \bibinfo {pages} {107001} (\bibinfo {year}
  {2013})}\BibitemShut {NoStop}%
\bibitem [{\citenamefont {Axline}\ \emph {et~al.}(2018)\citenamefont {Axline},
  \citenamefont {Burkhart}, \citenamefont {Pfaff}, \citenamefont {Zhang},
  \citenamefont {Chou}, \citenamefont {Campagne-Ibarcq}, \citenamefont
  {Reinhold}, \citenamefont {Frunzio}, \citenamefont {Girvin}, \citenamefont
  {Jiang}, \citenamefont {Devoret},\ and\ \citenamefont
  {Schoelkopf}}]{18_On-demand_Schoelkopf}%
  \BibitemOpen
  \bibfield  {author} {\bibinfo {author} {\bibfnamefont {C.~J.}\ \bibnamefont
  {Axline}}, \bibinfo {author} {\bibfnamefont {L.~D.}\ \bibnamefont
  {Burkhart}}, \bibinfo {author} {\bibfnamefont {W.}~\bibnamefont {Pfaff}},
  \bibinfo {author} {\bibfnamefont {M.}~\bibnamefont {Zhang}}, \bibinfo
  {author} {\bibfnamefont {K.}~\bibnamefont {Chou}}, \bibinfo {author}
  {\bibfnamefont {P.}~\bibnamefont {Campagne-Ibarcq}}, \bibinfo {author}
  {\bibfnamefont {P.}~\bibnamefont {Reinhold}}, \bibinfo {author}
  {\bibfnamefont {L.}~\bibnamefont {Frunzio}}, \bibinfo {author} {\bibfnamefont
  {S.~M.}\ \bibnamefont {Girvin}}, \bibinfo {author} {\bibfnamefont
  {L.}~\bibnamefont {Jiang}}, \bibinfo {author} {\bibfnamefont {M.~H.}\
  \bibnamefont {Devoret}}, \ and\ \bibinfo {author} {\bibfnamefont {R.~J.}\
  \bibnamefont {Schoelkopf}},\ }\href {\doibase 10.1038/s41567-018-0115-y}
  {\bibfield  {journal} {\bibinfo  {journal} {Nature Physics}\ }\textbf
  {\bibinfo {volume} {14}},\ \bibinfo {pages} {705} (\bibinfo {year}
  {2018})}\BibitemShut {NoStop}%
\bibitem [{\citenamefont {Kurpiers}\ \emph {et~al.}(2018)\citenamefont
  {Kurpiers}, \citenamefont {Magnard}, \citenamefont {Walter}, \citenamefont
  {Royer}, \citenamefont {Pechal}, \citenamefont {Heinsoo}, \citenamefont
  {Salath{\'e}}, \citenamefont {Akin}, \citenamefont {Storz}, \citenamefont
  {Besse} \emph {et~al.}}]{18_Deterministic_Wallraff}%
  \BibitemOpen
  \bibfield  {author} {\bibinfo {author} {\bibfnamefont {P.}~\bibnamefont
  {Kurpiers}}, \bibinfo {author} {\bibfnamefont {P.}~\bibnamefont {Magnard}},
  \bibinfo {author} {\bibfnamefont {T.}~\bibnamefont {Walter}}, \bibinfo
  {author} {\bibfnamefont {B.}~\bibnamefont {Royer}}, \bibinfo {author}
  {\bibfnamefont {M.}~\bibnamefont {Pechal}}, \bibinfo {author} {\bibfnamefont
  {J.}~\bibnamefont {Heinsoo}}, \bibinfo {author} {\bibfnamefont
  {Y.}~\bibnamefont {Salath{\'e}}}, \bibinfo {author} {\bibfnamefont
  {A.}~\bibnamefont {Akin}}, \bibinfo {author} {\bibfnamefont {S.}~\bibnamefont
  {Storz}}, \bibinfo {author} {\bibfnamefont {J.-C.}\ \bibnamefont {Besse}},
  \emph {et~al.},\ }\href@noop {} {\bibfield  {journal} {\bibinfo  {journal}
  {Nature}\ }\textbf {\bibinfo {volume} {558}},\ \bibinfo {pages} {264}
  (\bibinfo {year} {2018})}\BibitemShut {NoStop}%
\bibitem [{\citenamefont {Li}\ \emph {et~al.}(2018)\citenamefont {Li},
  \citenamefont {Ma}, \citenamefont {Han}, \citenamefont {Chen}, \citenamefont
  {Xu}, \citenamefont {Cai}, \citenamefont {Wang}, \citenamefont {Song},
  \citenamefont {Xue}, \citenamefont {Yin} \emph {et~al.}}]{18_Perfect_Sun}%
  \BibitemOpen
  \bibfield  {author} {\bibinfo {author} {\bibfnamefont {X.}~\bibnamefont
  {Li}}, \bibinfo {author} {\bibfnamefont {Y.}~\bibnamefont {Ma}}, \bibinfo
  {author} {\bibfnamefont {J.}~\bibnamefont {Han}}, \bibinfo {author}
  {\bibfnamefont {T.}~\bibnamefont {Chen}}, \bibinfo {author} {\bibfnamefont
  {Y.}~\bibnamefont {Xu}}, \bibinfo {author} {\bibfnamefont {W.}~\bibnamefont
  {Cai}}, \bibinfo {author} {\bibfnamefont {H.}~\bibnamefont {Wang}}, \bibinfo
  {author} {\bibfnamefont {Y.}~\bibnamefont {Song}}, \bibinfo {author}
  {\bibfnamefont {Z.-Y.}\ \bibnamefont {Xue}}, \bibinfo {author} {\bibfnamefont
  {Z.-q.}\ \bibnamefont {Yin}},  \emph {et~al.},\ }\href@noop {} {\bibfield
  {journal} {\bibinfo  {journal} {Physical Review Applied}\ }\textbf {\bibinfo
  {volume} {10}},\ \bibinfo {pages} {054009} (\bibinfo {year}
  {2018})}\BibitemShut {NoStop}%
\bibitem [{\citenamefont {Saeedi}\ \emph {et~al.}(2013)\citenamefont {Saeedi},
  \citenamefont {Simmons}, \citenamefont {Salvail}, \citenamefont {Dluhy},
  \citenamefont {Riemann}, \citenamefont {Abrosimov}, \citenamefont {Becker},
  \citenamefont {Pohl}, \citenamefont {Morton},\ and\ \citenamefont
  {Thewalt}}]{13_Room_Thewalt}%
  \BibitemOpen
  \bibfield  {author} {\bibinfo {author} {\bibfnamefont {K.}~\bibnamefont
  {Saeedi}}, \bibinfo {author} {\bibfnamefont {S.}~\bibnamefont {Simmons}},
  \bibinfo {author} {\bibfnamefont {J.~Z.}\ \bibnamefont {Salvail}}, \bibinfo
  {author} {\bibfnamefont {P.}~\bibnamefont {Dluhy}}, \bibinfo {author}
  {\bibfnamefont {H.}~\bibnamefont {Riemann}}, \bibinfo {author} {\bibfnamefont
  {N.~V.}\ \bibnamefont {Abrosimov}}, \bibinfo {author} {\bibfnamefont
  {P.}~\bibnamefont {Becker}}, \bibinfo {author} {\bibfnamefont {H.-J.}\
  \bibnamefont {Pohl}}, \bibinfo {author} {\bibfnamefont {J.~J.}\ \bibnamefont
  {Morton}}, \ and\ \bibinfo {author} {\bibfnamefont {M.~L.}\ \bibnamefont
  {Thewalt}},\ }\href@noop {} {\bibfield  {journal} {\bibinfo  {journal}
  {Science}\ }\textbf {\bibinfo {volume} {342}},\ \bibinfo {pages} {830}
  (\bibinfo {year} {2013})}\BibitemShut {NoStop}%
\bibitem [{\citenamefont {Zhong}\ \emph {et~al.}(2015)\citenamefont {Zhong},
  \citenamefont {Hedges}, \citenamefont {Ahlefeldt}, \citenamefont
  {Bartholomew}, \citenamefont {Beavan}, \citenamefont {Wittig}, \citenamefont
  {Longdell},\ and\ \citenamefont {Sellars}}]{15_Optically_Sellars}%
  \BibitemOpen
  \bibfield  {author} {\bibinfo {author} {\bibfnamefont {M.}~\bibnamefont
  {Zhong}}, \bibinfo {author} {\bibfnamefont {M.~P.}\ \bibnamefont {Hedges}},
  \bibinfo {author} {\bibfnamefont {R.~L.}\ \bibnamefont {Ahlefeldt}}, \bibinfo
  {author} {\bibfnamefont {J.~G.}\ \bibnamefont {Bartholomew}}, \bibinfo
  {author} {\bibfnamefont {S.~E.}\ \bibnamefont {Beavan}}, \bibinfo {author}
  {\bibfnamefont {S.~M.}\ \bibnamefont {Wittig}}, \bibinfo {author}
  {\bibfnamefont {J.~J.}\ \bibnamefont {Longdell}}, \ and\ \bibinfo {author}
  {\bibfnamefont {M.~J.}\ \bibnamefont {Sellars}},\ }\href@noop {} {\bibfield
  {journal} {\bibinfo  {journal} {Nature}\ }\textbf {\bibinfo {volume} {517}},\
  \bibinfo {pages} {177} (\bibinfo {year} {2015})}\BibitemShut {NoStop}%
\bibitem [{\citenamefont {Collins}\ \emph {et~al.}(2007)\citenamefont
  {Collins}, \citenamefont {Jenkins}, \citenamefont {Kuzmich},\ and\
  \citenamefont {Kennedy}}]{07_Multiplexed_Kennedy}%
  \BibitemOpen
  \bibfield  {author} {\bibinfo {author} {\bibfnamefont {O.}~\bibnamefont
  {Collins}}, \bibinfo {author} {\bibfnamefont {S.}~\bibnamefont {Jenkins}},
  \bibinfo {author} {\bibfnamefont {A.}~\bibnamefont {Kuzmich}}, \ and\
  \bibinfo {author} {\bibfnamefont {T.}~\bibnamefont {Kennedy}},\ }\href@noop
  {} {\bibfield  {journal} {\bibinfo  {journal} {Physical Review Letters}\
  }\textbf {\bibinfo {volume} {98}},\ \bibinfo {pages} {060502} (\bibinfo
  {year} {2007})}\BibitemShut {NoStop}%
\bibitem [{\citenamefont {Romanenko}\ \emph {et~al.}(2018)\citenamefont
  {Romanenko}, \citenamefont {Pilipenko}, \citenamefont {Zorzetti},
  \citenamefont {Frolov}, \citenamefont {Awida}, \citenamefont {Posen},\ and\
  \citenamefont {Grassellino}}]{18_3D_Grassellino}%
  \BibitemOpen
  \bibfield  {author} {\bibinfo {author} {\bibfnamefont {A.}~\bibnamefont
  {Romanenko}}, \bibinfo {author} {\bibfnamefont {R.}~\bibnamefont
  {Pilipenko}}, \bibinfo {author} {\bibfnamefont {S.}~\bibnamefont {Zorzetti}},
  \bibinfo {author} {\bibfnamefont {D.}~\bibnamefont {Frolov}}, \bibinfo
  {author} {\bibfnamefont {M.}~\bibnamefont {Awida}}, \bibinfo {author}
  {\bibfnamefont {S.}~\bibnamefont {Posen}}, \ and\ \bibinfo {author}
  {\bibfnamefont {A.}~\bibnamefont {Grassellino}},\ }\href@noop {} {\bibfield
  {journal} {\bibinfo  {journal} {arXiv:1810.03703}\ } (\bibinfo {year}
  {2018})}\BibitemShut {NoStop}%
\bibitem [{\citenamefont {D{\"u}r}\ and\ \citenamefont
  {Briegel}(2007)}]{07_Entanglement_Briegel}%
  \BibitemOpen
  \bibfield  {author} {\bibinfo {author} {\bibfnamefont {W.}~\bibnamefont
  {D{\"u}r}}\ and\ \bibinfo {author} {\bibfnamefont {H.~J.}\ \bibnamefont
  {Briegel}},\ }\href@noop {} {\bibfield  {journal} {\bibinfo  {journal}
  {Reports on Progress in Physics}\ }\textbf {\bibinfo {volume} {70}},\
  \bibinfo {pages} {1381} (\bibinfo {year} {2007})}\BibitemShut {NoStop}%
\bibitem [{\citenamefont {Gottesman}\ \emph {et~al.}(2001)\citenamefont
  {Gottesman}, \citenamefont {Kitaev},\ and\ \citenamefont
  {Preskill}}]{01_Encoding_Preskill}%
  \BibitemOpen
  \bibfield  {author} {\bibinfo {author} {\bibfnamefont {D.}~\bibnamefont
  {Gottesman}}, \bibinfo {author} {\bibfnamefont {A.}~\bibnamefont {Kitaev}}, \
  and\ \bibinfo {author} {\bibfnamefont {J.}~\bibnamefont {Preskill}},\
  }\href@noop {} {\bibfield  {journal} {\bibinfo  {journal} {Physical Review
  A}\ }\textbf {\bibinfo {volume} {64}},\ \bibinfo {pages} {012310} (\bibinfo
  {year} {2001})}\BibitemShut {NoStop}%
\bibitem [{\citenamefont {Leghtas}\ \emph {et~al.}(2013)\citenamefont
  {Leghtas}, \citenamefont {Kirchmair}, \citenamefont {Vlastakis},
  \citenamefont {Schoelkopf}, \citenamefont {Devoret},\ and\ \citenamefont
  {Mirrahimi}}]{13_Hardware_Mirrahimi}%
  \BibitemOpen
  \bibfield  {author} {\bibinfo {author} {\bibfnamefont {Z.}~\bibnamefont
  {Leghtas}}, \bibinfo {author} {\bibfnamefont {G.}~\bibnamefont {Kirchmair}},
  \bibinfo {author} {\bibfnamefont {B.}~\bibnamefont {Vlastakis}}, \bibinfo
  {author} {\bibfnamefont {R.~J.}\ \bibnamefont {Schoelkopf}}, \bibinfo
  {author} {\bibfnamefont {M.~H.}\ \bibnamefont {Devoret}}, \ and\ \bibinfo
  {author} {\bibfnamefont {M.}~\bibnamefont {Mirrahimi}},\ }\href@noop {}
  {\bibfield  {journal} {\bibinfo  {journal} {Physical Review Letters}\
  }\textbf {\bibinfo {volume} {111}},\ \bibinfo {pages} {120501} (\bibinfo
  {year} {2013})}\BibitemShut {NoStop}%
\bibitem [{\citenamefont {Mirrahimi}\ \emph {et~al.}(2014)\citenamefont
  {Mirrahimi}, \citenamefont {Leghtas}, \citenamefont {Albert}, \citenamefont
  {Touzard}, \citenamefont {Schoelkopf}, \citenamefont {Jiang},\ and\
  \citenamefont {Devoret}}]{14_Dynamically_Devoret}%
  \BibitemOpen
  \bibfield  {author} {\bibinfo {author} {\bibfnamefont {M.}~\bibnamefont
  {Mirrahimi}}, \bibinfo {author} {\bibfnamefont {Z.}~\bibnamefont {Leghtas}},
  \bibinfo {author} {\bibfnamefont {V.~V.}\ \bibnamefont {Albert}}, \bibinfo
  {author} {\bibfnamefont {S.}~\bibnamefont {Touzard}}, \bibinfo {author}
  {\bibfnamefont {R.~J.}\ \bibnamefont {Schoelkopf}}, \bibinfo {author}
  {\bibfnamefont {L.}~\bibnamefont {Jiang}}, \ and\ \bibinfo {author}
  {\bibfnamefont {M.~H.}\ \bibnamefont {Devoret}},\ }\href@noop {} {\bibfield
  {journal} {\bibinfo  {journal} {New Journal of Physics}\ }\textbf {\bibinfo
  {volume} {16}},\ \bibinfo {pages} {045014} (\bibinfo {year}
  {2014})}\BibitemShut {NoStop}%
\bibitem [{\citenamefont {Leghtas}\ \emph {et~al.}(2015)\citenamefont
  {Leghtas}, \citenamefont {Touzard}, \citenamefont {Pop}, \citenamefont {Kou},
  \citenamefont {Vlastakis}, \citenamefont {Petrenko}, \citenamefont {Sliwa},
  \citenamefont {Narla}, \citenamefont {Shankar}, \citenamefont {Hatridge}
  \emph {et~al.}}]{15_Confining_Devoret}%
  \BibitemOpen
  \bibfield  {author} {\bibinfo {author} {\bibfnamefont {Z.}~\bibnamefont
  {Leghtas}}, \bibinfo {author} {\bibfnamefont {S.}~\bibnamefont {Touzard}},
  \bibinfo {author} {\bibfnamefont {I.~M.}\ \bibnamefont {Pop}}, \bibinfo
  {author} {\bibfnamefont {A.}~\bibnamefont {Kou}}, \bibinfo {author}
  {\bibfnamefont {B.}~\bibnamefont {Vlastakis}}, \bibinfo {author}
  {\bibfnamefont {A.}~\bibnamefont {Petrenko}}, \bibinfo {author}
  {\bibfnamefont {K.~M.}\ \bibnamefont {Sliwa}}, \bibinfo {author}
  {\bibfnamefont {A.}~\bibnamefont {Narla}}, \bibinfo {author} {\bibfnamefont
  {S.}~\bibnamefont {Shankar}}, \bibinfo {author} {\bibfnamefont {M.~J.}\
  \bibnamefont {Hatridge}},  \emph {et~al.},\ }\href@noop {} {\bibfield
  {journal} {\bibinfo  {journal} {Science}\ }\textbf {\bibinfo {volume}
  {347}},\ \bibinfo {pages} {853} (\bibinfo {year} {2015})}\BibitemShut
  {NoStop}%
\bibitem [{\citenamefont {Ofek}\ \emph {et~al.}(2016)\citenamefont {Ofek},
  \citenamefont {Petrenko}, \citenamefont {Heeres}, \citenamefont {Reinhold},
  \citenamefont {Leghtas}, \citenamefont {Vlastakis}, \citenamefont {Liu},
  \citenamefont {Frunzio}, \citenamefont {Girvin}, \citenamefont {Jiang} \emph
  {et~al.}}]{16_Extending_Schoelkopf}%
  \BibitemOpen
  \bibfield  {author} {\bibinfo {author} {\bibfnamefont {N.}~\bibnamefont
  {Ofek}}, \bibinfo {author} {\bibfnamefont {A.}~\bibnamefont {Petrenko}},
  \bibinfo {author} {\bibfnamefont {R.}~\bibnamefont {Heeres}}, \bibinfo
  {author} {\bibfnamefont {P.}~\bibnamefont {Reinhold}}, \bibinfo {author}
  {\bibfnamefont {Z.}~\bibnamefont {Leghtas}}, \bibinfo {author} {\bibfnamefont
  {B.}~\bibnamefont {Vlastakis}}, \bibinfo {author} {\bibfnamefont
  {Y.}~\bibnamefont {Liu}}, \bibinfo {author} {\bibfnamefont {L.}~\bibnamefont
  {Frunzio}}, \bibinfo {author} {\bibfnamefont {S.}~\bibnamefont {Girvin}},
  \bibinfo {author} {\bibfnamefont {L.}~\bibnamefont {Jiang}},  \emph
  {et~al.},\ }\href@noop {} {\bibfield  {journal} {\bibinfo  {journal}
  {Nature}\ }\textbf {\bibinfo {volume} {536}},\ \bibinfo {pages} {441}
  (\bibinfo {year} {2016})}\BibitemShut {NoStop}%
\bibitem [{\citenamefont {Cohen}\ \emph {et~al.}(2017)\citenamefont {Cohen},
  \citenamefont {Smith}, \citenamefont {Devoret},\ and\ \citenamefont
  {Mirrahimi}}]{17_Degeneracy_Mirrahimi}%
  \BibitemOpen
  \bibfield  {author} {\bibinfo {author} {\bibfnamefont {J.}~\bibnamefont
  {Cohen}}, \bibinfo {author} {\bibfnamefont {W.~C.}\ \bibnamefont {Smith}},
  \bibinfo {author} {\bibfnamefont {M.~H.}\ \bibnamefont {Devoret}}, \ and\
  \bibinfo {author} {\bibfnamefont {M.}~\bibnamefont {Mirrahimi}},\ }\href
  {\doibase 10.1103/PhysRevLett.119.060503} {\bibfield  {journal} {\bibinfo
  {journal} {Physical Review Letters}\ }\textbf {\bibinfo {volume} {119}},\
  \bibinfo {pages} {060503} (\bibinfo {year} {2017})}\BibitemShut {NoStop}%
\bibitem [{\citenamefont {Touzard}\ \emph {et~al.}(2018)\citenamefont
  {Touzard}, \citenamefont {Grimm}, \citenamefont {Leghtas}, \citenamefont
  {Mundhada}, \citenamefont {Reinhold}, \citenamefont {Axline}, \citenamefont
  {Reagor}, \citenamefont {Chou}, \citenamefont {Blumoff}, \citenamefont
  {Sliwa}, \citenamefont {Shankar}, \citenamefont {Frunzio}, \citenamefont
  {Schoelkopf}, \citenamefont {Mirrahimi},\ and\ \citenamefont
  {Devoret}}]{18_Coherent_Devoret}%
  \BibitemOpen
  \bibfield  {author} {\bibinfo {author} {\bibfnamefont {S.}~\bibnamefont
  {Touzard}}, \bibinfo {author} {\bibfnamefont {A.}~\bibnamefont {Grimm}},
  \bibinfo {author} {\bibfnamefont {Z.}~\bibnamefont {Leghtas}}, \bibinfo
  {author} {\bibfnamefont {S.~O.}\ \bibnamefont {Mundhada}}, \bibinfo {author}
  {\bibfnamefont {P.}~\bibnamefont {Reinhold}}, \bibinfo {author}
  {\bibfnamefont {C.}~\bibnamefont {Axline}}, \bibinfo {author} {\bibfnamefont
  {M.}~\bibnamefont {Reagor}}, \bibinfo {author} {\bibfnamefont
  {K.}~\bibnamefont {Chou}}, \bibinfo {author} {\bibfnamefont {J.}~\bibnamefont
  {Blumoff}}, \bibinfo {author} {\bibfnamefont {K.~M.}\ \bibnamefont {Sliwa}},
  \bibinfo {author} {\bibfnamefont {S.}~\bibnamefont {Shankar}}, \bibinfo
  {author} {\bibfnamefont {L.}~\bibnamefont {Frunzio}}, \bibinfo {author}
  {\bibfnamefont {R.~J.}\ \bibnamefont {Schoelkopf}}, \bibinfo {author}
  {\bibfnamefont {M.}~\bibnamefont {Mirrahimi}}, \ and\ \bibinfo {author}
  {\bibfnamefont {M.~H.}\ \bibnamefont {Devoret}},\ }\href {\doibase
  10.1103/PhysRevX.8.021005} {\bibfield  {journal} {\bibinfo  {journal}
  {Physical Review X}\ }\textbf {\bibinfo {volume} {8}},\ \bibinfo {pages}
  {021005} (\bibinfo {year} {2018})}\BibitemShut {NoStop}%
\bibitem [{\citenamefont {Rosenblum}\ \emph {et~al.}(2018)\citenamefont
  {Rosenblum}, \citenamefont {Reinhold}, \citenamefont {Mirrahimi},
  \citenamefont {Jiang}, \citenamefont {Frunzio},\ and\ \citenamefont
  {Schoelkopf}}]{18_Fault_Schoelkopf}%
  \BibitemOpen
  \bibfield  {author} {\bibinfo {author} {\bibfnamefont {S.}~\bibnamefont
  {Rosenblum}}, \bibinfo {author} {\bibfnamefont {P.}~\bibnamefont {Reinhold}},
  \bibinfo {author} {\bibfnamefont {M.}~\bibnamefont {Mirrahimi}}, \bibinfo
  {author} {\bibfnamefont {L.}~\bibnamefont {Jiang}}, \bibinfo {author}
  {\bibfnamefont {L.}~\bibnamefont {Frunzio}}, \ and\ \bibinfo {author}
  {\bibfnamefont {R.~J.}\ \bibnamefont {Schoelkopf}},\ }\href {\doibase
  10.1126/science.aat3996} {\bibfield  {journal} {\bibinfo  {journal}
  {Science}\ }\textbf {\bibinfo {volume} {361}},\ \bibinfo {pages} {266}
  (\bibinfo {year} {2018})}\BibitemShut {NoStop}%
\bibitem [{\citenamefont {Puri}\ \emph {et~al.}(2018)\citenamefont {Puri},
  \citenamefont {Grimm}, \citenamefont {Campagne-Ibarcq}, \citenamefont
  {Eickbusch}, \citenamefont {Noh}, \citenamefont {Roberts}, \citenamefont
  {Jiang}, \citenamefont {Mirrahimi}, \citenamefont {Devoret},\ and\
  \citenamefont {Girvin}}]{18_Stabilized_Girvin}%
  \BibitemOpen
  \bibfield  {author} {\bibinfo {author} {\bibfnamefont {S.}~\bibnamefont
  {Puri}}, \bibinfo {author} {\bibfnamefont {A.}~\bibnamefont {Grimm}},
  \bibinfo {author} {\bibfnamefont {P.}~\bibnamefont {Campagne-Ibarcq}},
  \bibinfo {author} {\bibfnamefont {A.}~\bibnamefont {Eickbusch}}, \bibinfo
  {author} {\bibfnamefont {K.}~\bibnamefont {Noh}}, \bibinfo {author}
  {\bibfnamefont {G.}~\bibnamefont {Roberts}}, \bibinfo {author} {\bibfnamefont
  {L.}~\bibnamefont {Jiang}}, \bibinfo {author} {\bibfnamefont
  {M.}~\bibnamefont {Mirrahimi}}, \bibinfo {author} {\bibfnamefont {M.~H.}\
  \bibnamefont {Devoret}}, \ and\ \bibinfo {author} {\bibfnamefont {S.~M.}\
  \bibnamefont {Girvin}},\ }\href@noop {} {\bibfield  {journal} {\bibinfo
  {journal} {arXiv:1807.09334}\ } (\bibinfo {year} {2018})}\BibitemShut
  {NoStop}%
\bibitem [{\citenamefont {Sun}\ \emph {et~al.}(2014)\citenamefont {Sun},
  \citenamefont {Petrenko}, \citenamefont {Leghtas}, \citenamefont {Vlastakis},
  \citenamefont {Kirchmair}, \citenamefont {Sliwa}, \citenamefont {Narla},
  \citenamefont {Hatridge}, \citenamefont {Shankar}, \citenamefont {Blumoff}
  \emph {et~al.}}]{14_Tracking_Schoelkopf}%
  \BibitemOpen
  \bibfield  {author} {\bibinfo {author} {\bibfnamefont {L.}~\bibnamefont
  {Sun}}, \bibinfo {author} {\bibfnamefont {A.}~\bibnamefont {Petrenko}},
  \bibinfo {author} {\bibfnamefont {Z.}~\bibnamefont {Leghtas}}, \bibinfo
  {author} {\bibfnamefont {B.}~\bibnamefont {Vlastakis}}, \bibinfo {author}
  {\bibfnamefont {G.}~\bibnamefont {Kirchmair}}, \bibinfo {author}
  {\bibfnamefont {K.}~\bibnamefont {Sliwa}}, \bibinfo {author} {\bibfnamefont
  {A.}~\bibnamefont {Narla}}, \bibinfo {author} {\bibfnamefont
  {M.}~\bibnamefont {Hatridge}}, \bibinfo {author} {\bibfnamefont
  {S.}~\bibnamefont {Shankar}}, \bibinfo {author} {\bibfnamefont
  {J.}~\bibnamefont {Blumoff}},  \emph {et~al.},\ }\href@noop {} {\bibfield
  {journal} {\bibinfo  {journal} {Nature}\ }\textbf {\bibinfo {volume} {511}},\
  \bibinfo {pages} {444} (\bibinfo {year} {2014})}\BibitemShut {NoStop}%
\bibitem [{\citenamefont {Wang}\ \emph {et~al.}(2016)\citenamefont {Wang},
  \citenamefont {Gao}, \citenamefont {Reinhold}, \citenamefont {Heeres},
  \citenamefont {Ofek}, \citenamefont {Chou}, \citenamefont {Axline},
  \citenamefont {Reagor}, \citenamefont {Blumoff}, \citenamefont {Sliwa} \emph
  {et~al.}}]{16_Schrodinger_Schoelkopf}%
  \BibitemOpen
  \bibfield  {author} {\bibinfo {author} {\bibfnamefont {C.}~\bibnamefont
  {Wang}}, \bibinfo {author} {\bibfnamefont {Y.~Y.}\ \bibnamefont {Gao}},
  \bibinfo {author} {\bibfnamefont {P.}~\bibnamefont {Reinhold}}, \bibinfo
  {author} {\bibfnamefont {R.~W.}\ \bibnamefont {Heeres}}, \bibinfo {author}
  {\bibfnamefont {N.}~\bibnamefont {Ofek}}, \bibinfo {author} {\bibfnamefont
  {K.}~\bibnamefont {Chou}}, \bibinfo {author} {\bibfnamefont {C.}~\bibnamefont
  {Axline}}, \bibinfo {author} {\bibfnamefont {M.}~\bibnamefont {Reagor}},
  \bibinfo {author} {\bibfnamefont {J.}~\bibnamefont {Blumoff}}, \bibinfo
  {author} {\bibfnamefont {K.}~\bibnamefont {Sliwa}},  \emph {et~al.},\
  }\href@noop {} {\bibfield  {journal} {\bibinfo  {journal} {Science}\ }\textbf
  {\bibinfo {volume} {352}},\ \bibinfo {pages} {1087} (\bibinfo {year}
  {2016})}\BibitemShut {NoStop}%
\bibitem [{\citenamefont {Michael}\ \emph {et~al.}(2016)\citenamefont
  {Michael}, \citenamefont {Silveri}, \citenamefont {Brierley}, \citenamefont
  {Albert}, \citenamefont {Salmilehto}, \citenamefont {Jiang},\ and\
  \citenamefont {Girvin}}]{16_New_Girvin}%
  \BibitemOpen
  \bibfield  {author} {\bibinfo {author} {\bibfnamefont {M.~H.}\ \bibnamefont
  {Michael}}, \bibinfo {author} {\bibfnamefont {M.}~\bibnamefont {Silveri}},
  \bibinfo {author} {\bibfnamefont {R.~T.}\ \bibnamefont {Brierley}}, \bibinfo
  {author} {\bibfnamefont {V.~V.}\ \bibnamefont {Albert}}, \bibinfo {author}
  {\bibfnamefont {J.}~\bibnamefont {Salmilehto}}, \bibinfo {author}
  {\bibfnamefont {L.}~\bibnamefont {Jiang}}, \ and\ \bibinfo {author}
  {\bibfnamefont {S.~M.}\ \bibnamefont {Girvin}},\ }\href {\doibase
  10.1103/PhysRevX.6.031006} {\bibfield  {journal} {\bibinfo  {journal}
  {Physical Review X}\ }\textbf {\bibinfo {volume} {6}},\ \bibinfo {pages}
  {031006} (\bibinfo {year} {2016})}\BibitemShut {NoStop}%
\bibitem [{\citenamefont {Albert}\ \emph {et~al.}(2018)\citenamefont {Albert},
  \citenamefont {Noh}, \citenamefont {Duivenvoorden}, \citenamefont {Young},
  \citenamefont {Brierley}, \citenamefont {Reinhold}, \citenamefont {Vuillot},
  \citenamefont {Li}, \citenamefont {Shen}, \citenamefont {Girvin} \emph
  {et~al.}}]{18_Performance_Jiang}%
  \BibitemOpen
  \bibfield  {author} {\bibinfo {author} {\bibfnamefont {V.~V.}\ \bibnamefont
  {Albert}}, \bibinfo {author} {\bibfnamefont {K.}~\bibnamefont {Noh}},
  \bibinfo {author} {\bibfnamefont {K.}~\bibnamefont {Duivenvoorden}}, \bibinfo
  {author} {\bibfnamefont {D.~J.}\ \bibnamefont {Young}}, \bibinfo {author}
  {\bibfnamefont {R.}~\bibnamefont {Brierley}}, \bibinfo {author}
  {\bibfnamefont {P.}~\bibnamefont {Reinhold}}, \bibinfo {author}
  {\bibfnamefont {C.}~\bibnamefont {Vuillot}}, \bibinfo {author} {\bibfnamefont
  {L.}~\bibnamefont {Li}}, \bibinfo {author} {\bibfnamefont {C.}~\bibnamefont
  {Shen}}, \bibinfo {author} {\bibfnamefont {S.}~\bibnamefont {Girvin}},  \emph
  {et~al.},\ }\href@noop {} {\bibfield  {journal} {\bibinfo  {journal}
  {Physical Review A}\ }\textbf {\bibinfo {volume} {97}},\ \bibinfo {pages}
  {032346} (\bibinfo {year} {2018})}\BibitemShut {NoStop}%
\end{thebibliography}%
\end{document}